\documentclass[trackchanges]{aastex701}

\usepackage{makecell}
\usepackage{graphicx}
\usepackage{graphicx}   
\usepackage{subcaption} 
\usepackage{graphicx}	
\usepackage{amsmath}	

\usepackage{graphicx}
\usepackage{subcaption}
\usepackage{longtable}

\usepackage{amsmath}
\usepackage{array}
\usepackage{caption}

\begin{document}

\title[Physics of 1 keV line in X-ray binaries]{Physics of 1 keV line in X-ray binaries}

\author{Priyanka Chakraborty}
\affiliation{University of Arkansas, Fayetteville, AR, USA}
\affiliation{Center for Astrophysics $\vert$ Harvard \& Smithsonian, Cambridge, MA, USA}
\email{priyanka.chakraborty@cfa.harvard.edu}

\author{Gary Ferland}
\affiliation{University of Kentucky, Lexington, KY, USA}\email{}

\author{Andrew Fabian}
\affiliation{Institute of Astronomy, Madingley Road, Cambridge CB3 0HA, UK}\email{}

\author{Arnab Sarkar}
\affiliation{Kavli Institute for Astrophysics and Space Research, Massachusetts Institute of Technology, Cambridge, MA, USA}\email{}

\author{Renee Ludlam}
\affiliation{Department of Physics and Astronomy, Wayne State University, 666 W. Hancock St., Detroit, MI 48201, USA}\email{}

\author{Stefano Bianchi}
\affiliation{Dipartimento di Matematica e Fisica, Universit\`a degli Studi Roma Tre, Via della Vasca Navale 84, I-00146 Roma, Italy}\email{}

\author{Hayden Hall}
\affiliation{Department of Physics and Astronomy, Wayne State University, 666 W. Hancock St., Detroit, MI 48201, USA}\email{}

\author{Peter Kosec}
\affiliation{Center for Astrophysics $\vert$ Harvard \& Smithsonian, Cambridge, MA, USA}\email{}

\begin{abstract}
  X-ray binaries (XRBs) often exhibit spectral residuals in the 0.5 to 2 keV range, known as the ``1 keV residual/1 keV feature", with variable centroid and intensity across different systems. Yet a comprehensive scientific explanation of the variability of the 1 keV feature has remained largely elusive.  In this paper, we explain for the first time the origin and variability of the 1 keV feature in XRBs using the spectral synthesis code \textsc{Cloudy}. We constructed line blends for the emission and absorption lines and study the variability of these blends with ionization parameters, temperature, and column density. 
We conducted a sample study involving five XRBs including two ultraluminous X-ray sources (ULXs): NGC 247 ULX-1, NGC 1313 X-1, a  binary X-ray pulsar : Hercules X-1, and two typical  low-mass X-ray binaries (LMXBs): Cygnus X-2, and Serpens X-1. Our analysis establishes a self-consistent framework explaining the variability of the 1 keV spectral feature, attributing its diversity to differences in spectral energy distribution, ionization parameter, temperature, column density, and disk reflection properties. This framework provides a comprehensive explanation for the observed 1 keV feature across these diverse XRB systems, offering insights into the underlying physical mechanisms at play.

\end{abstract}

\section{Introduction}

X-ray binaries (XRBs) feature a compact object (white dwarf, neutron star, or black hole) accreting from a stellar companion, harnessing gravitational energy release for power. 
XRBs often exhibit spectral
features, commonly referred to as 
the ``1 keV residuals,''
within the energy range of 
0.5 to 2 keV. 
One of the earliest and well-studied instances of this feature was reported in Hercules X-1,
where a broad emission structure between 0.8 and 1.4 keV was initially attributed to unresolved iron (Fe) L-shell emission \citep{1994nhxr.conf..419M, 1997A&A...327..215O}. Modeling efforts varied, with \citet{1994nhxr.conf..419M} employing two narrow Gaussian components, while \citet{1997A&A...327..215O} utilized a single broad Gaussian. \citet{1982ApJ...262..301M} proposed that this feature results from the reprocessing of the hard X-ray beam by the inner disk at the magnetospheric boundary of the neutron star. Over the decades, similar spectral features have been ubiquitously observed in a wide range of X-ray binaries, including ultraluminous X-ray sources (ULXs) \citep{2006MNRAS.368..397S,2006MNRAS.371.1877R, 2015MNRAS.454.3134M, 2020MNRAS.492.4646P},  X-ray pulsars \citep{2023A&A...669A..38M, 2023A&A...674L...9S}, and typical low-mass X-ray binaries (LMXBs) \citep{2000ApJS..131..571A, 2016A&A...596A..21I, 2018ApJ...858L...5L, 2021MNRAS.508.3569K}.  Interestingly, both the centroid and intensity of this feature exhibit variations, not only across different binaries but also over time within the same binary \citep{2002ApJ...579..411P, 2020MNRAS.494.6012W}. Despite scattered attempts to model the 1 keV feature using various approaches, a comprehensive explanation for the origin and variability of these residuals has proven challenging to ascertain. In this paper, we present a comprehensive model aimed at understanding the physics of this feature. We investigate various scenarios, including collisionally-ionized, photoionized, and reflective origins of the feature, while also examining the variability of their centroids and intensity.
We demonstrate the application of our model on a diverse array of X-ray binaries, including ULXs (NGC 1313 X-1 and NGC 247 ULX-1), typical  LMXBs (Cygnus X-2 and Serpens X-1), and X-ray pulsars (Hercules X-1).

\section{Sample selection overview}

\textit{NGC 1313 X-1}:
NGC 1313 X-1 is one of  the ULX in the starburst galaxy NGC 1313, exhibiting isotropic luminosity of  $L_{X}$ = 2.0 $\times$ 10$^{40}$ erg s$^{-1}$ (0.2-10.0 keV) and unusual spectral variability \citep{2020MNRAS.494.6012W}.  
Strong rest-frame emission lines around 1 keV have been detected along with  blue-shifted  atomic absorption features through XMM-Newtn observations of NGC 1313 X-1 \citep{2016Natur.533...64P}, which was later confirmed to vary in time \citep{2020MNRAS.492.4646P}.

\textit{NGC 247 ULX-1}:
NGC 247 ULX-1 was first identified  as an ULX from a short XMM-Newton observation \citep{2006ApJ...649..730W}, while a second XMM-Newton observation confirmed its soft nature \citep{2011ApJ...737...87J}. Later, \citet{2016ApJ...831..117F} found that NGC 247 ULX-1 switches between
supersoft ultraluminous (SSUL) and soft ultraluminous (SUL) regime \citep{2016ApJ...831..117F}.  \citet{2021MNRAS.505.5058P} analyzed XMM-Newton observations of NGC 247 ULX-1 for eight observations taken between 2019 and 2020 and detected 1 keV residuals that seemed to vary with each observation, both the centroid and intensity, and identified the need for a complex atomic model for correctly explaining the residuals. 

\textit{Hercules X-1}:
Hercules X-1, commonly referred to as Her X-1,  is one of the extensively observed accreting binary X-ray pulsars with intriguing properties.
The X-ray emissions from the system exhibit three distinct timescales: the 1.24s X-ray pulsation \citep{1975Natur.253..250F},  the 1.7-day orbital period \citep{1972ApJ...174L.143T}, and a super-orbital 35-day cycle \citep{1985SSRv...40..347O}. The  35-day cycle of high and low flux state is believed to be induced by the precession of a twisted accretion disc. The first 1 keV residual in Hercules X-1 was detected  through BeppoSAX observations and exhibited variability \citep{1997A&A...327..215O, 2001A&A...375..922O}. A more prominent detection of 1 keV residual in both emission and absorption was achieved through a series of XMM-Newton observations in August 2020 \citep{2022ApJ...936..185K}.

\textit{Serpens X-1}:
The luminous persistent LMXB Serpens X-1, abbreviated as Ser X-1, was first identified in 1965 \citep{1965Sci...147..394B}.  Type I X-ray bursts were detected from the source in
1976, confirming the presence of its neutron star accretor \citep{1976IAUC.3000....5S}.
Serpens X-1 offers a unique chance to identify numerous reflection features owing to the minimal amount of absorbing material along its line of sight. The first reflection feature detected was the  Fe K complex through XMM-Newton observations \citep{2007ApJ...664L.103B}, with  detection of the Fe L complex reflection feature partly contributing to the 1 keV residual in Serpens X-1 \citep{2018ApJ...858L...5L, 2025ApJ...980..234H} and emission residuals between 0.5- 0.9 keV which remained unaccounted for.

\textit{Cygnus X-2}:
Cygnus X-2, often referred to as Cyg X-2, has long been recognized as a typical low-mass X-ray binary hosting a neutron star with a relatively extended orbital period of P = 9.84 days \citep{1979ApJ...231..539C, 1998ApJ...498L.141S}. Over the years, several studies have identified a 1 keV feature in Cyg X-2  through Einstein,  BeppoSAX, and Suzaku observatories \citep{1986ApJ...307..698V, 2002A&A...386..535D, 2010ApJ...720..205C},  with a recent study by \citet{2022ApJ...927..112L} detecting a prominent emission feature at 1 keV with minimal absorption using \textit{NICER} observations carried out in 2019.

We used the spectral synthesis code \textsc{Cloudy} \citep{2017RMxAA..53..385F} for all our simulations.
We adopted a cosmology of
$H_0$ = 70 km s$^{-1}$ Mpc$^{-1}$, 
$\Omega_{\Lambda}$ = 0.7.
Unless otherwise stated, all 
reported error bars 
are at 90\% confidence level.

\begin{figure}
\centering
\includegraphics[width=0.5\textwidth]{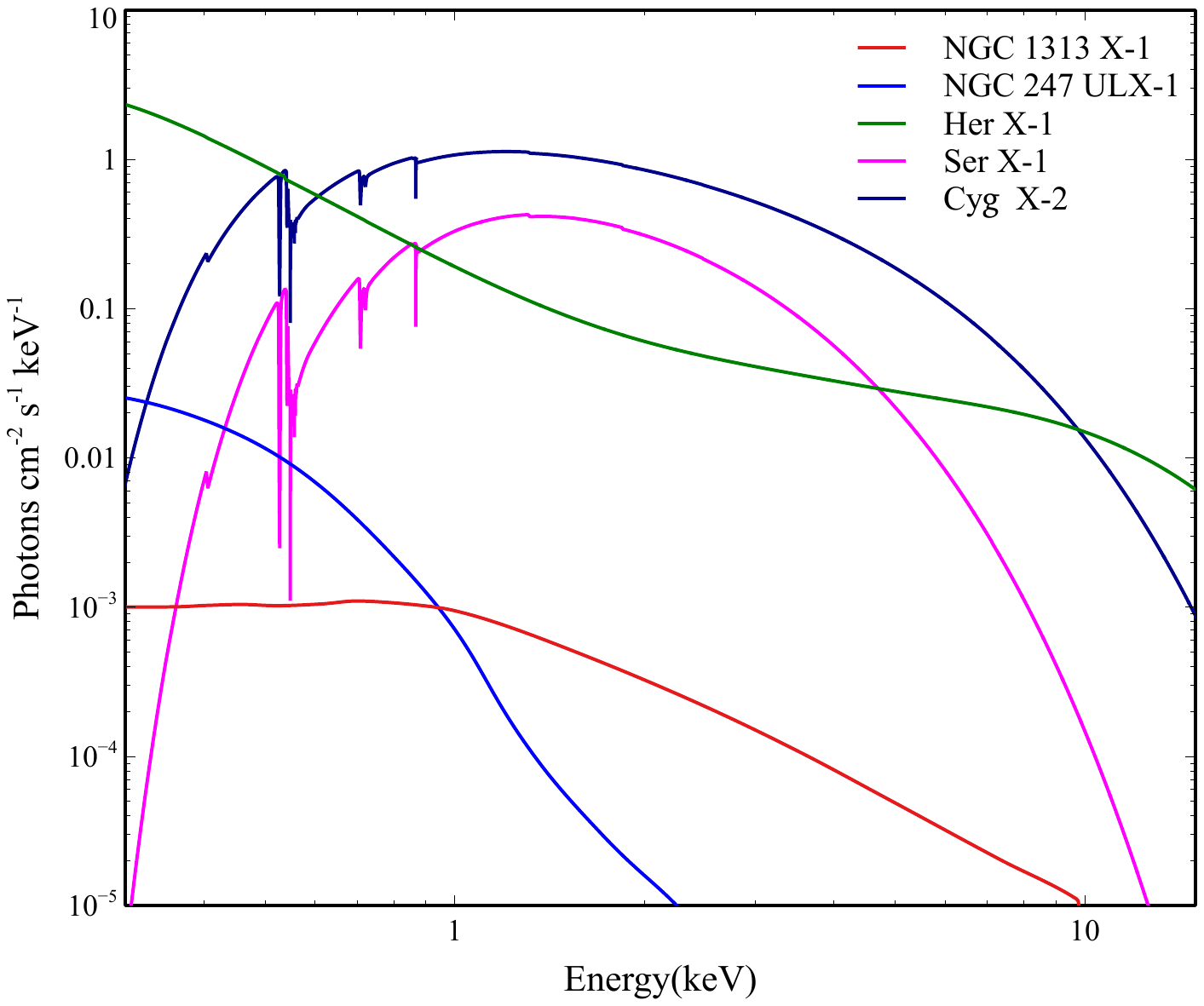}
    \caption{Spectral energy distributions (SEDs) used in the \textsc{Cloudy} simulation for each X-ray binary, with their sources detailed in Section \ref{data}.
   }
\label{fig:sed}
\end{figure}

\section{Data Reduction}\label{data}

\subsection{NGC 1313 X-1 and NGC 247 ULX-1}

NGC 1313 X-1 was observed
using XMM-Newton
during nine distinct periods 
spanning from 2014 to 2017.
These observations utilized
both the European Photon Imaging Camera \citep[EPIC, ][]{2001A&A...365L..27T} and 
Reflection Grating Spectrometer  \citep[RGS, ][]{2001A&A...365L...7D}  onboard XMM-Newton. 
Five out of the nine observations were reported to be in the intermediate-bright state \citep{2020MNRAS.492.4646P}. To gain a detailed view of the 1 keV emission line complex, we utilized the RGS observations of these  five brightest observations (obs ids: 0742590301, 0803990101, 0803990201,
0803990501, 0803990601), with exposure time totaling 592 ks. In our analysis of NGC 1313 X-1, we used the Spectral Energy Distribution (SED) from \citet{2020MNRAS.492.4646P},  along with a hydrogen density of  =$10^{10} \, \rm cm^{-3}$ from the same source.


NGC 247 ULX-1 was observed 
with XMM-Newton
between 2019 and 2020,
with the
exposure time totaling 886 ks.
For this work,
we utilized all eight RGS
observations
of NGC 247 ULX-1, identified by observation IDs: 0844860101, 0844860201, 0844860301, 
0844860401, 0844860501,
0844860601, 0844860701, and 0844860801. 
We adopted the SED derived from \citet{2021MNRAS.505.5058P} and a hydrogen density of $10^{10} \, \rm cm^{-3}$. 
For both targets, 
The RGS data reduction was 
performed using the \texttt{rgsproc} 
pipeline from the 
Science Analysis System (\textsc{sas})
software package. 
The periods affected by
contamination from
solar flare 
were filtered using 
\texttt{rgsfilter},
by selecting background-quiescent 
intervals in the lightcurves of 
the RGS 1 and 2 from CCD number 9.
A count rate threshold of $\sim$ 0.2 c/s
were applied. 
The clean exposure times are listed
in Table \ref{t:o1}.

For each source,
we extracted the first and 
second ordered
RGS spectra from a cross-dispersion 
region of 0.8$\arcmin$.
Both cross-dispersion
regions were centered on the 
respective source coordinates--
RA = 49.58$^{\circ}$, 
DEC = -66.48 $^{\circ}$ for 
NGC~1313 X-1 and 
RA = 11.76$^{\circ}$,
DEC = -20.79$^{\circ}$ for 
NGC~247 ULX-1.
The background spectra for each
source were extracted by selecting
photons that extended beyond 
the 98\% boundary of the source 
point-spread-function.
We note that the background regions
never overlapped the bright
source under consideration.
The {\tt rgsproc} tool
also generates the 
corresponding response files
for RGS 1 and 2.
For each source,
we used the \texttt{rgscombine}
tool to stack RGS 1 and 
2
first-order source 
spectra, background
spectra
and corresponding response files
from 
all the obs ids.

\subsection{Hercules X-1}

We utilized the RGS spectrum 
of Hercules X-1 
during its Main High state \citep[precession phase $0.063-0.098$, ][]{2021MNRAS.508.3569K}
from an observation with a 126.5 ks raw exposure (Obs Id: 0865440401).
The data were obtained
from the XSA archive and
reduced using a standard pipeline with \textsc{sas} v19, 
\textsc{caldb} as of 2021 June. 
The RGS data reduction was performed using the \texttt{rgsproc} task. A cross-dispersion region with a width of = 1.4$\arcmin$  was adopted, centered at RA =254.46$^{\circ}$ and DEC = 35.34$^{\circ}$.  We only used periods during which Her X-1 is in a high flux state to create the RGS spectra. For this reason, we excluded a time interval at the beginning of the observation during which Her X-1 was in eclipse, and a time interval at the end of the observation during which the source exhibited absorption dipping. We identified no background flares during this exposure. The remaining net RGS exposure time (per detector) is 104 ks.  We only used 1st-order RGS data and stacked the RGS 1 and 2 spectra using the \texttt{rgscombine} routine.  Energies below 0.56 keV were excluded from the fitting due to the detection of a potential instrumental residual in the $0.52-0.56$ keV energy band, as detailed in \citet{2021MNRAS.508.3569K}, Appendix B. We referenced the SED specific to this observation from \citet{2022ApJ...936..185K} and a hydrogen density of $ 10^{10} \, \rm cm^{-3}$.




\subsection{Cyg X-2 and Ser X-1}\label{cygser}
 
 For Cyg X-2,  we utilized the 12.1 ks observation with observation ID 2631010201 from \textit{NICER} for observations carried out in 2019. The data reduction was carried out utilizing CALDB version 20210427, and the source and background spectrum for the NICER observation were created using the  `3C50' tool \citep{2022AJ....163..130R}. The detailed data reduction method is described in \citet{2022ApJ...927..112L} under horizontal branch (HB), obs 3. The SED specific to this observation was taken from \citet{2022ApJ...927..112L} as well as a hydrogen density of  $10^{15} \, \rm cm^{-3}$ from the same source.
For Ser X-1, our analysis incorporated \textit{NICER} data collected from July to November 2017, across ObsIDs 1050320101–1050320113, totaling an exposure of 39.9 ks.
The data for Ser X-1 was processed using the same methods as Cyg X-2, using the updated NICER CALDB version xti20221001. The SED for this observation was adopted from \citet{2018ApJ...858L...5L} and the hydrogen density was assumed to be  $10^{15} \, \rm cm^{-3}$.

\begin{table*}
\centering{
\caption{\label{t:o1} Observation log detailing the XRB sample utilized in this paper.}}
\begin{tabular}{ccccc}
\hline
Target & Instrument & Observation ID  & Start Date/Time  &  Exposure(ksec)\\
\hline

NGC 1313 X-1 & \textit{XMM-Newton}/RGS &0742590301& 2014-07-05 22:37:13  & 62.99 \\
 && 0803990101 & 2017-06-14 20:40:21  & 137.10 \\
 && 0803990201& 2017-06-20 21:06:00  & 133.80\\
 && 0803990501 & 2017-12-07 10:28:21 & 128.90 \\
 && 0803990601 & 2017-12-09 10:20:09  & 128.90 \\
\hline
NGC 247 ULX-1 & \textit{XMM-Newton}/RGS &0844860101 & 2019-12-03 12:41:51 & 115.50 \\
 && 0844860201 & 2019-12-09 12:48:29 & 118.60 \\
 && 0844860301 & 2019-12-09 12:48:29 & 119.60 \\
 && 0844860401& 2020-01-02 10:43:35 & 120.00 \\
 && 0844860501 & 2020-01-04 10:52:05 & 119.00 \\
 && 0844860601 & 2020-01-06 04:36:08 & 115.00 \\
 && 0844860701 & 2020-01-06 04:36:08 & 115.00 \\
 && 0844860801 & 2020-01-12 03:59:00 & 63.20\\
\hline
Her X-1 &  \textit{XMM-Newton}/RGS &0865440401 & 2020-08-12 16:06:27 & 126.5\\
\hline
 Ser X-1 & \textit{NICER} & 1050320101–1050320113 &  2017-7 - 2017-11 & 39.9 \\

\hline
Cyg X-2 & \textit{NICER} & 2631010201 & 2019-09-12 02:09:44 & 12.1\\
 
 \hline
\end{tabular}
\end{table*}

\section{Physics of the blend: \textsc{Cloudy} models}

In this work,
we conducted a detailed investigation into the origins of the 1 keV lines, exploring photoionized equilibrium (PIE), collisional ionization equilibrium (CIE), and reflection mechanisms.
Previous observations using
high-resolution spectra
have shown that
 XRBs often exhibit strong emission residuals in the energy range
between 0.6 and 1.4 keV
\citep{2014MNRAS.438L..51M, 2021AstBu..76....6F}. 
To model the emission feature\footnote{\textsc{Cloudy} self-consistently computes emission features by modeling thousands of spectral lines, incorporating spectral energy distribution, ionization state and gradients, chemical composition, level populations, and optical depth effects. Optical depth, proportional to the hydrogen column density \citep{2020ApJ...901...69C}, governs absorption and scattering, directly shaping the observed emission features \citep{2022ApJ...935...70C}. This dependence on column density is explicitly accounted for through optical depth effects, ensuring a rigorous and self-consistent treatment in the simulations.}
within this full energy range
in \textsc{Cloudy},
we constructed an
emission line blend \footnote{Within \textsc{Cloudy}, line blends represent the total intensity from all emission lines within a defined energy range. This method offers the most direct way to quantify the combined intensity of the 1 keV feature (0.6–1.4 keV) and investigate how it varies with SED and key physical parameters such as ionization parameter, temperature, and column density. Additionally, by analyzing the combined intensity of emission lines on either side of 1 keV—specifically in the 0.6–1.0 keV and 1.0–1.4 keV bands—it is possible to track how the 1 keV feature shifts leftward or rightward in response to changing physical conditions.}, which we 
call Em$_{\rm blend}$ 
including  emission lines in the
0.6--1.4 keV energy band. 
Further, we constructed two 
emission line 
blends by including all the emission lines
between 0.6 to 1 keV and 1 keV to 
1.4 keV, which we call Em$_{\rm left}$ 
and  Em$_{\rm right}$, respectively. 
This decomposition above and below 1 keV allows for a precise tracking of the centroid shift of the 1 keV feature under varying conditions, providing a more accurate characterization than Gaussian modeling. Broad Gaussians, commonly used in lower-resolution studies, often oversimplify complex line structures. In contrast, the line blend approach preserves the intrinsic shape of the feature, which frequently deviates from a true Gaussian profile, ensuring a more physically robust representation.

Previous observations
also showed that two
absorption features on 
either sides of the emission feature
are ubiquitous among XRB spectra, particularly ULXs
\citep{2016Natur.533...64P, 2019ApJ...883...44W}. 
We, therefore,
constructed two
absorption line blends in \textsc{Cloudy}: 
one comprising 
all the absorption lines located in the 
left of the 1 keV energy, 
designated as Abs$_{\rm left}$, and 
another encompassing all the absorption
lines
at the right of the 1 keV energy, 
denoted as Abs$_{\rm right}$.
The full absorption line blend, 
including all absorption lines within 
the energy band 0.5--2.0 keV, 
is referred to as Abs$_{\rm blend}$. These absorption blends collectively represent the summed equivalent widths of the individual spectral lines, providing a quantitative measure of the total absorption strength relative to the continuum.

XRBs might display noticeable reflection characteristics, often characterized by the presence of the Fe L complex \citep{2010ApJ...720..205C, 2018ApJ...858L...5L} . We created two reflection line blends within the energy interval 0.5-2.0 keV using \textsc{Cloudy}: one containing all the reflection lines to the left of the 1 keV energy, labeled as Reflect$_{\rm left}$, and another comprising all the reflection lines positioned to the right of the 1 keV energy, designated as Reflect$_{\rm right}$. The reflection line blend covering the energy range of 0.5 to 2.0 keV was designated as Reflect$_{\rm blend}$. Table \ref{t:p2} in the appendix lists all the lines 
from Em$_{\rm blend}$,  Abs$_{\rm blend}$, and Reflect$_{\rm blend}$, along with instructions for creating these blends within \textsc{Cloudy} in the appendix.

The emission, absorption, and reflection line blends
are expected to vary among different
systems and a successful model should
predict these variation to make it 
useful in understanding the evolution
of 1 keV feature in different systems.
To study how the shift of
the centroid and intensity of the
1 keV emission feature vary with the
different physical properties of 
XRBs, 
we  
examined the variation of  
Em$_{\rm left}$/Em$_{\rm right}$ 
and Em$_{\rm blend}$ with 
a range of ionization 
parameters ($\xi$) and 
hydrogen column densities ($N$$_{\rm H}$) 
in a PIE plasma.
For CIE plasma, we investigated
the variation Em$_{\rm left}$/Em$_{\rm right}$ 
and Em$_{\rm blend}$ with
a range of temperatures ($T$) 
and hydrogen column densities ($N$$_{\rm H}$).
Similarly, for absorption line blends,
we examined the variation
in Abs$_{\rm left}$/Abs$_{\rm right}$ 
and Abs$_{\rm blend}$ with $\xi$ and
$N$$_{\rm H}$ to trace the shifts in the
centroid and intensity of the
1 keV absorption feature. 
Since the absorption features have  been consistently linked to a PIE wind origin \citep{2008ApJ...677.1233C, 2017AN....338..234P}, we did not investigate the 
variation of  the absorption features (Abs$_{\rm blend}$
and Abs$_{\rm left}$/Abs$_{\rm right}$)
with $T$ or $N$$_{\rm H}$ for  CIE plasma. 

Note that, SEDs also play a crucial role in shaping emission and absorption lines in photoionized plasma by influencing the ionization state of the gas. The rate at which atoms or ions are ionized by incoming photons is given by the photoionization rate:

\noindent

\[
\Gamma_n = \int_{\nu_o}^{\infty} \frac{4\pi J_{\nu}}{h\nu} \alpha_{\nu} \, d\nu \quad \text{[s}^{-1}\text{]}
\]

where \( J_{\nu} \) [erg cm\(^{-2}\) s\(^{-1}\) sr\(^{-1}\) Hz\(^{-1}\)] is the mean intensity of the incident radiation per unit frequency and per unit solid angle, and \( \alpha_{\nu} \) [cm\(^2\)] is the photoionization cross-section for the atom or ion at photon energy \( h\nu \) \citep{2021ApJ...912...26C}. The shape of the SED determines the distribution of photon energies, directly affecting \( \Gamma_n \) and, consequently, the ionization balance. Variations in SEDs directly shape spectral lines and, collectively, the 1 keV feature.

We utilized target-specific SEDs from Section \ref{data} as radiation sources in \textsc{Cloudy}, which were then used in Section \ref{sec4} for spectral fitting.  In the following subsections on PIE and reflection features, we illustrate how line blends vary by highlighting two extreme cases: NGC 247 ULX-1, which exhibits the softest SED, and Cyg X-2, which exhibits the hardest SED.

\begin{figure*}
\centering
\begin{tabular}{cc}
    \includegraphics[width=0.3\textwidth]{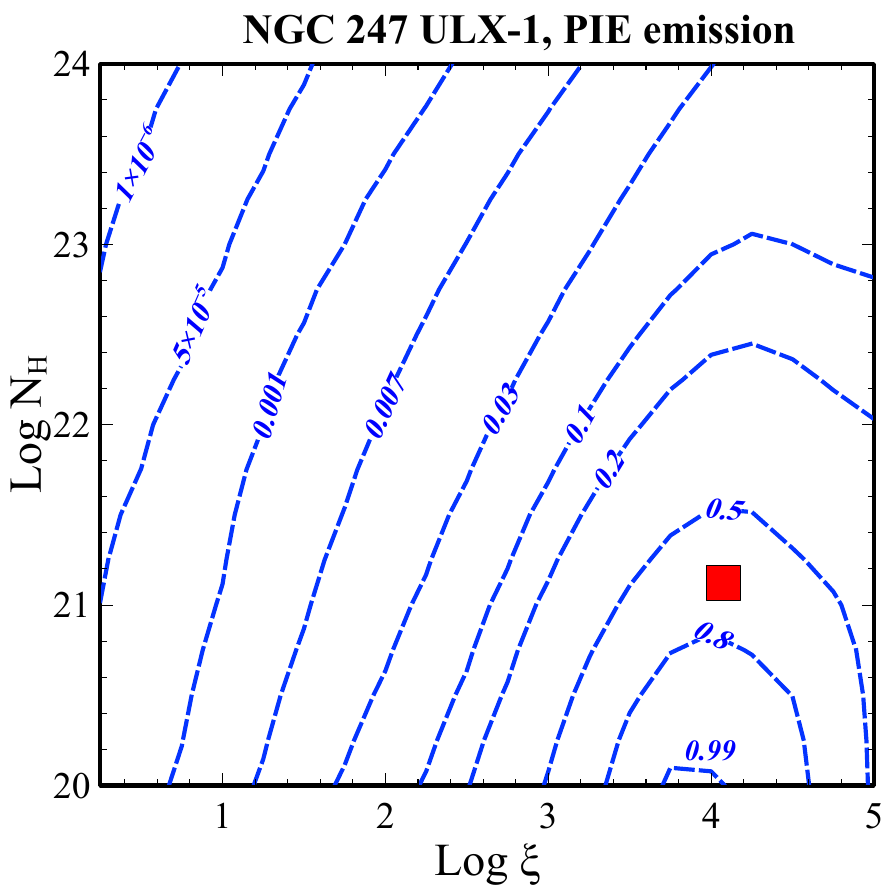} &\includegraphics[width=0.3\textwidth]{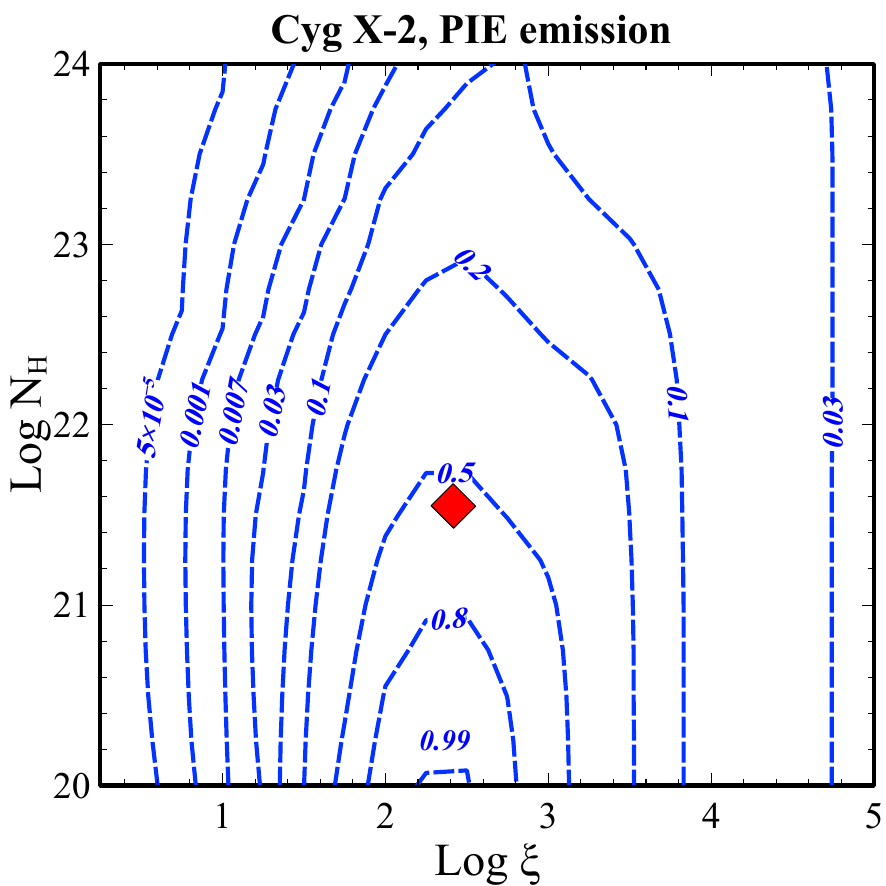} \\

     \end{tabular}

\caption{{
Variation of the normalized 1 keV emission blend emissivity (dimensionless) with log $\xi$ and log $N$$_{\rm H}$ for PIE plasma in   NGC 247 ULX-1 and  Cyg X-2.  
The  square and inverted square red data points indicate the best-fit parameters for the PIE emission model for  NGC 247 ULX-1 and Cyg X-2, respectively, as also discussed in Section \ref{sec4}.}
}
\label{fig:em1}
\end{figure*}

\begin{figure*}
\centering

\begin{tabular}{cc}
    \includegraphics[width=0.3\textwidth]{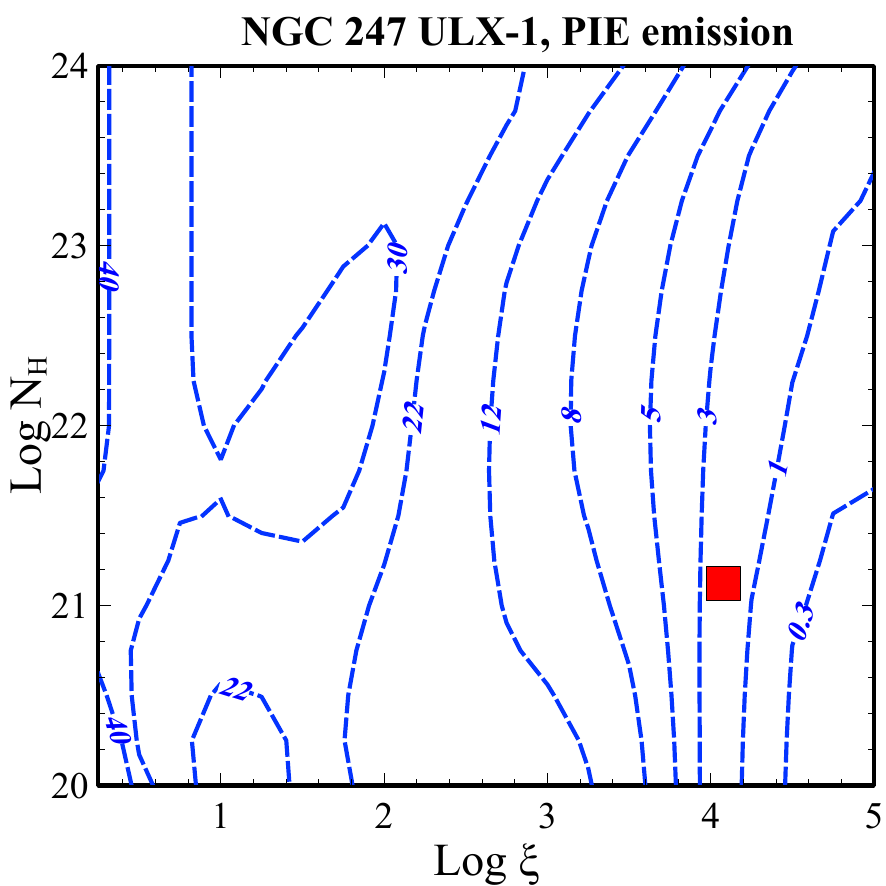} &\includegraphics[width=0.3\textwidth]{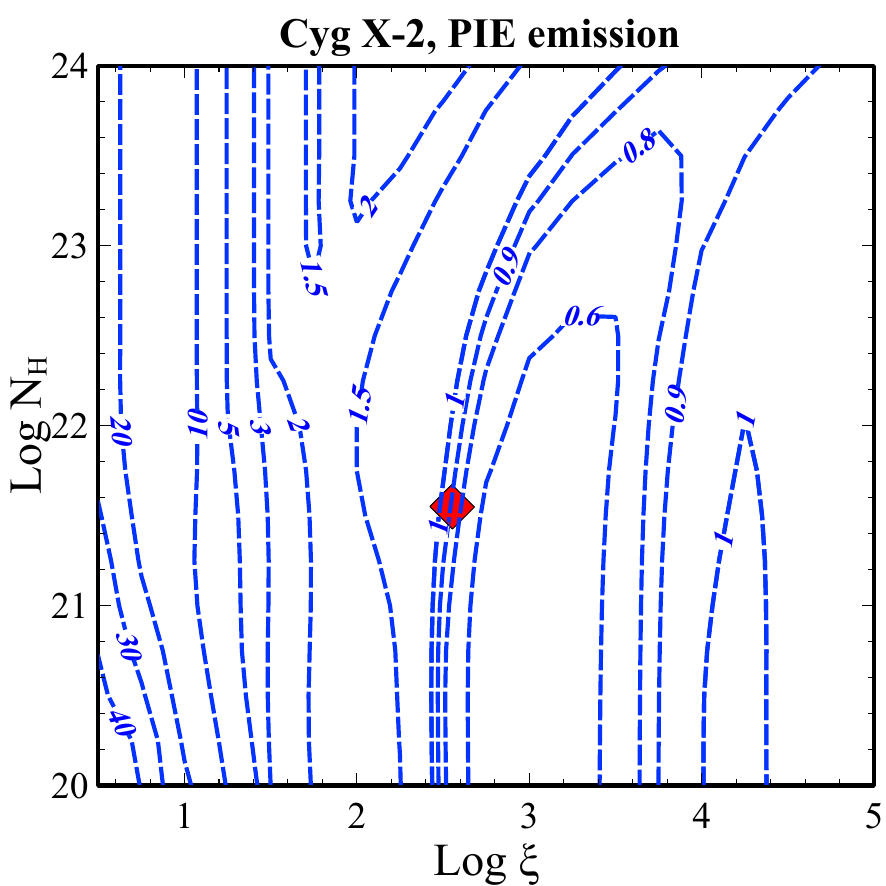} \\

     \end{tabular}

\caption{
 Variation of  Em$_{\rm left}$/Em$_{\rm right}$ ratio with log $\xi$ and log $N$$_{\rm H}$ for PIE plasma in  NGC 247 ULX-1 and Cyg X-2.
The  square and inverted square red data points indicate the best-fit parameters for the PIE emission model for  NGC 247 ULX-1 and Cyg X-2, respectively, as discussed in Section \ref{sec4}.
}
\label{fig:em2}
\end{figure*}

\begin{figure*}
\centering

     \begin{tabular}{cc}
   \includegraphics[width=0.3\textwidth]{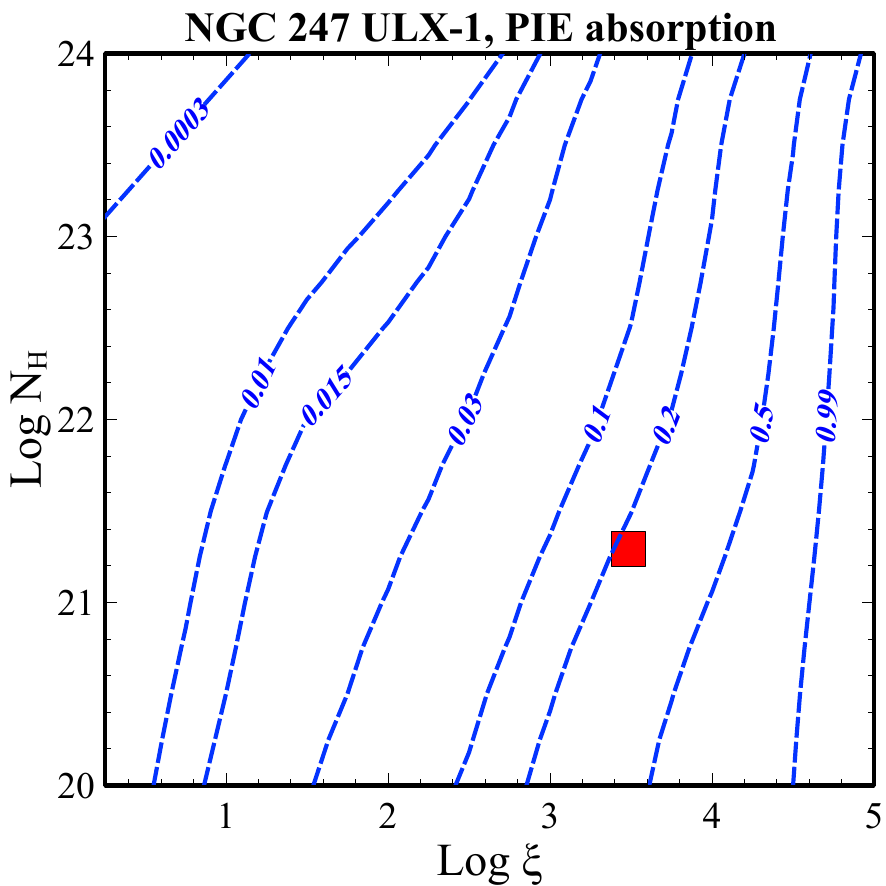} &
   \includegraphics[width=0.3\textwidth]{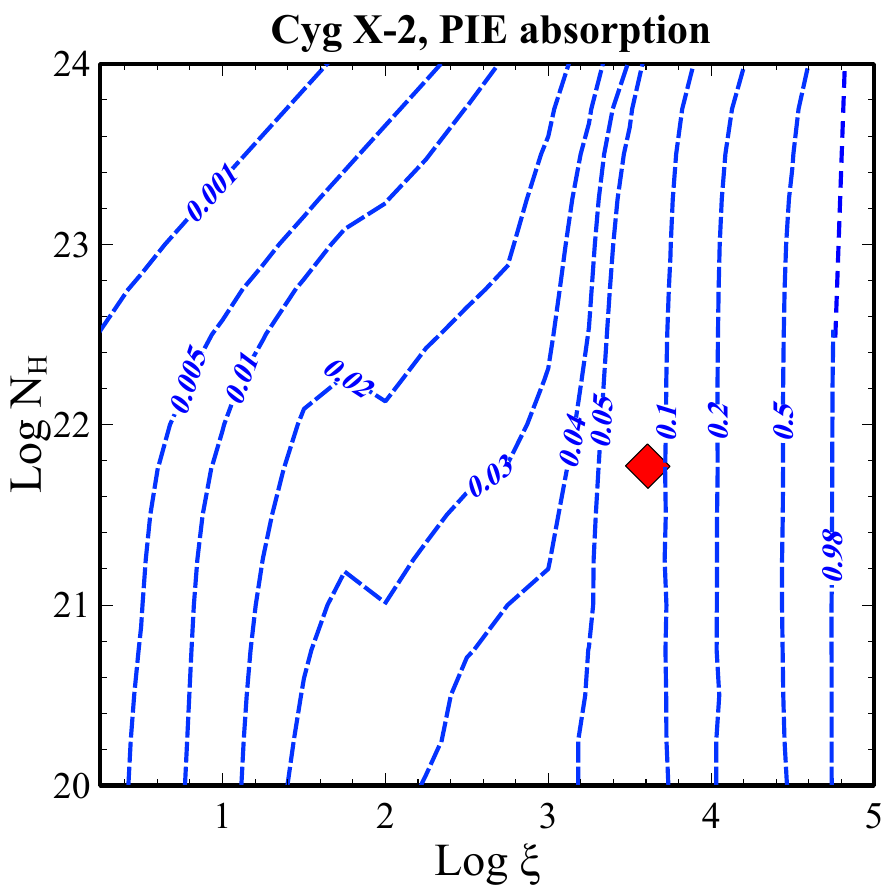}  \\

     \end{tabular}

\caption{
 Variation of the normalized 1 keV absorption blend emissivity with log $\xi$ and log $N$$_{\rm H}$ for PIE plasma in   NGC 247 ULX-1 and Cyg X-2.   The square and inverted square data points represent the best-fit parameters for  NGC 247 ULX-1 and Cyg X-2, respectively, for the PIE absorption model detailed in Section \ref{sec4}.
}
\label{fig:abs1}
\end{figure*}

\begin{figure*}
\centering

\begin{tabular}{cc}
     \includegraphics[width=0.3\textwidth]{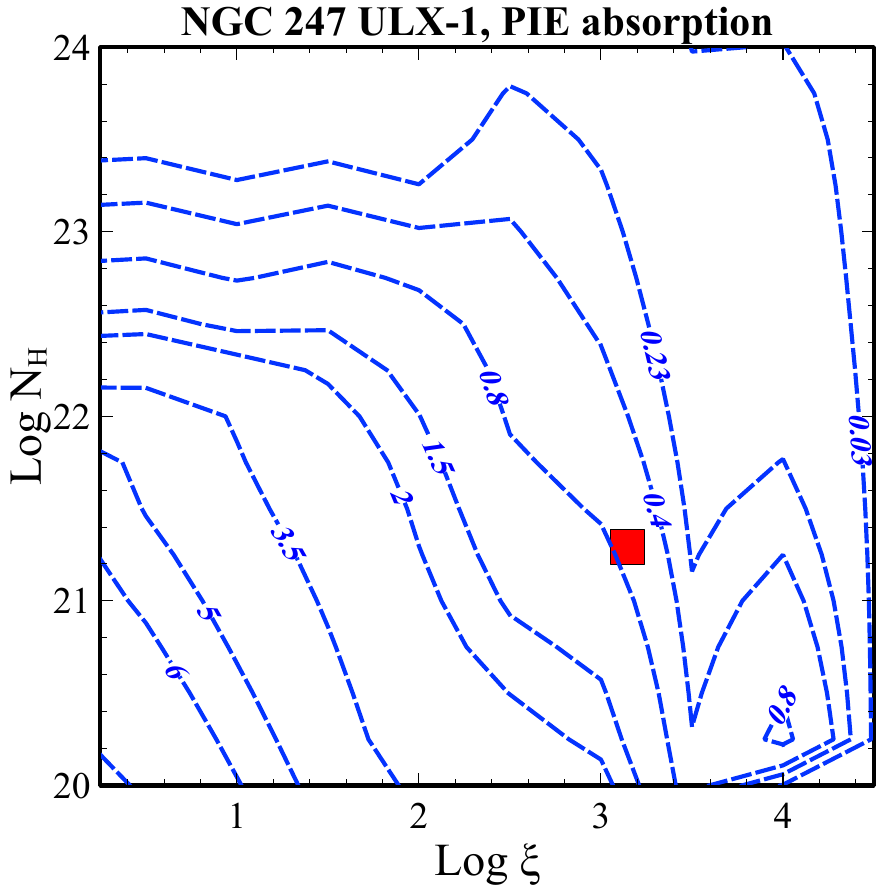}&
     \includegraphics[width=0.3\textwidth]{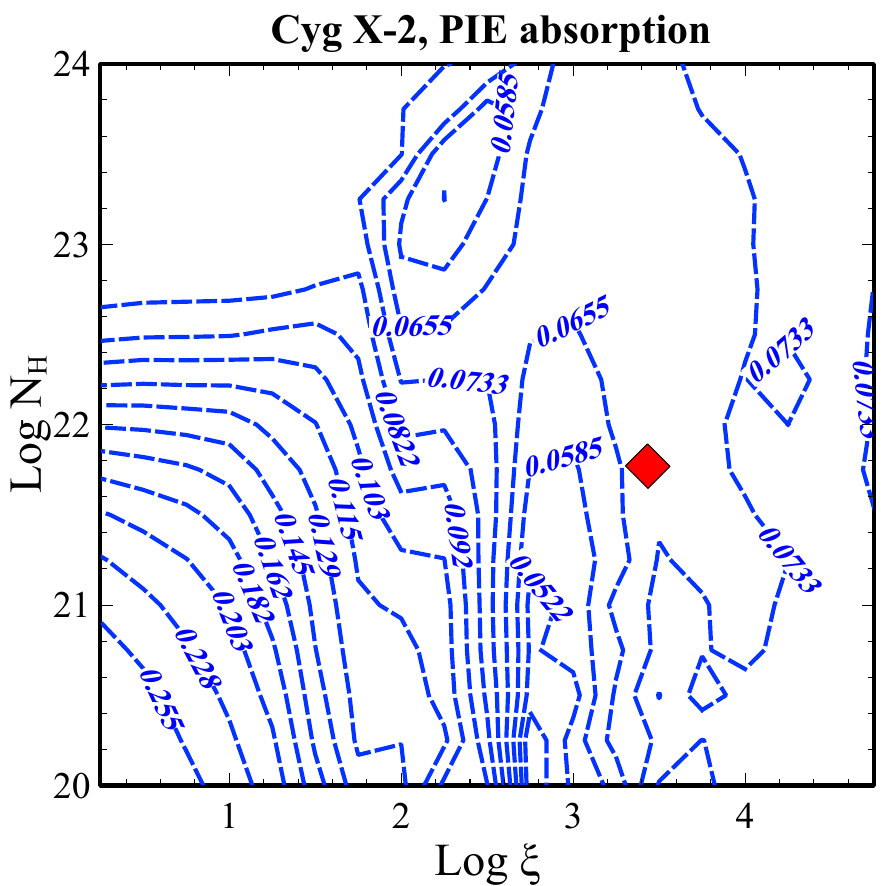}  \\

     \end{tabular}

\caption{
 Variation of  Abs$_{\rm left}$/Abs$_{\rm right}$ ratio with log $\xi$ and log $N$$_{\rm H}$ for PIE plasma in   NGC 247 ULX-1 and Cyg X-2.   The red  square and inverted square data points represent the best-fit parameters for NGC 247 ULX-1 and Cyg X-2, respectively, for the PIE absorption model detailed in Section \ref{sec4}.
}
\label{fig:abs2}
\end{figure*}

\begin{figure*}
\centering
\begin{tabular}{cc}
    \includegraphics[width=0.3\textwidth]{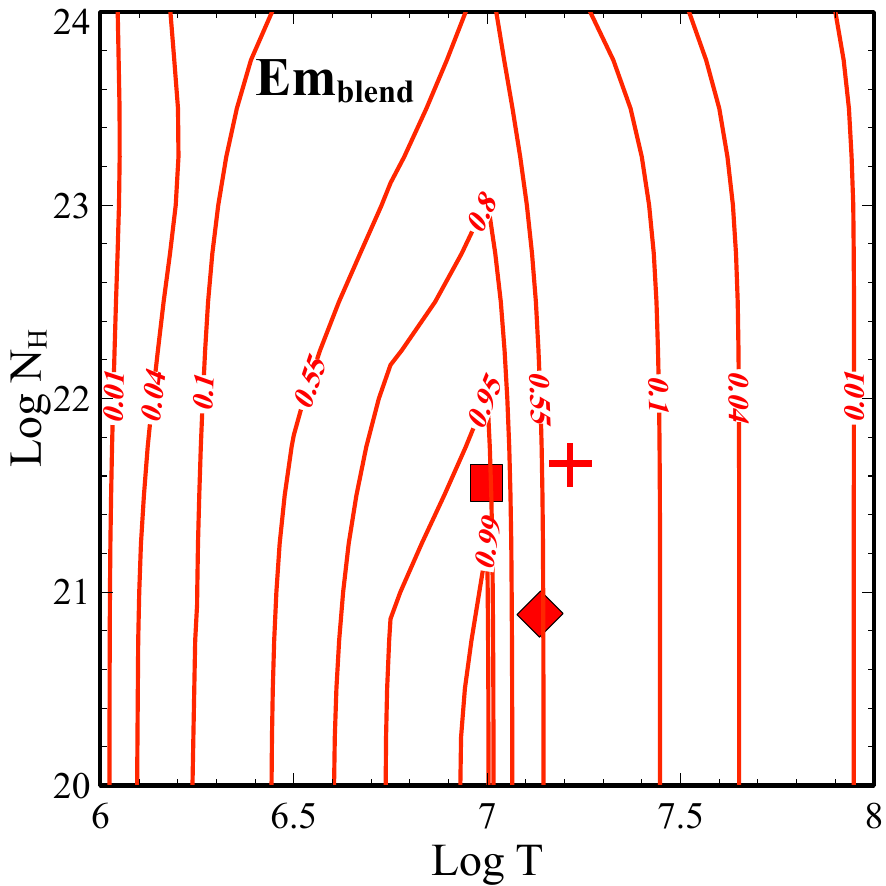} &
    \includegraphics[width=0.3\textwidth]{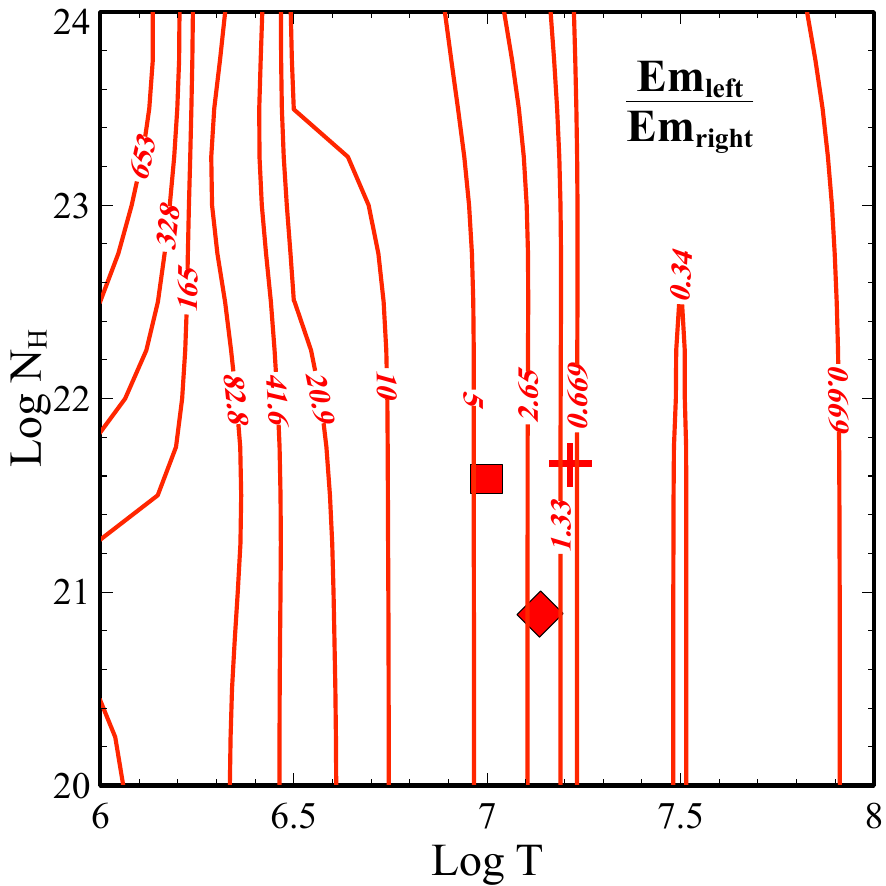}\\
    \end{tabular}
\caption{
Left panel: Variation of the normalized 1 keV emission line blend emissivity in CIE plasma. Right panel: Variation of  Em$_{\rm left}$/Em$_{\rm right}$ ratio with log $T$ and log $N$$_{\rm H}$ for CIE plasma. The red square, cross, and inverted square data points represent the best-fit parameters for NGC 247 ULX-1, Her X-1, and Cyg X-2, respectively, for the CIE emission model detailed in Section \ref{sec4}.}
\label{fig:col}
\end{figure*}

\begin{figure*}
\centering

    \begin{tabular}{cc}
    \includegraphics[width=0.4\textwidth]{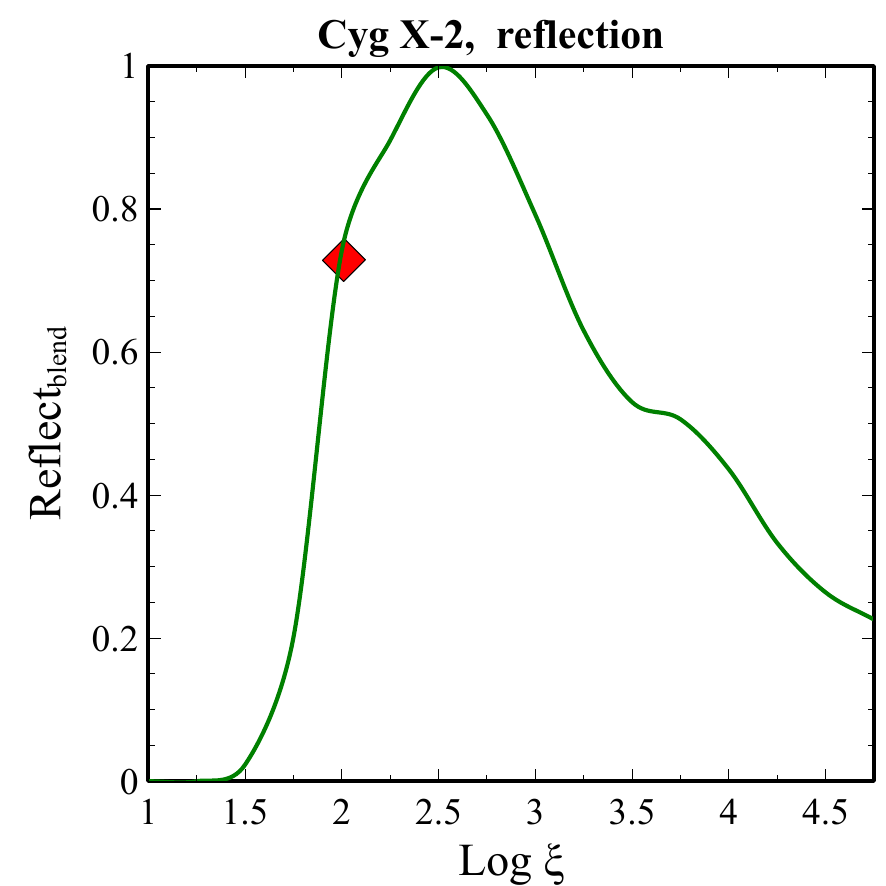} &
    \includegraphics[width=0.4\textwidth]{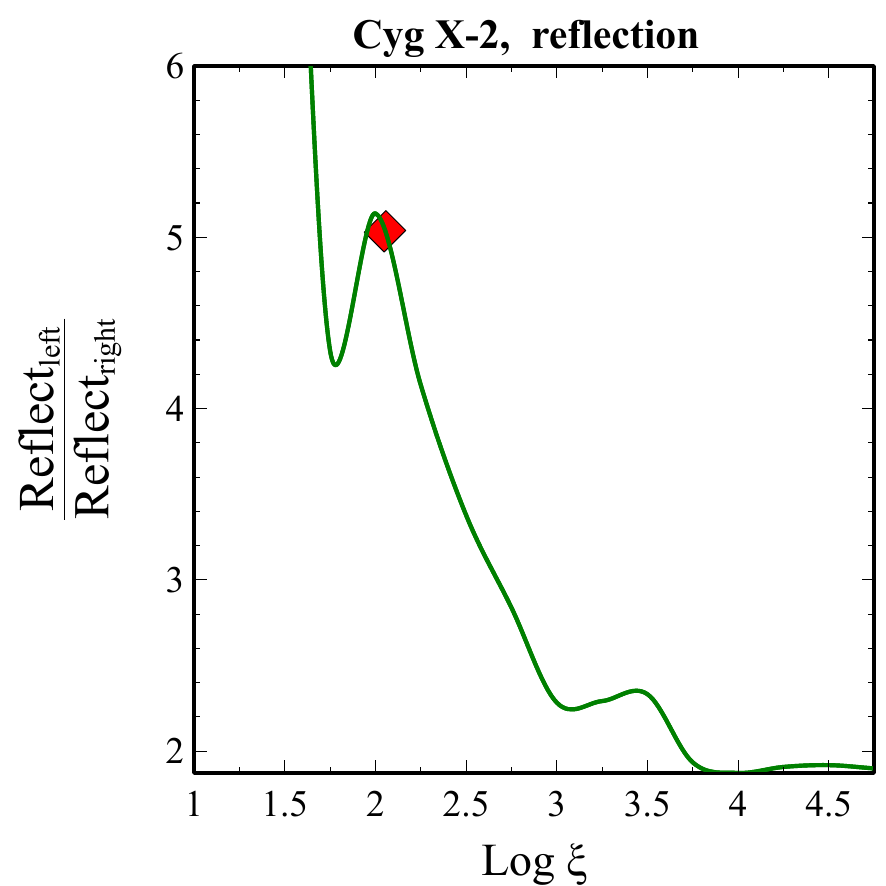}\\
    \end{tabular}
\caption{
Left panels: Variation of the Reflect$_{\rm blend}$ with log $\xi$ for Cyg X-2.
Right panels: Variation of  Reflect$_{\rm left}$/Reflect$_{\rm right}$ ratio with log $\xi$ for Cyg X-2. The  inverted square data points represent the best-fit log $\xi$ values for Cyg X-2 as detailed in Section \ref{sec4}.}
\label{fig:ref}
\end{figure*}

\subsection {Emission lines from PIE}\label{emisspie}

Our investigation focused on understanding
how Em$_{\rm blend}$ and  
Em$_{\rm left}$/Em$_{\rm right}$
vary with two key parameters:
ionization parameter ($\xi$) and
hydrogen column density ($N$$_{\rm H}$).
For each XRB under consideration,
we constructed a logarithmic grid 
for $\xi$, spanning log $\xi$ values 
from
0.5 to 5.0, with increments of 0.1 dex.
Similarly, we created
a logarithmic
grid for $N$$_{\rm H}$, ranging between 
log $N$$_{\rm H}$ values
of 20 to 24, also in 0.1 dex increments.
In total, this resulted in 
2000 distinct grids for our analysis.

Figure \ref{fig:em1} and \ref{fig:em2}
illustrates the observed changes in
Em$_{\rm blend}$ and  
Em$_{\rm left}$/Em$_{\rm right}$ highlighting two extreme cases: NGC 247 ULX-1 with the softest SED, and Cyg X-2 with the hardest SED.
In Figure
\ref{fig:em1},
we displayed how Em$_{\rm blend}$ 
varies
with 
log $\xi$ and log $N$$_{\rm H}$. 
This variation in Em$_{\rm blend}$ represents
the alterations in
the intensity of the 1 keV emission
feature.  The normalization of \( \text{Em}_{\text{blend}} \) is executed as follows:  each value of \( \text{Em}_{\text{blend}} \) is scaled by dividing it by the maximum \( \text{Em}_{\text{blend}} \) observed within the same parameter space. This method ensures that \( \text{Em}_{\text{blend}} \) values are dimensionless, ranging from 0 to 1, thereby standardizing the data for comparative analysis across different parameter sets. Specifically, 
as Em$_{\rm blend}$ approaches  1 , 
it indicates a stronger 1 keV 
emission feature, 
whereas Em$_{\rm blend}$ $\ll$ 1 
signifies a diminished 1 keV
emission feature.
In  Figure \ref{fig:em2}, we depicted
the variation of 
Em$_{\rm left}$/Em$_{\rm right}$ with
log $\xi$ 
and 
log $N$$_{\rm H}$, 
which characterizes the shift in 
the centroid
of the 1 keV emission feature. 
A ratio above 1 indicates
a leftward shift
of the centroid
of the 1 keV feature, while a 
ratio below 1 suggests a rightward shift.
This captures
the variation of the centroid shift of 
the 1 keV emission feature.

\subsection {Absorption  lines from PIE plasma:}
For absorption line generation from PIE plasma, we used target-specific SEDs outlined in Section \ref{data}.
We  
examined the variation of 
Abs$_{\rm blend}$
and Abs$_{\rm left}$/Abs$_{\rm right}$
in response to variations in $\xi$ 
and $N$$_{\rm H}$.
We used the same logarithmic 
grids described earlier in  section \ref{emisspie} to explore these 
variations.

In  Figure 
\ref{fig:abs1}, we presented
the variation 
of Abs$_{\rm blend}$ with log $\xi$ and 
log $N$$_{\rm H}$ for NGC 247 ULX-1 and  Cyg X-2. 
These variations in 
Abs$_{\rm blend}$ illustrate changes 
in the strength of the 1 keV 
absorption feature.
When Abs$_{\rm blend}$ approaches 1, 
it indicates that strong 1 keV 
absorption residuals are expected, 
while a value of Abs$_{\rm blend}$ $\ll$ 1 
suggests weaker absorption residuals.

In Figure 
\ref{fig:abs2}, we showed
the variation of 
Abs$_{\rm left}$/Abs$_{\rm right}$ with 
log $\xi$  and 
log $N$$_{\rm H}$. 
This ratio, 
Abs$_{\rm left}$/Abs$_{\rm right}$, 
characterizes 
the shift 
in the location of the 1 keV 
absorption residuals.  
A ratio exceeding 
1
signifies
a more pronounced absorption residual to the
left of the central emission feature, 
whereas a ratio below 1 indicates a more 
pronounced dip on the right. 
This ratio captures the variability
in the centroid of the absorption residuals.

\subsection {Emission lines from CIE plasma:}

For studying the emission from a CIE plasma, we investigated the variation in
Em$_{\rm blend}$ and Em$_{\rm left}$/Em$_{\rm right}$  
with temperature ($T$), 
and hydrogen column density ($N$$_{\rm H}$). 
A logarithmic grid was created for
$T$ with log $T$ values ranging from 
6.0 to 8.0, and for log $N$$_{\rm H}$
with values ranging from 20 to 24, 
both with increments of 0.1 dex, resulting in a total of 800 grids.
For CIE plasma, Em$_{\rm blend}$ and 
Em$_{\rm left}$/Em$_{\rm right}$ for
NGC 247 ULX-1, NGC
1313 X-1, Her X-1, Cyg X-2, and Ser X-1 overlaps
as the only difference between emission
from these systems stems from different 
electron density,
which gets canceled out while
calculating normalized emissivities and 
emission line ratios.
The overlapping contours are
displayed with solid red lines.

The left panel of Figure
\ref{fig:col} shows the variation
of Em$_{\rm blend}$ with
log $T$ and log $N$$_{\rm H}$. 
Em$_{\rm blend}$ approaches values close 
to 1 for temperatures around 
log $T$ $\sim$ 7.0 and for column 
densities spanning from 
log $N$$_{\rm H}$ = 20 to log $N$$_{\rm H}$ = 21.5
when the 1 keV emission feature is the 
strongest. 
At higher or lower temperatures, 
the 1 keV emission feature diminishes
in intensity.
The right panel of 
Figure \ref{fig:col} displays
the variation of Em$_{\rm left}$/Em$_{\rm right}$
with log $T$ and 
log $N$$_{\rm H}$, characterizing the 
variation in the centroid of the 
1 keV emission feature in a CIE plasma.
Around log $T$$\sim$ 7.2, the emission
feature is centered at 1 keV. 
For lower temperatures, it shifts to the left, and for higher temperatures, it shifts to the right.

\subsection {Reflection from accretion disk:}\label{accr}

Out of the five XRBs in our sample, Ser X-1 and Cyg X-2 displayed reflection features near 1 keV,  prompting us to investigate their variability further. The reflection from the accretion disk has been associated with high column densities \citep{2020A&A...639A..13K}, we therefore set  $N$$_{\rm H}$ to 10$^{23}$ cm$^{-2}$ for the reflection blends. The incident continuum SED\footnote{The default settings for reflection modeling in \textsc{Cloudy} utilize a power-law continuum as the primary radiation source (see the \href{https://gitlab.nublado.org/cloudy/cloudy}{\texttt{hazy1}} manual).} illuminating the accretion disk is derived from the observed SEDs of Ser X-1 and Cyg X-2, after subtracting the \texttt{diskbb} continuum component (see sections \ref{serx1} and \ref{cyg x-1}), which represents the contribution from the accretion disk itself.
The illuminating SED interacts with the accretion disk, where it is absorbed and subsequently re-emitted, producing characteristic emission lines, including those in the Fe L complex.

For studying the variation in the reflection features, we created a logarithmic grid for $\xi$, covering a range of log $\xi$ values from 0.5 to 5.0, with increments of 0.1 dex.  Figure
\ref{fig:ref} shows the variation
of Reflect$_{\rm blend}$ and Reflect$_{\rm left}$/Reflect$_{\rm right}$
with log $\xi$ in Cyg X-2. The reflection features are additionally influenced by the viewing angle and inner disk radius, both of which were treated as free parameters, as elaborated in Sections  \ref{serx1} and \ref{cyg x-1}.

\section{Fitting the 1 keV features with \textsc{Cloudy}}\label{sec4}
We performed spectral fitting in
XSPEC version 12.01 \citep{1996ASPC..101...17A}.
The SEDs of the
respective XRBs
 were incorporated 
into \textsc{Cloudy} for spectral modeling of the  PIE emission/absorption and reflection features. 
To model the PIE line emission/absorption
for both targets, 
we utilized logarithmic grids for
two key parameters:  $\xi$, which 
ranged from 0 to 5 in increments 
of 0.1, and  $N$$_{\rm H}$,
which spanned from 20 to 24 in 
increments of 0.1. 
For the CIE line emission,
we ran logarithmic grids on 
temperature ($T$),
ranging from 6 to 8 in increments of 0.1, 
and on $N$$_{\rm H}$, which
spanned from 
20 to 24 in increments of 0.1. 
To model the reflection lines, we utilized a logarithmic grid for  $\xi$, ranging from 0 to 5 in increments of 0.1, a logarithmic grid for inner radius (r) spanning from 6 to 8 in increments of 0.1, a linear grid for the inclination angle ranging from 0 to 90 degrees in increments of 5 degrees, and kept the hydrogen column density fixed at $N$$_{\rm H}$ = 10$^{23}$  cm$^{-2}$. 
The output files from these calculations were
converted into FITS format using the \textsc{Cloudy}–\textsc{xspec}
interface \citep{2006PASP..118..920P} to ensure compatibility
with \textsc{xspec}. CIE emission, PIE emission, and reflection components were included using the \texttt{atable} format for additive tabular models \footnote{By default, the \texttt{atable} format includes a normalization component that scales the additive model to fit the observed spectrum.}, while absorption was applied using the \texttt{mtable} format for multiplicative tabular models reading in the FITS files produced using the \textsc{Cloudy}-\textsc{xspec} interface.
By default,  \textsc{Cloudy} does not include turbulent or bulk velocity broadening in its line profiles.
Velocity broadening to the line profiles was implemented using the \texttt{gsmooth} convolution model in \textsc{xspec}, applied multiplicatively to the FITS files generated using \textsc{Cloudy}.
 The resulting best-fit broadening values are reported for each system  later in this section, for both the CIE and PIE models.
The best-fit parameters were determined by minimizing the C-statistics \citep{1979ApJ...228..939C}. To ensure continuity with previous literature \citep{2020MNRAS.492.4646P, 2021MNRAS.505.5058P, 2022ApJ...936..185K}, residuals are presented in the (data--model)/error format.

\subsection{NGC 1313 X-1:}

\begin{figure*}
\centering
\begin{subfigure}{0.5\textwidth}
    \centering
    \includegraphics[width=\linewidth]{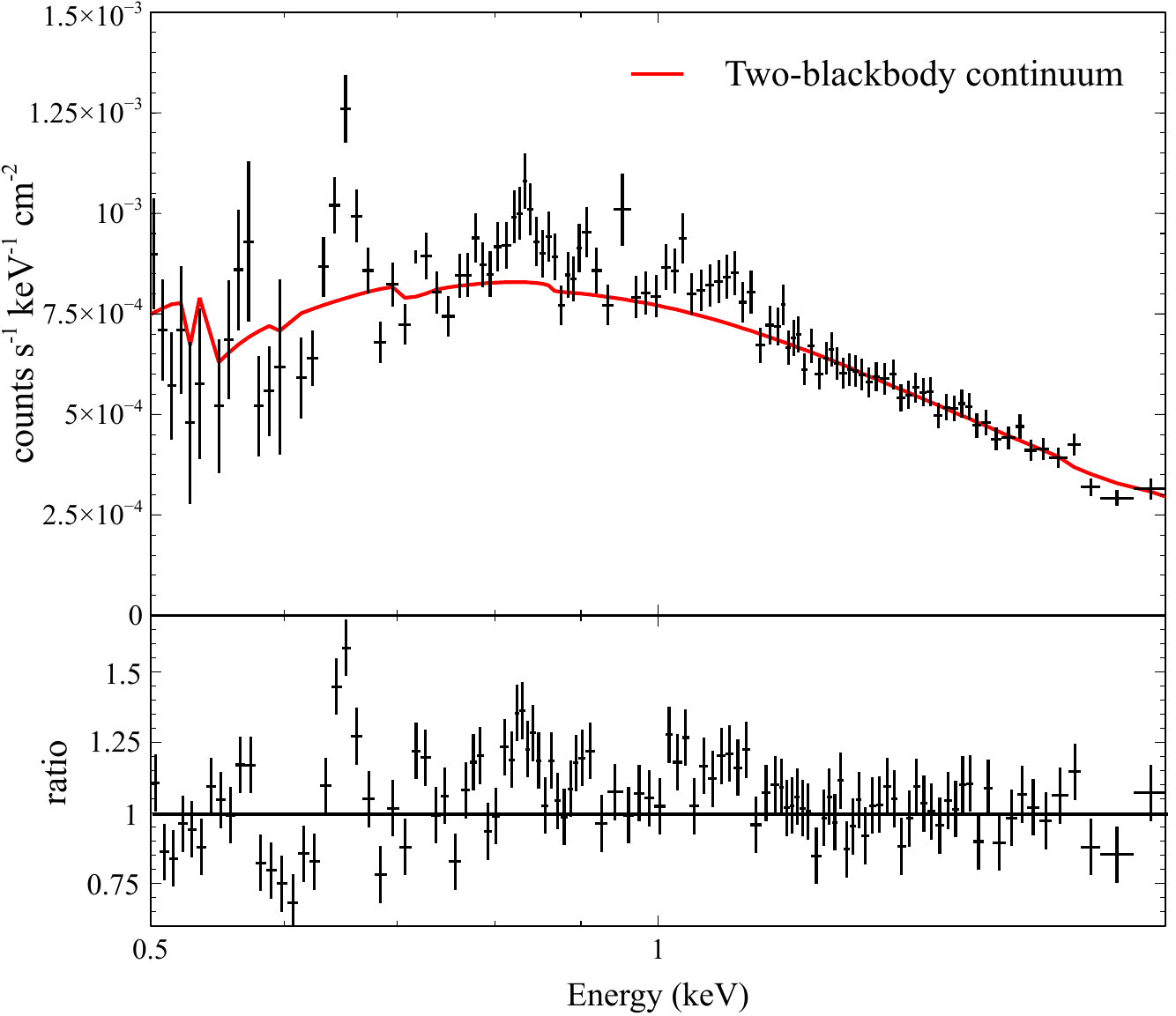}
    \caption{}
    \label{fig:a}
\end{subfigure}%
\begin{subfigure}{0.5\textwidth}
    \centering
    \includegraphics[width=\linewidth]{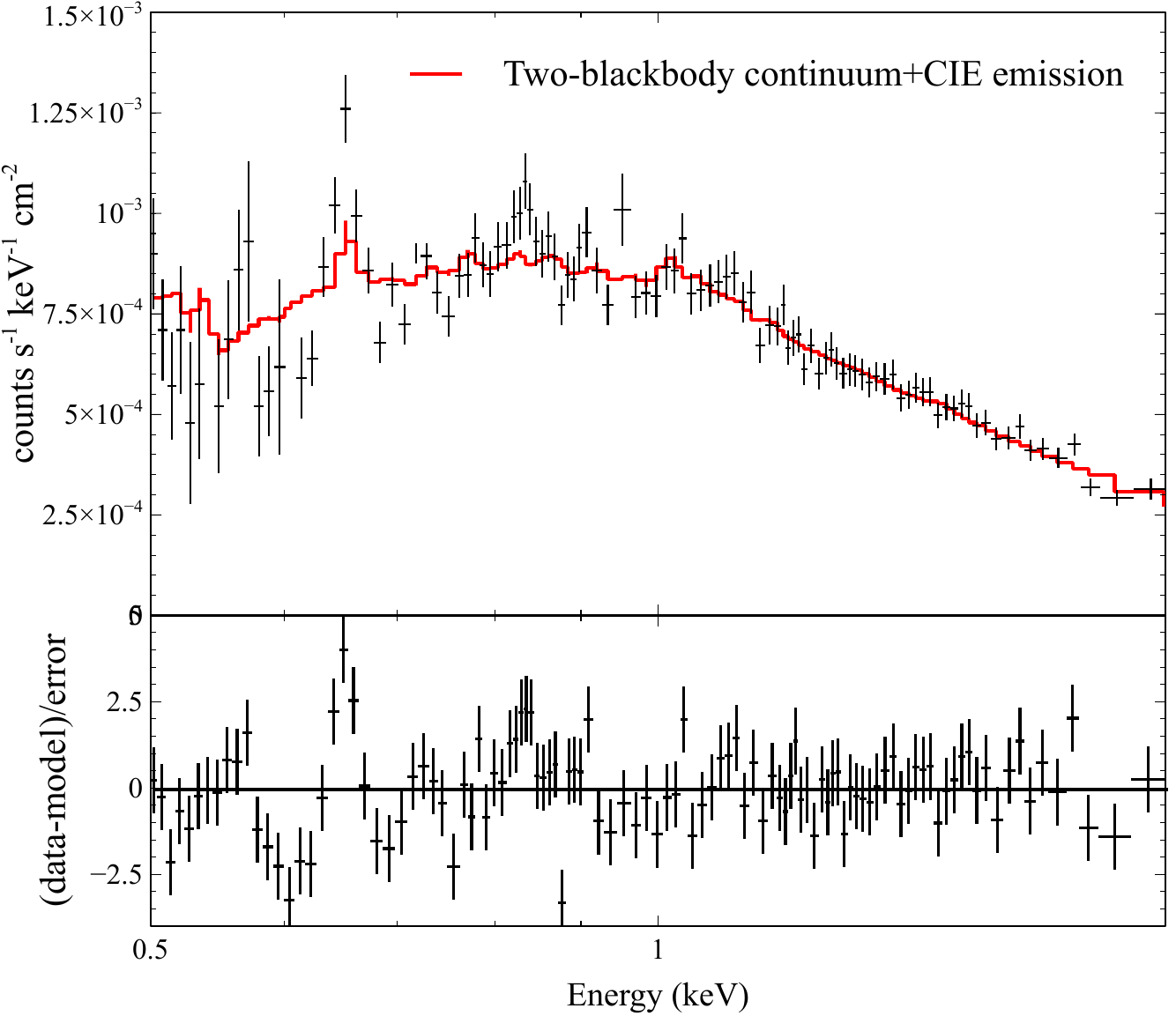}
    \caption{}
    \label{fig:b}
\end{subfigure}
\begin{subfigure}{0.5\textwidth}
    \centering
    \includegraphics[width=\linewidth]{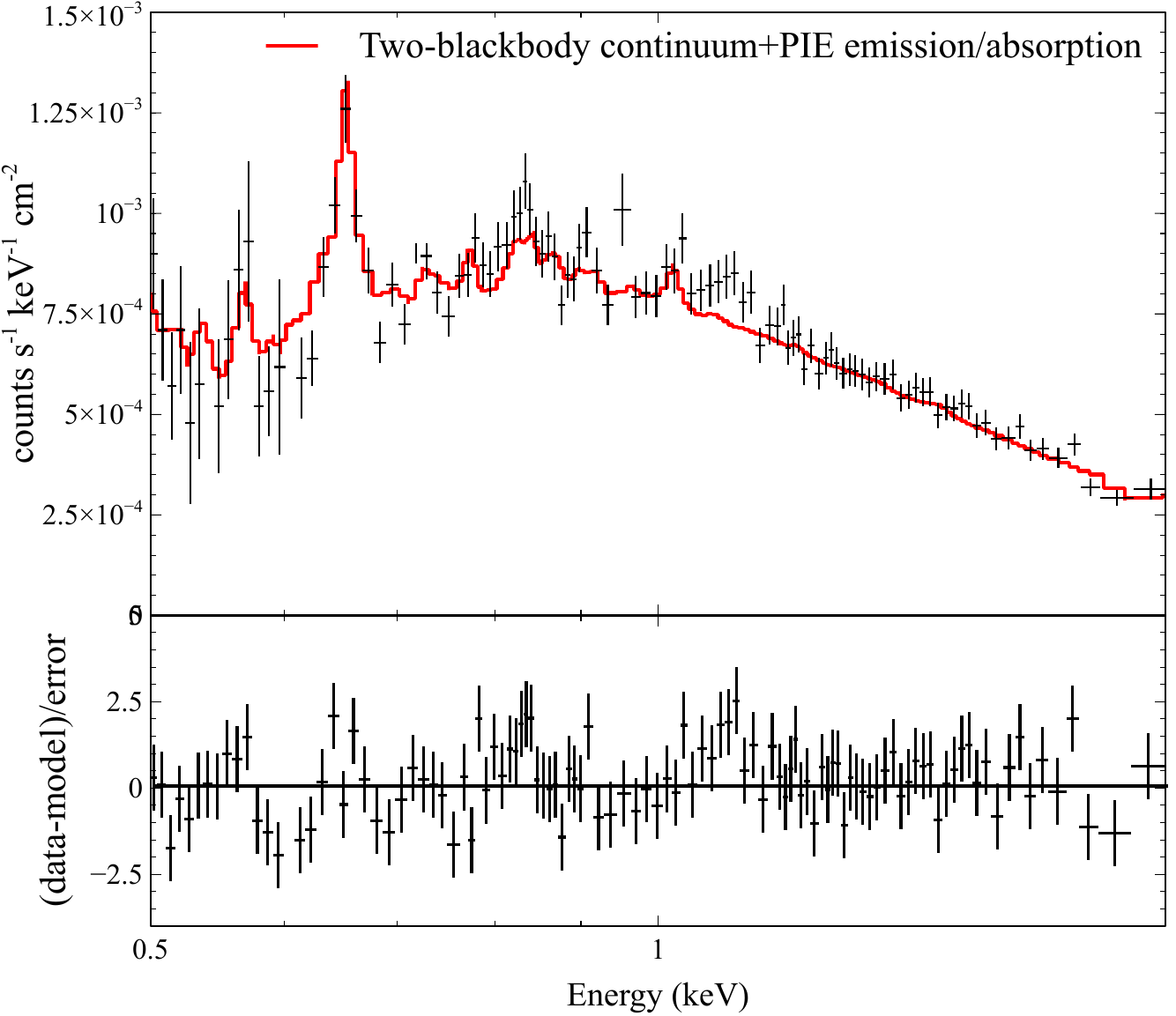}
    \caption{}
    \label{fig:c}
\end{subfigure}
\caption{{a) Combined first-order spectra of NGC 1313 X-1, overplotted with best-fitting two blackbody continuum model, the 1 keV residual is visible within 0.6 and 1.25 keV. b) Spectra of NGC 1313 X-1 overlaid with the continuum model + line emission from CIE plasma, with the 1 keV emission/absorption residuals persisting. c) The same spectra overlaid with the continuum model + PIE line emission, effectively resolving the 1 keV emission/absorption residuals, yielding the most complete and statistically preferred fit.}}
\label{fig:247fit_2}
\end{figure*}

In our initial analysis, we aimed 
to fit the observed spectrum of 
NGC 1313 X-1 using a two-blackbody
continuum model, specifically, 
\texttt{tbabs*(bbody+bbody)} in the 0.5--2.0 keV energy band.
The {\tt tbabs} model 
was employed to account
for Galactic absorption, 
with the absorbing hydrogen 
column density set at 
7$\times$10$^{20}$ cm$^{-2}$  \citep{2016A&A...594A.116H}\footnote{https://heasarc.gsfc.nasa.gov/cgi-bin/Tools/w3nh/w3nh.pl}.
We found the best-fit temperature
of 0.35$\pm$ 0.02 keV 
for the lower-energy blackbody 
component
and of 1.55 $\pm$ 0.03 keV 
for the higher-energy blackbody
component. 
Figure \ref{fig:247fit_2}, Panel a,
displays the best-fitted 
two-blackbody
continuum model
along with the 
observed spectra of NGC 1313 X-1
in the 0.5--2 keV energy band. 
The sub-figures at the bottom
of each panel in 
Figure \ref{fig:247fit_2}
show the residual
of the observed spectra 
after fitting, which
estimated as
(data - model)/error.
Notably, we observed a significant
emission residual within the energy
range of 0.6 to 1.25 keV. 
This emission residual displayed a 
leftward shift from 1 keV,
primarily concentrating in  
the 0.6 to 1.0 keV energy range.
Furthermore, the absorption residuals 
were more prominent in the
0.5 to 0.8 keV energy range, 
with fainter residuals observed 
for energies above 1 keV.

Following that, we adopted a 
 spectral-fitting approach that 
combined the two-blackbody 
continuum model with a CIE 
plasma emission
model,
as illustrated in Panel b of Figure 
\ref{fig:247fit_2}. 
The fitting process resulted in the
best-fit temperature for the 
CIE emission model, yielding 
log$(T$/K) = 7.1 $\pm$ 0.1, along with 
a best-fit column density of 
log$(N$$_{\rm H}$/cm$^{-2}$) = 21.0 $\pm$ 0.1.
The blackbody temperatures remained unchanged.
Although adding the CIE model with
the blackbody
continuum improved 
the overall fit to the 
observed spectra 
slightly with $\Delta$C-stat $>$ 30, 
there was not a substantial 
improvement in fitting the residuals. 
Notably, the strong O VIII line at
$\sim$ 0.65 keV remained 
challenging to fit adequately.  
This simulation did not yield a
significantly stronger O VIII emission 
feature in comparison to the lines
in the vicinity of 1 keV.
Given the presence
of a stronger O VIII emission 
feature in NGC 1313 X-1, particularly
in contrast to the lines around 1 keV,
it is evident that relying solely 
on the CIE emission line feature 
is inadequate for modeling
the 1 keV blend emission
in NGC 1313 X-1. 

Next, we attempted to 
model the 1 keV  feature
 by combining the two-blackbody continuum 
model with a PIE plasma 
emission/absorption model,
as depicted in Panel c of Figure 
\ref{fig:247fit_2}.
We adjusted the absorption model for NGC 1313 X-1 to include a blueshift of 0.07c, based on previous observations that linked the absorption residuals to a slower wind component in the intermediate-bright state \citep{2020MNRAS.492.4646P}.
The blackbody temperatures were determined to be 0.36 $\pm$ 0.03 keV and 1.55 $\pm$ 0.04 keV, respectively.
This combined model provided a good 
fit for both the 1 keV emission blend,
which includes the prominent 
O VIII emission feature, 
as well as the absorption dips with $\Delta$C-stat $>$ 80 compared to the two-blackbody model.
The best-fit parameters for the PIE
emission model were found to be
log $\xi$ = 1.93 $\pm$ 0.05, accompanied
by a best-fit column density of
log$(N$$_{\rm H}$/cm$^{-2}$) = 21.1 $\pm$ 0.1.  
Additionally, the best-fit parameters 
for the PIE absorption model were 
log $\xi$ = 0.7 $\pm$ 0.1,
and  
log$(N$$_{\rm H}$/cm$^{-2}$) = 21.2 $\pm$ 0.1. 
The line broadening (v$_{\rm broad}$) measured at $\sim$ 3000 km/s.


\subsection{NGC 247 ULX-1:}\label{ngc247}

\begin{figure*}
    \centering
    \begin{tabular}{cc}
        
        \includegraphics[width=0.45\textwidth]{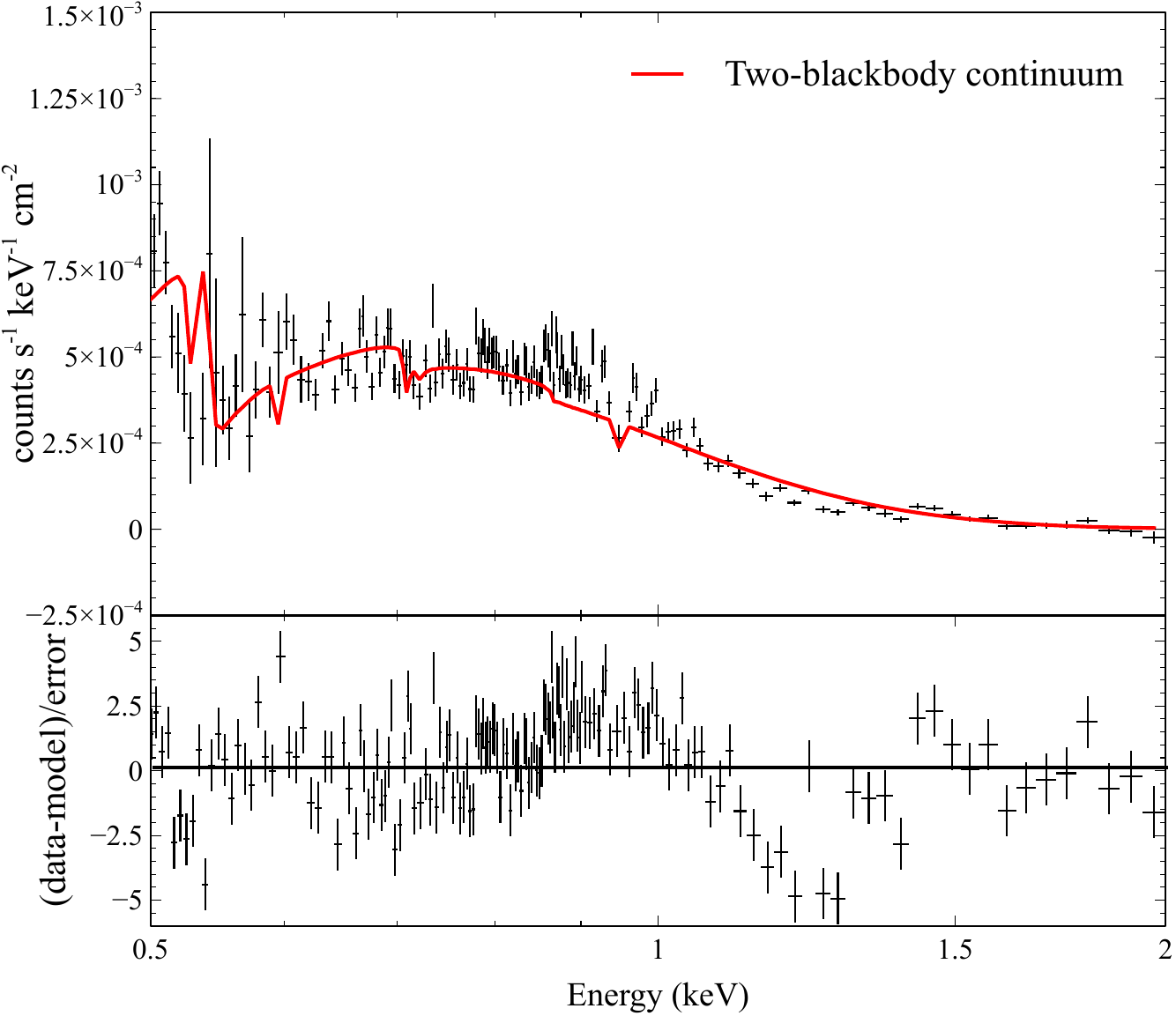} & 
        \includegraphics[width=0.45\textwidth]{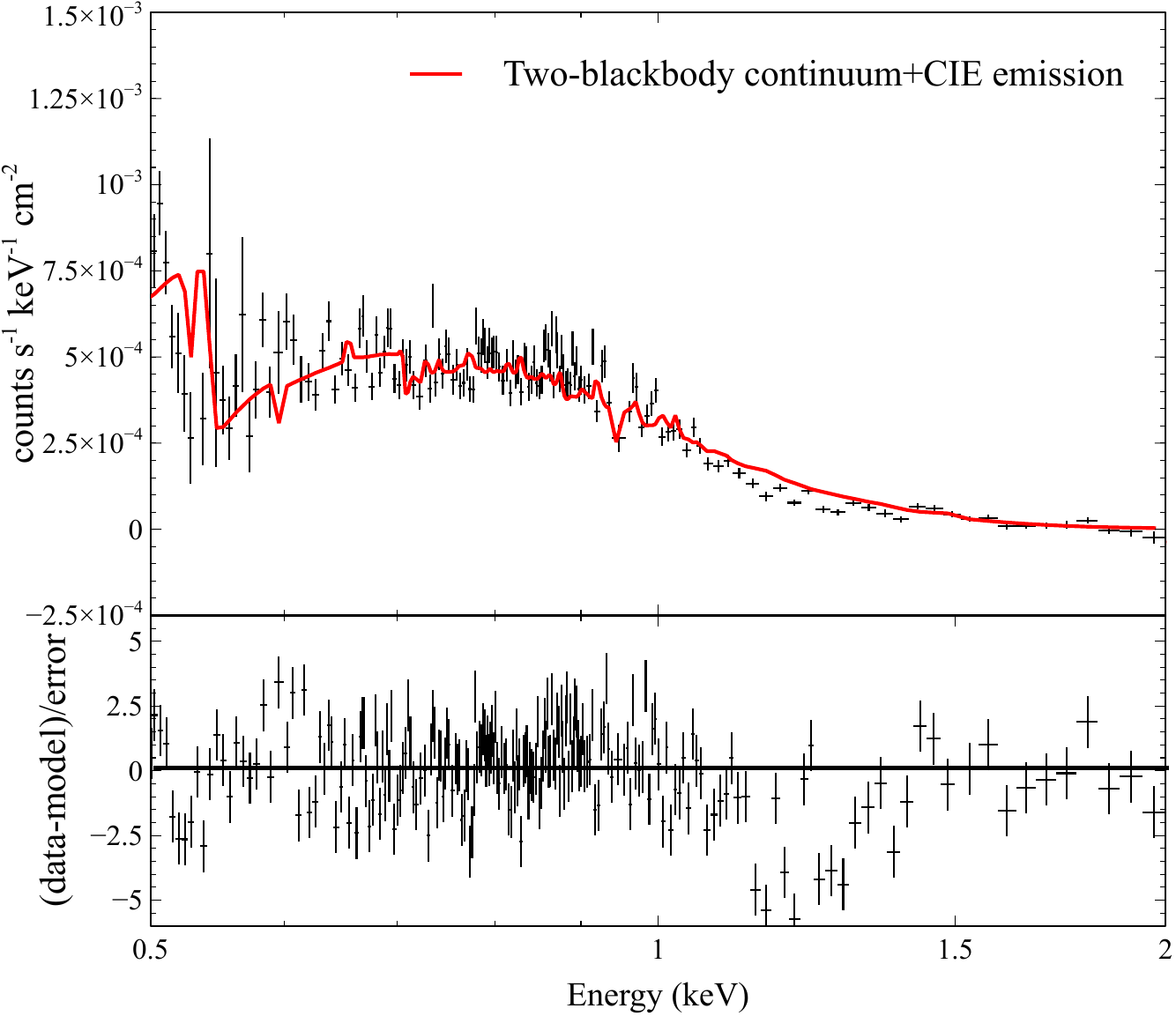} \\
         (a) & (b) \\
        
        \includegraphics[width=0.45\textwidth]{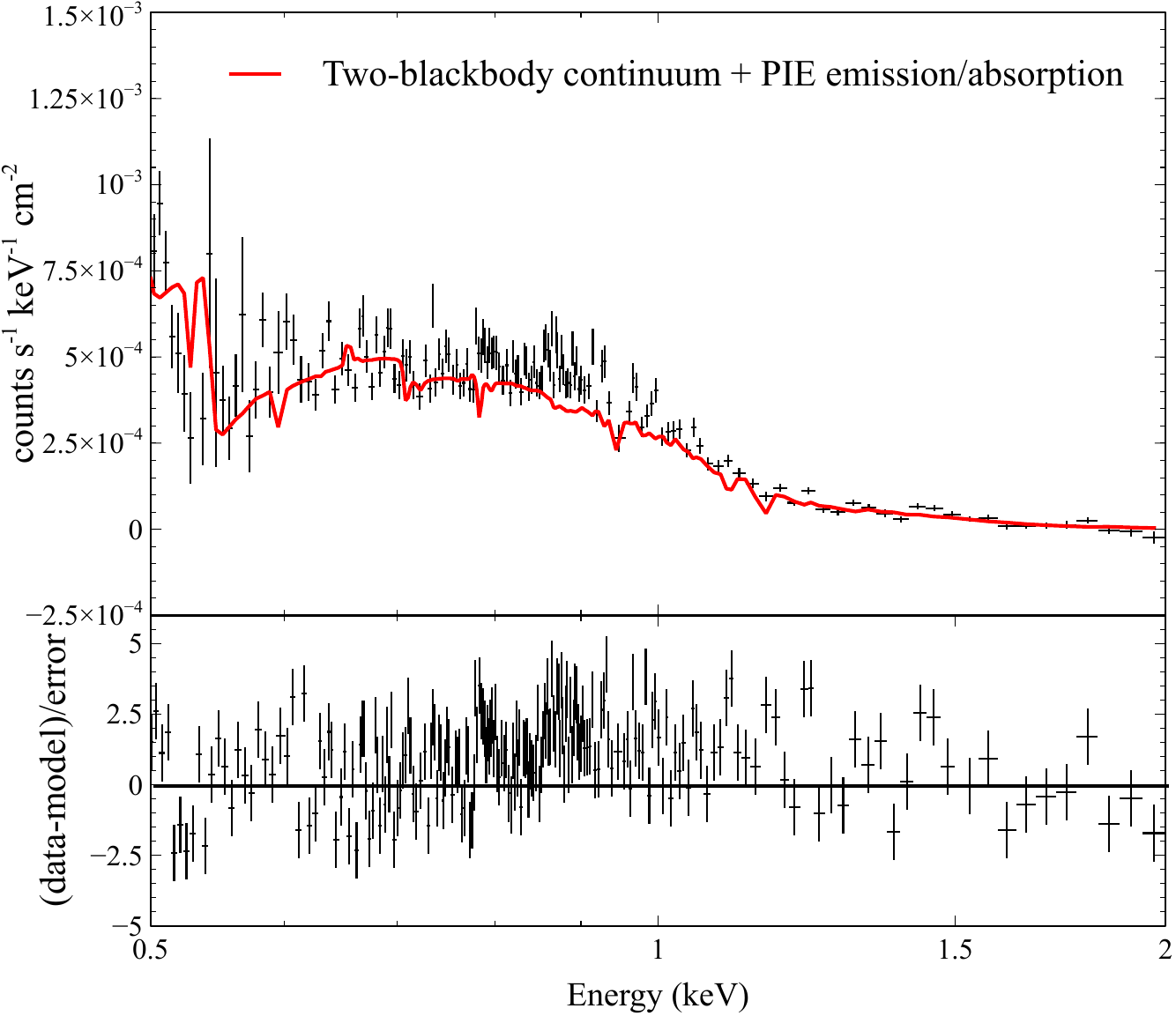} & 
        \includegraphics[width=0.45\textwidth]{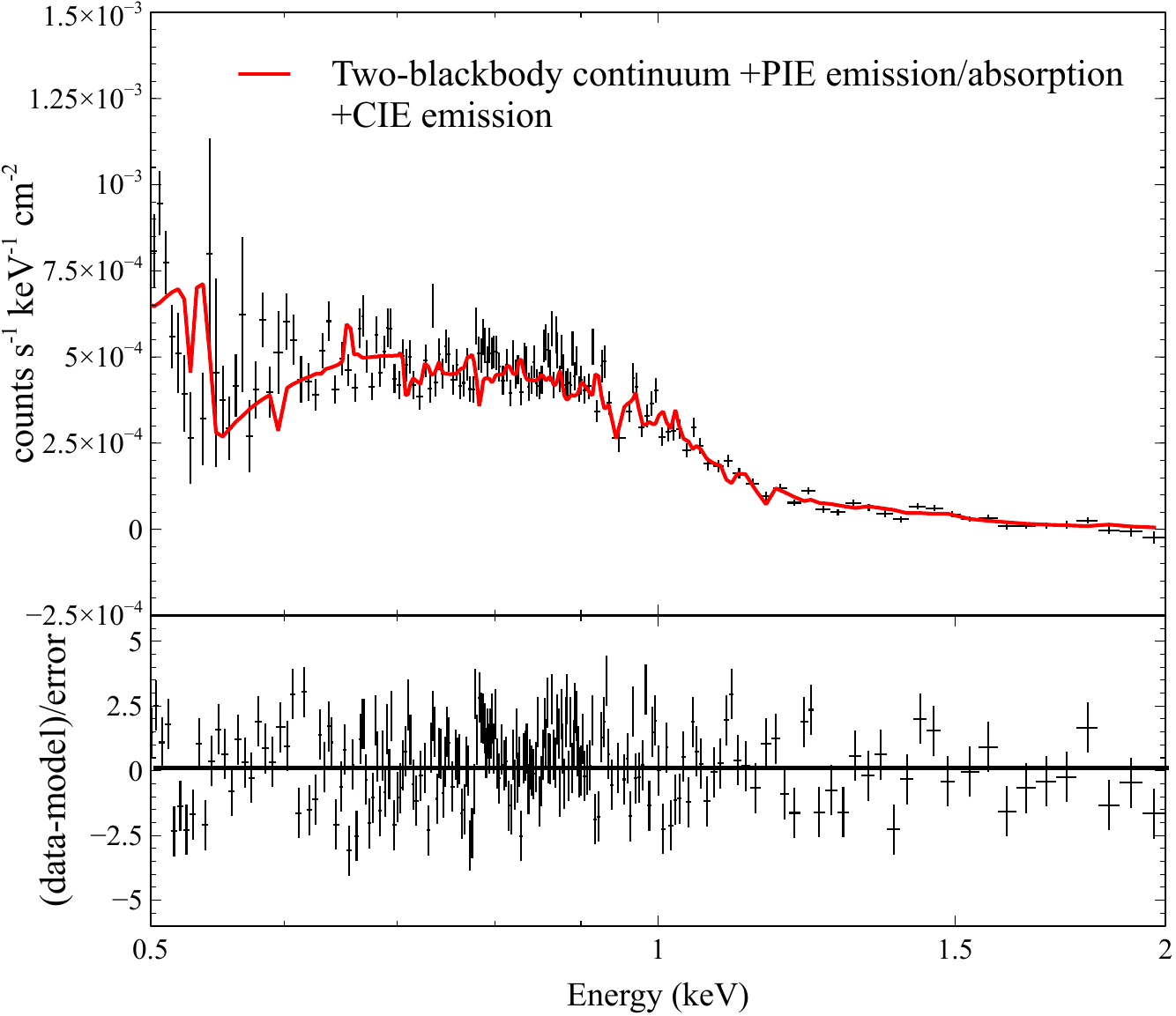} \\
        (c) & (d) \\
    \end{tabular}
    \caption{{a) Combined first-order spectra of NGC 247 ULX-1, overlaid with best-fitting two blackbody continuum model, the 1 keV residual is visible. b) NGC 247 ULX-1 spectra, featuring the continuum model with line emission derived from a CIE model, successfully accounting for the 1 keV emission residual, while the 1 keV absorption residual remains. c) Same spectra of NGC 247 ULX-1, overlaid with the continuum model along with line emission/absorption produced by a PIE model, effectively addressing the 1 keV absorption residual while the 1 keV emission residual persists. d) Spectra of NGC 247 ULX-1 overlaid with the continuum model, incorporating PIE and CIE emission/absorption, fully resolving the 1 keV emission and absorption features for the most complete and accurate fit.}}
    \label{fig:247fit_1}
\end{figure*}

Our initial analysis aimed to fit the observed spectrum of NGC 247 ULX-1 within the energy band of 0.5–2.0 keV.  To achieve this, we utilized a two-blackbody continuum model: \texttt{tbabs*(bbody+bbody)}. Within this model, \texttt{tbabs} was used to account for Galactic absorption, with the absorbing hydrogen column density fixed at 6 $\times$ 10$^{20}$ cm$^{-2}$ \citep{2016A&A...594A.116H}.
Our analysis yielded the best-fit temperatures of 0.14 $\pm$ 0.01  keV for the lower-energy blackbody component and 0.44 $\pm$ 0.02 keV for the higher-energy blackbody component. Figure \ref{fig:247fit_1}, Panel a, shows the best-fitted two-blackbody continuum model overlaid with the observed spectra of NGC 247 ULX-1 within the 0.5–2 keV energy range. The corresponding residual is displayed in the bottom sub-figure of the panel. Significant emission residuals were observed within the energy range of 0.8 to 1.1 keV. These emission residuals displayed a leftward shift from 1 keV and primarily concentrating between 0.8 and 1.0 keV. In addition, significant absorption residuals were identified in two distinct energy intervals: between 0.55 to 0.85 keV and 1.1 to 1.5 keV. Notably, the absorption residuals in the latter energy range exhibited a more pronouced absorption dip.

Next, we combined the two-blackbody continuum model with a CIE plasma emission model in our spectral analysis, as displayed in Panel b of Figure \ref{fig:247fit_1}. The blackbody temperatures remained unchanged.
The fitting process yielded a best-fit temperature of log$(T$/K) = 6.99 $\pm$ 0.02, along with 
a best-fit column density of log$(N$$_{\rm H}$/cm$^{-2}$) = 21.5 $\pm$ 0.1. 
This modified model marginally improved the fit with a $\Delta$C-stat $>$ 60 compared to the two-blackbody model. The inclusion of the CIE emission component effectively resolved the emission residuals, eliminating them (refer to the bottom subfigure in Panel b). However, the absorption residuals persisted in the spectrum.

Following that, we aimed to model the 1 keV feature by merging the two-blackbody continuum model with a PIE plasma emission/absorption model, as illustrated in panel c of Figure \ref{fig:247fit_1}.
We adjusted the absorption model for NGC 247 ULX-1 to include a blueshift of  0.17c, reflecting the previously observed absorption residuals linked to a strong outflow \citep{2021MNRAS.505.5058P}.
The fitting showed further improvement compared to both the two-blackbody and two-blackbody+CIE emission models, resulting in $\Delta$C-stat $>$ 80 compared to the continuum-only fit.
The fitting process resulted in a best-fit ionization parameter of log $\xi$ = 4.1 $\pm$ 0.1 and a best-fit column density of log$(N$$_{\rm H}$/cm$^{-2}$) = 21.0 $\pm$ 0.1 for the emission lines and log $\xi$ = 3.4 $\pm$ 0.1 and a best-fit column density of log$(N$$_{\rm H}$/cm$^{-2}$) = 21.7 $\pm$ 0.1 for the absorption lines.  The blackbody temperatures remained unchanged.
The two-blackbody+PIE emission/absorption model effectively fit the absorption dips but did not provide a good fit for the emission features. The robust fit of the absorption feature by the two-blackbody+PIE model strongly indicates its origin in the PIE plasma. Furthermore, the effective fit of the emission feature with the two-blackbody+CIE emission model implies that the emission predominantly arises from the CIE plasma.

\begin{figure*}
    \centering
    \begin{tabular}{cc}
        
        \includegraphics[width=0.45\textwidth]{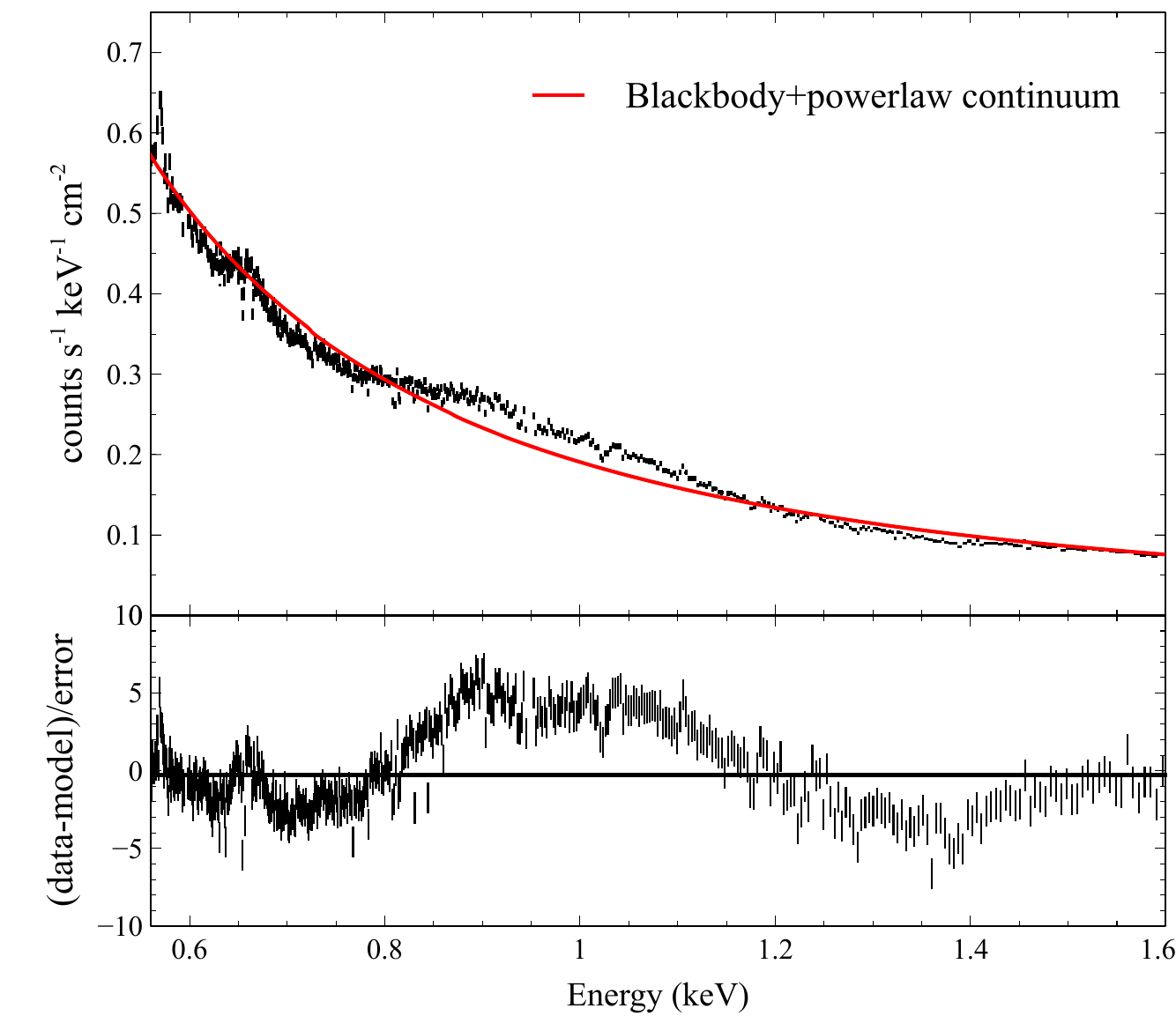} & 
        \includegraphics[width=0.45\textwidth]{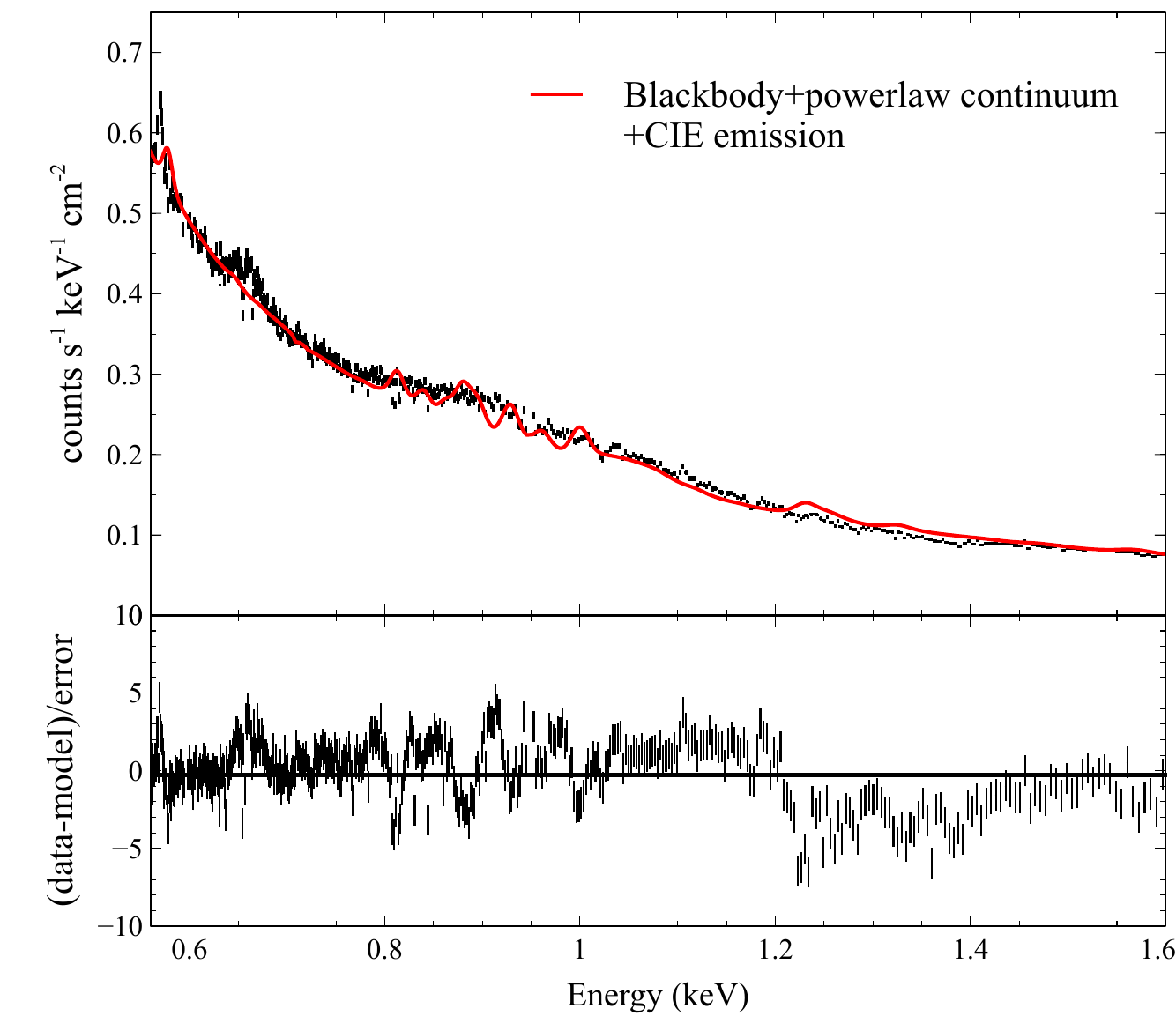} \\
        (a) & (b) \\
        \includegraphics[width=0.45\textwidth]{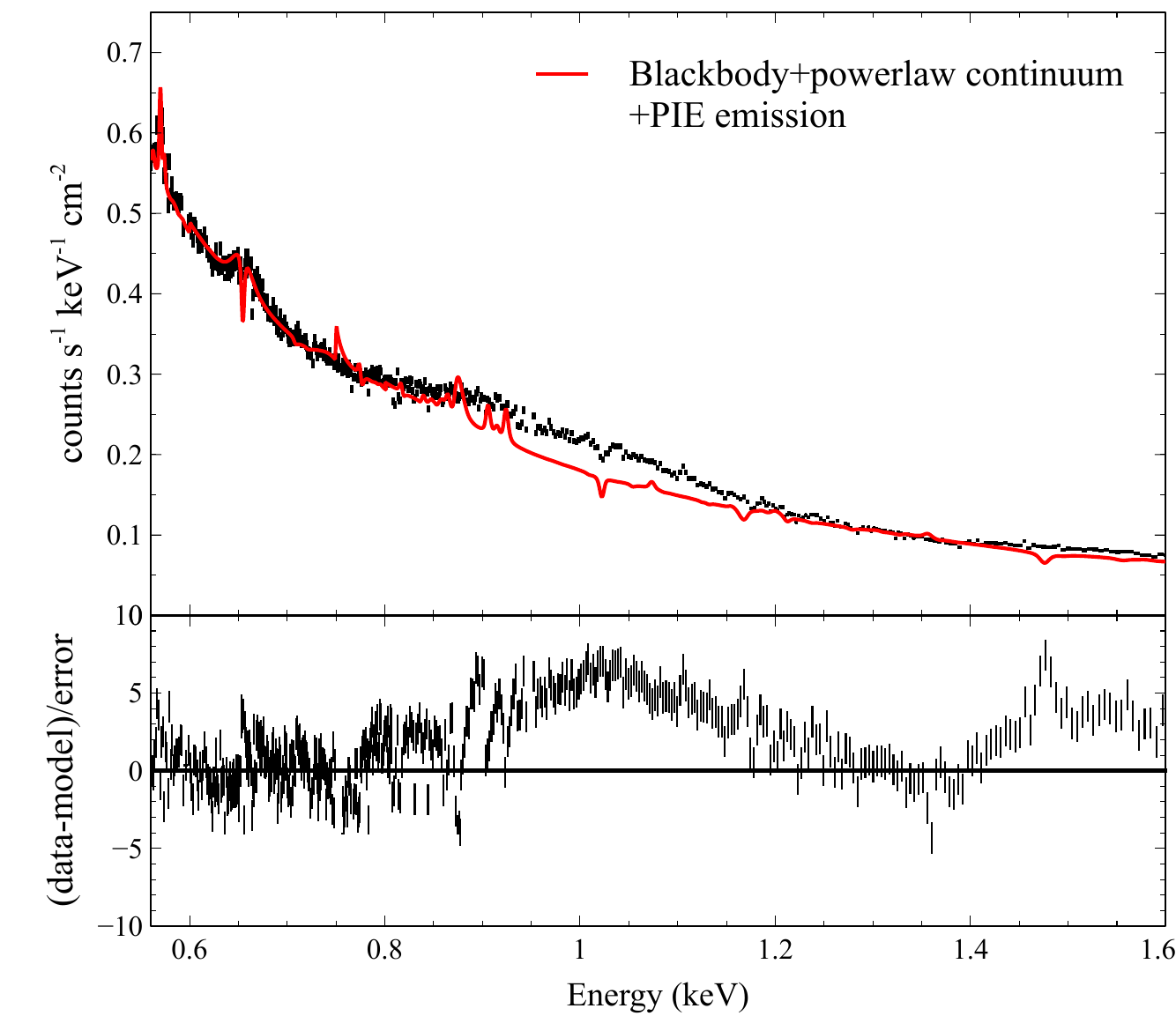} & 
        \includegraphics[width=0.45\textwidth]{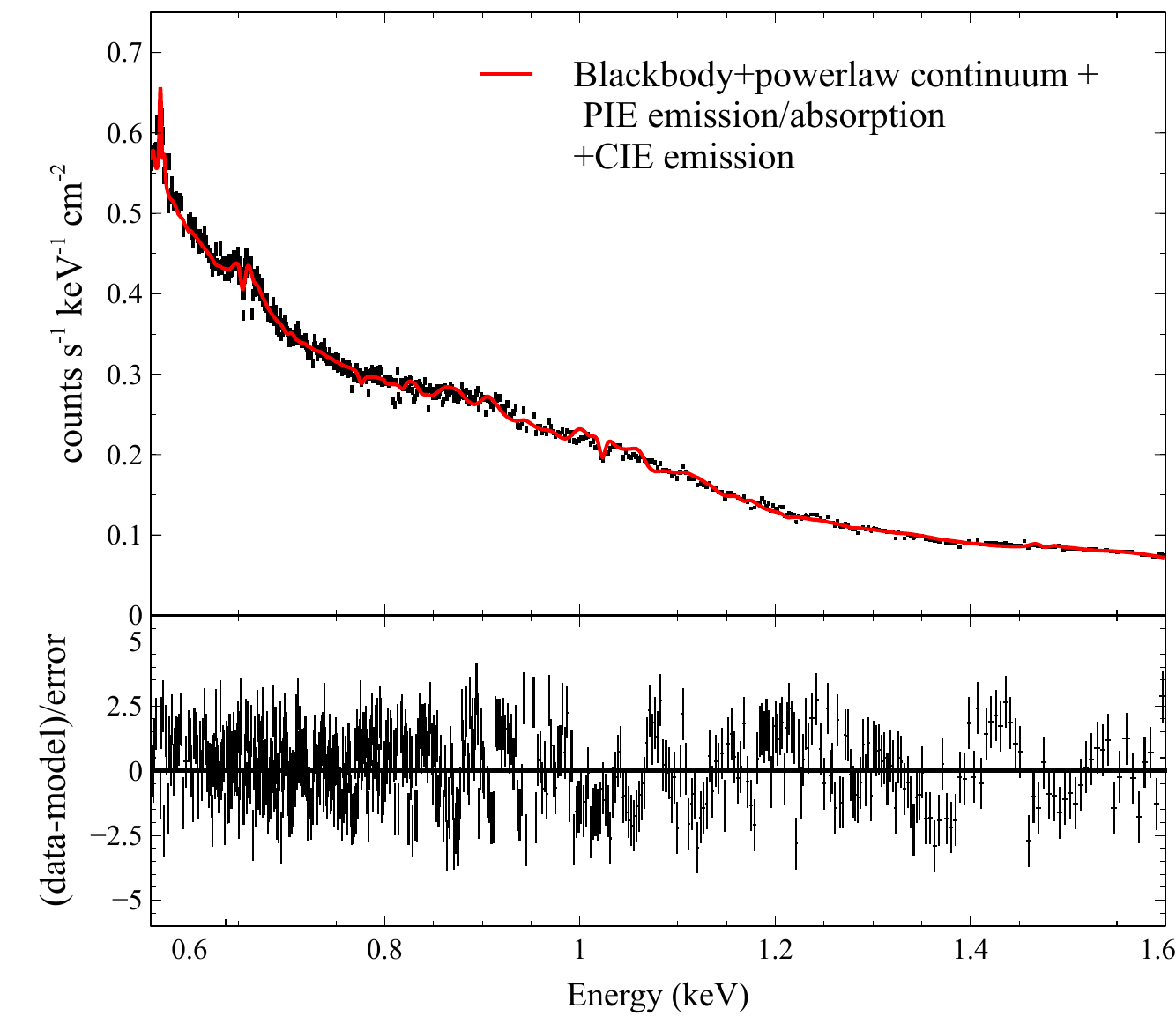} \\
        (c) & (d) \\
    \end{tabular}
    \caption{{a) Combined first-order spectrum of Hercules X-1, overlaid with best-fitting blackbody+powerlaw continuum model, the 1 keV residual is visible. b) Adding CIE emission to the  continuum model significantly improves the emission residuals between 0.8-1.2 keV, while the absorption residual remains. c)  The continuum model along with line emission/absorption produced by a PIE model improves the fit below 0.8 keV, but significant emission/absorption residuals persist for energies $>$ 0.8 keV.  d) The full model, combining the continuum, CIE emission, and PIE emission/absorption, eliminates nearly all residuals across 0.56–1.6 keV, representing the best fit.}}
    \label{fig:herx1}
\end{figure*}

Finally, to effectively model both the absorption and emission features, we used a two-blackbody+CIE emission+PIE emission/absorption plasma model to fit the observed spectrum, as shown in panel d of Figure \ref{fig:247fit_1}, obtaining a $\Delta$C-stat $>$ 110 compared to the continuum-only fit.
The best-fit temperatures were determined to be  0.15 $\pm$ 0.02  keV and 0.44 $\pm$ 0.04 keV. 
The best-fit parameters for the CIE model were log$(T$/K)= 7.00 $\pm$ 0.02 and log$(N$$_{\rm H}$/cm$^{-2}$) = 21.5 $\pm$ 0.1, with line broadening of $\sim$ 1200 km/s. For the PIE emission model, the best-fit parameters were log $\xi$ = 4.1 $\pm$ 0.1 and log$(N$$_{\rm H}$/cm$^{-2}$) = 21.0 $\pm$ 0.1, while for the PIE absorption model, the best-fit parameters were log $\xi$ = 3.5 $\pm$ 0.1 and log$(N$$_{\rm H}$/cm$^{-2}$) = 21.3 $\pm$ 0.1. We found the line broadening to be $\sim$ 1500 km/s. 
Both emission and absorption residuals were effectively eliminated, as evident in the bottom subfigure of panel d. In Figures \ref{fig:em1}, \ref{fig:em2}, \ref{fig:abs1},  \ref{fig:abs2}, and \ref{fig:col}, the values of the best-fit parameters for the two-blackbody+CIE emission+PIE emission/absorption plasma model are indicated by red squares on the contour plots.
\\

\begin{figure*}
\centering
\begin{subfigure}{0.5\textwidth}
    \centering
    \includegraphics[width=\linewidth]{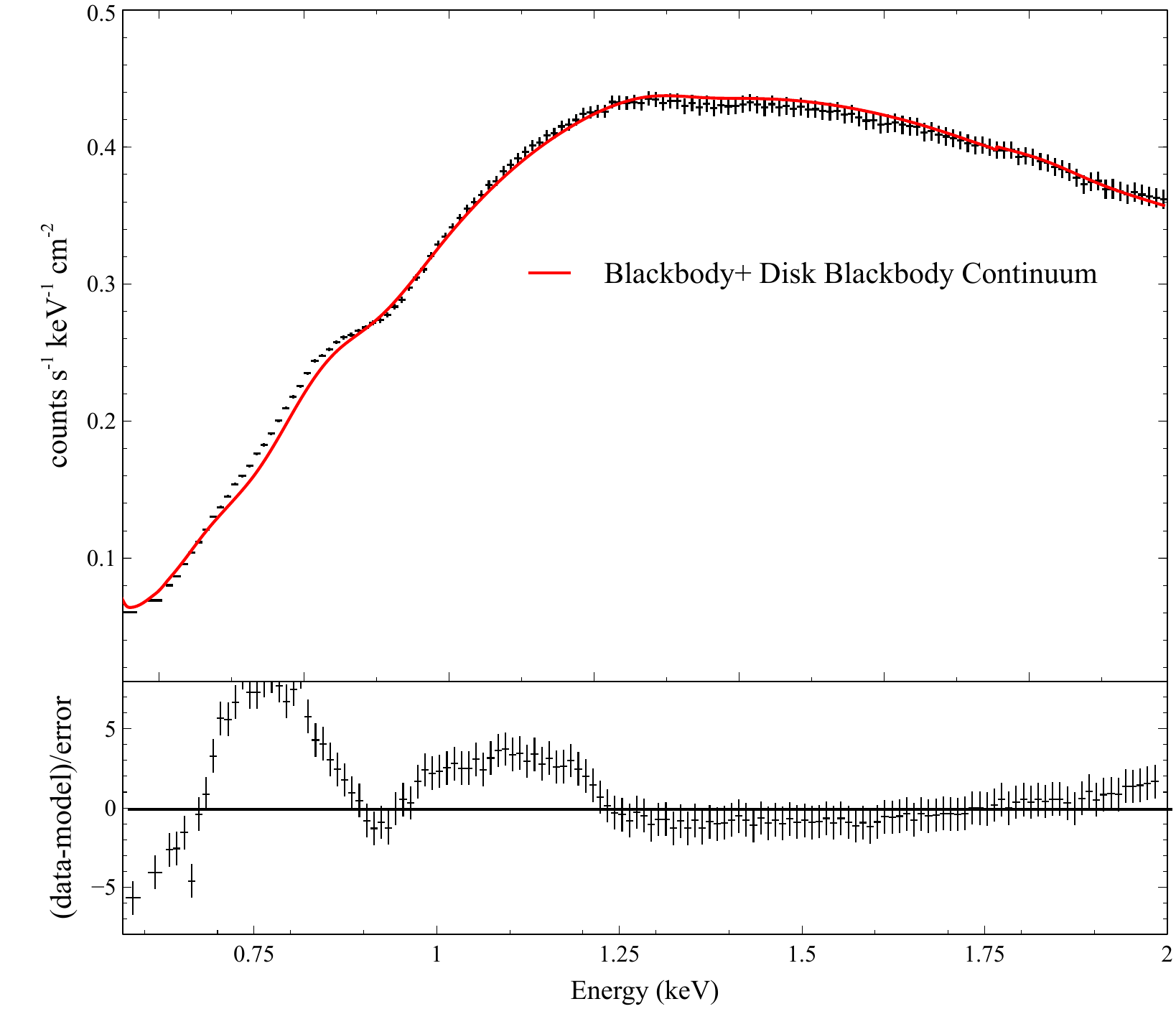}
    \caption{}
    \label{fig:a}
\end{subfigure}%
\begin{subfigure}{0.5\textwidth}
    \centering
    \includegraphics[width=\linewidth]{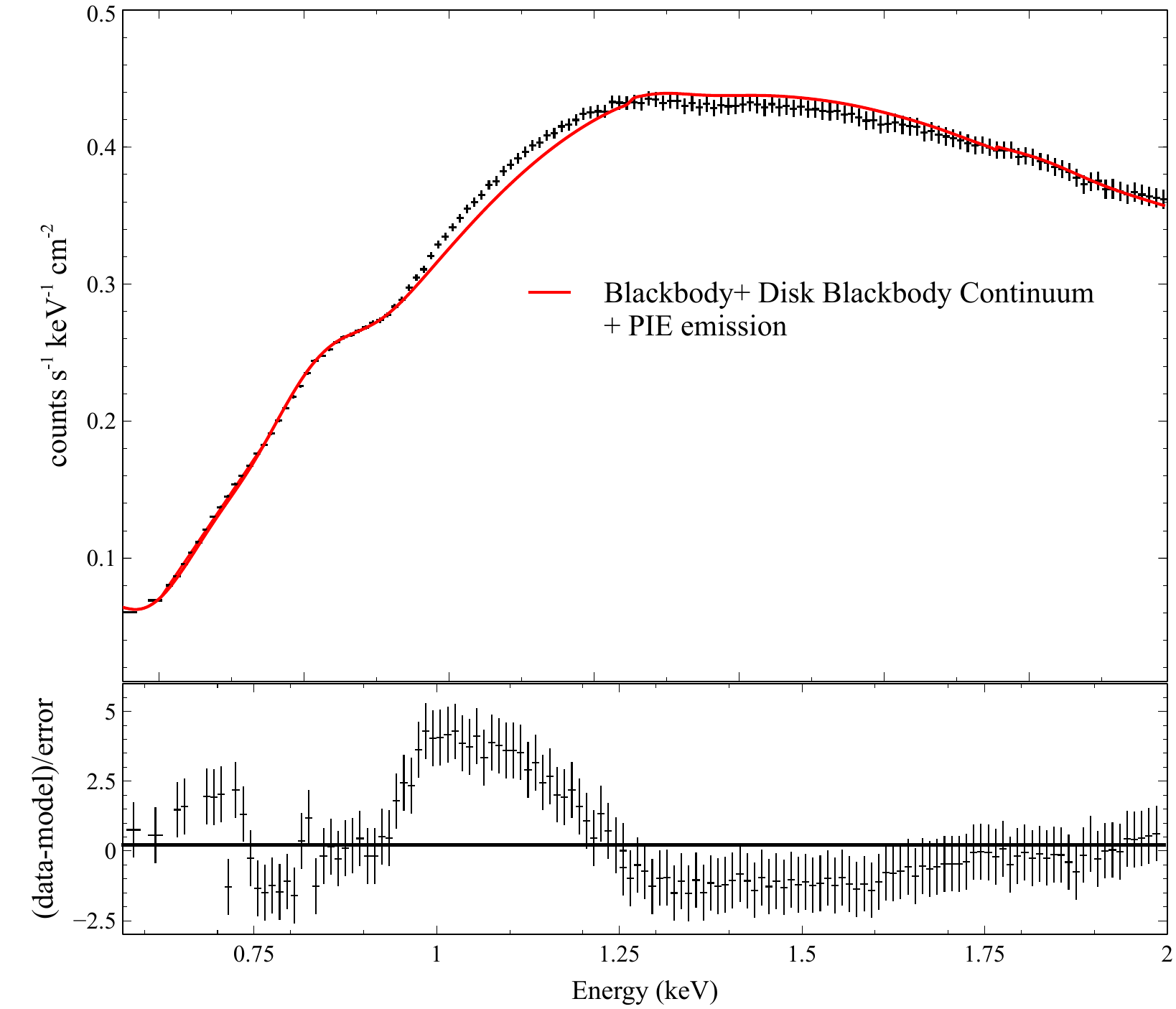}
    \caption{}
    \label{fig:b}
\end{subfigure}
\begin{subfigure}{0.5\textwidth}
    \centering
    \includegraphics[width=\linewidth]{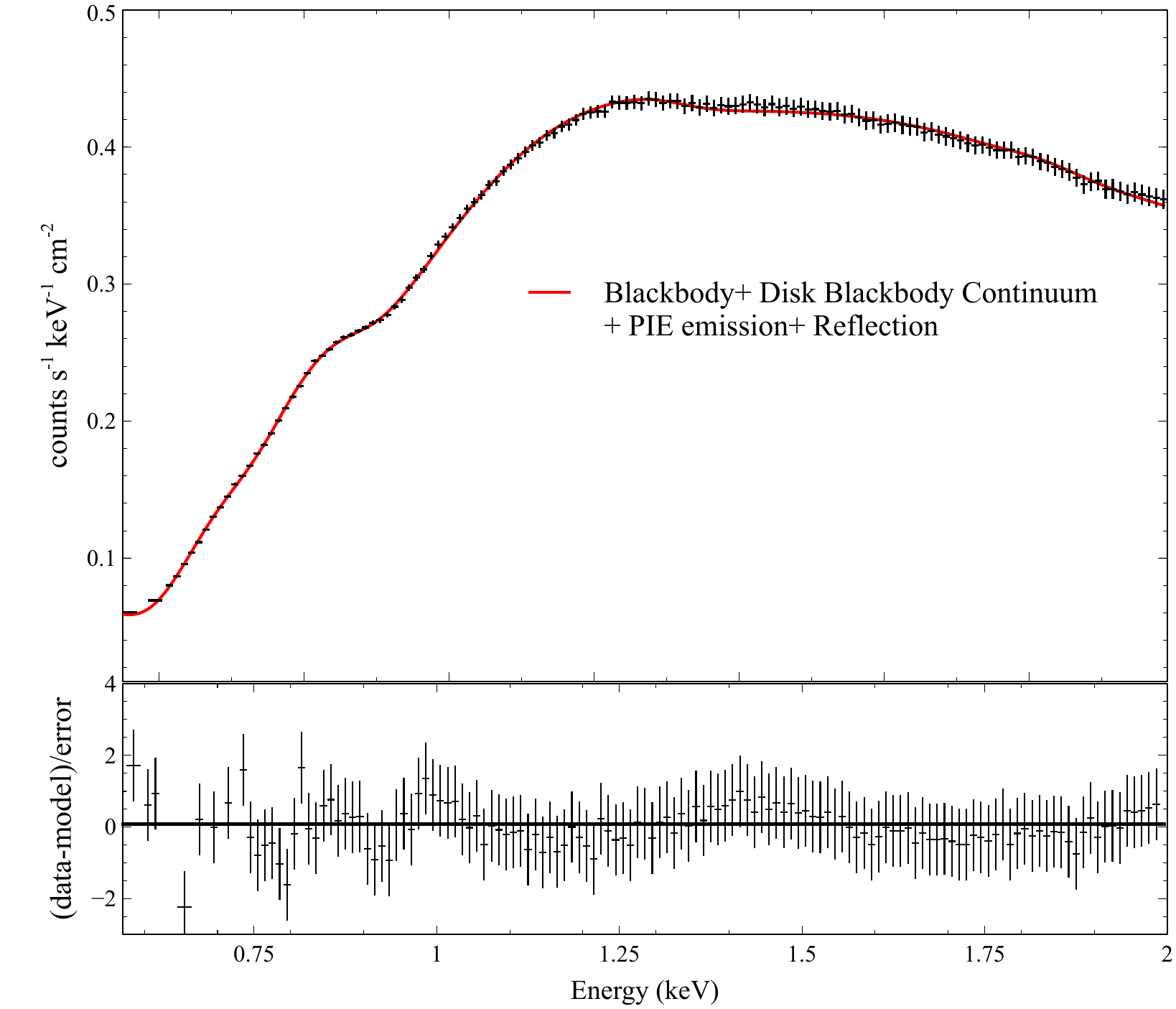}
    \caption{}
    \label{fig:c}
\end{subfigure}
\caption{a) {\textit{NICER} spectrum of Ser X-1, overlaid with best-fitting continuum model, a more pronounced emission residual is found (0.5-0.9 keV) with a less pronounced emission residual (0.9-1.3 keV). b) The same spectrum overlaid with the continuum model + line emission from PIE plasma. The residual between 0.5-0.9 keV have largely diminished, while the residual between 0.9-1.3 keV persists.
(c) The same spectrum overlaid with the continuum model, PIE line emission, and reflection emission. Panel (c) shows the most complete and statistically favored fit, eliminating all spectral residuals.}}
\label{fig:serx1}
\end{figure*}

\begin{table*}
\centering
\setlength{\tabcolsep}{1pt}
\scriptsize
\caption{Number of model components ($N_{\rm comp}$) and list of best-fit parameters associated with the 1~keV spectral feature across all sources. The table reports, for each source, whether CIE, PIE, and/or reflection components were used in the model. For each component type, the corresponding best-fit physical parameters are shown. Column units are as follows: $N_{\rm H}$ in cm$^{-2}$, $v_{\rm broad}$ in km/s, $T$ in K, and $r_{\rm in}$ in cm.}
\begin{tabular}{|c|c||c|c|c|c||c|c|c|c||c|c|c|c|}
\hline\hline
\textbf{Source} & \textbf{$N_{\rm comp}$} & \textbf{CIE?} &
\textbf{log\,T} & \textbf{log$(N$$_{\rm H}$)} & \textbf{v$_{\rm broad}$}  &
\textbf{PIE?} & \textbf{log\,$\xi$} & \textbf{log$(N$$_{\rm H}$)}  & \textbf{v$_{\rm broad}$} &
\textbf{Refl.?} & \textbf{log\,$\xi$} & \textbf{Incl} &  \textbf{log($r_{in}$)}  \\
\hline
NGC 1313 X-1 & 1 & No &  -- & -- & -- & Yes & $\sim$ 0.7--1.93  &  $\sim$ 21.1-21.2 & $\sim$ 3000  & No  & -- & -- & -- \\
NGC 247 ULX-1 & 2 & Yes  & $\sim$ 7.0     & $\sim$ 21.5        & $\sim$ 1200      & Yes & $\sim$ 3.5--4.1 & $\sim$ 21.0--21.3 &  $\sim$ 1500 & No & -- & -- & -- \\
Hercules X-1 & 2 & Yes & $\sim$ 7.2 & $\sim$ 21.6 & $\sim$ 900  & Yes  & $\sim$ 1.3--3.0     & $\sim$  21.4-22.1        &  $\sim$ 1200--6000     & No & -- & -- & -- \\
Serpens X-1 & 2 & No  & --      & --       & --      & Yes & $\sim$ 1.5 & $\sim$ 20.2 & $\sim$ 4200 & Yes & $\sim$ 3.2 & $\sim$ 29 & $\sim$ 6.2 \\
Cyg X-2 & 3 & Yes & $\sim$ 7.1 & $\sim$ 20.8 & $\sim$ 3400 & Yes  &   $\sim$ 2.5--3.5   & $\sim$  21.5--21.7        & $\sim$ 3400     & Yes & $\sim$  2.1 & $\sim$ 54 & $\sim$ 6.5\\
\hline
\end{tabular}
\label{tab:1keV_fullsummary}
\end{table*}

\subsection{Hercules X-1:}\label{herx1}
 We initially fitted the spectrum in the 0.56–1.6 keV energy band using the XSPEC model \texttt{tbabs*(bbody+powerlaw)}, with the Galactic absorption hydrogen column density fixed at 1.5 $\times$ 10$^{20}$ cm$^{-2}$ \citep{2016A&A...594A.116H}. The best-fit temperature obtained for the blackbody was 0.05 $\pm$ 0.01 keV and the powerlaw index was 1.992 $\pm$ 0.003. Figure \ref{fig:herx1}, Panel a, illustrates the overlaid best-fitted  continuum model and observed spectra of Hercules X-1 in the 0.56–1.6 keV energy range. The residual, shown in the lower sub-figure, reveals significant emissions between 0.8 and 1.18 keV, concentrated mainly between 0.8 and 1.0 keV, and for energies higher than 1.6 keV. 
Absorption residuals were found in two distinct energy intervals: 0.5 to 0.8 keV and 1.2 to 1.6 keV. \\
Following this, we integrated the continuum model with a CIE plasma emission model in our spectral analysis, as illustrated in panel b of Figure \ref{fig:herx1}. 
For the CIE component, we obtained a best-fit temperature of log$(T$/K)= 7.2 $\pm$ 0.1, and a best-fit column density of log$(N$$_{\rm H}$/cm$^{-2}$) = 21.6 $\pm$ 0.1. The best-fit blackbody temperatures remained unchanged. 
While the inclusion of the CIE emission component notably diminished the emission residuals between 0.8 and 1.18 keV, substantial residuals persisted above 1.2 keV, and some remained below 0.8 keV,  as shown in the bottom subfigure in panel b. The best-fit velocity of the CIE emission lines were found to be $\sim$ 900  km/s. The addition of a CIE model significantly improved the fit quality  ($\Delta$C-stat $>$ 170).\\
Subsequently, we tried to model the residuals by combining the  continuum model with a PIE plasma emission/absorption model, as depicted in panel c of Figure \ref{fig:herx1}. 
Modeling the PIE component required the inclusion of plasma at multiple velocities. This included a narrow O\,\textsc{vii} emission line at 0.57\,keV with a velocity broadening of approximately 1200 km/s, and a broad O\,\textsc{viii} emission line at 0.65\,keV with a velocity broadening of around 6000 km/s. The best-fit blackbody temperature and the power-law index remained unchanged. The narrow emission component was characterized by log\,$\xi$\,=\,1.3\,$\pm$\,0.1 and log\,($N_{\rm H}$/cm$^{-2}$)\,=\,21.6\,$\pm$\,0.1; the broad component by log\,$\xi$\,=\,2.0\,$\pm$\,0.2 and log\,($N_{\rm H}$/cm$^{-2}$)\,=\,22.1\,$\pm$\,0.1. The blended emission feature near 1\,keV was found to be associated with log\,$\xi$\,=\,2.7\,$\pm$\,0.2 and log\,($N_{\rm H}$/cm$^{-2}$)\,=\,21.4\,$\pm$\,0.1, and a velocity broadening of $\sim$ 1400 km/s for the  PIE lines within the blend. Additionally, we identify slightly blueshifted ($\sim$0.003\,$c$) absorption lines with a velocity broadening of $\sim$1250 km/s, corresponding to log\,$\xi$\,=\,3.0\,$\pm$\,0.2 and log\,($N_{\rm H}$/cm$^{-2}$)\,=\,22.1\,$\pm$\,0.1.
Velocity broadening for each PIE component was implemented using the \texttt{gsmooth} convolution model in \textsc{xspec}. 
 Despite substantial improvement in the residuals below 0.8 keV, significant residuals remained between 0.8 and 1.6 keV, as indicated in the bottom sub-panel of Figure \ref{fig:herx1} c. The inclusion of a PIE model improved the fit quality  compared to the continuum model ($\Delta$C-stat $>$ 80) but was inferior to the fit achieved with the continuum+CIE model.\\
In our final approach to model both absorption and emission features comprehensively, we used a continuum+CIE emission+PIE emission/absorption plasma model, as displayed in Figure \ref{fig:herx1} d.
All residuals within the range of 0.56 - 1.6 keV were successfully eliminated including the emission/absorption 1 keV residual with $\Delta$C-stat $>$ 270 compared to the continuum-only model (see the bottom sub-panel of Figure \ref{fig:herx1} d). The best-fit hydrogen column density was determined to be log$(N$$_{\rm H}$/cm$^{-2}$) of 21.6 $\pm$ 0.1 for the CIE plasma, and log$(N$$_{\rm H}$/cm$^{-2}$) = 22.1 $\pm$ 0.1 for the PIE plasma. The blackbody temperature, power-law index, ionization parameter values for the PIE plasma, temperature for the CIE plasma, and all the velocity broadenings remained unchanged.

\begin{figure*}
    \centering
    \begin{tabular}{cc}
        
        \hspace{-0.3in}\includegraphics[width=0.55\textwidth]{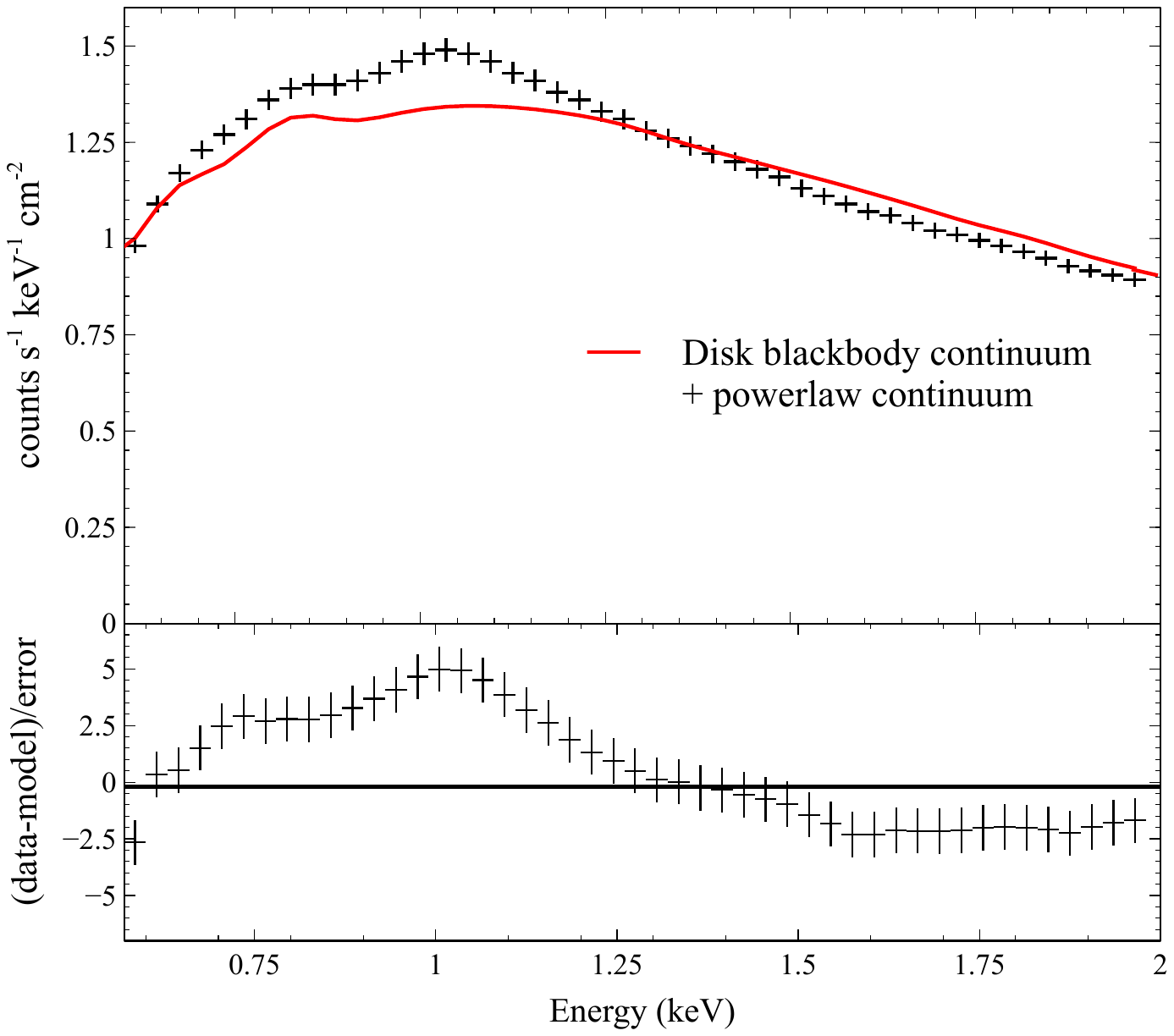} &  \hspace{-0.5in}\includegraphics[width=0.55\textwidth]{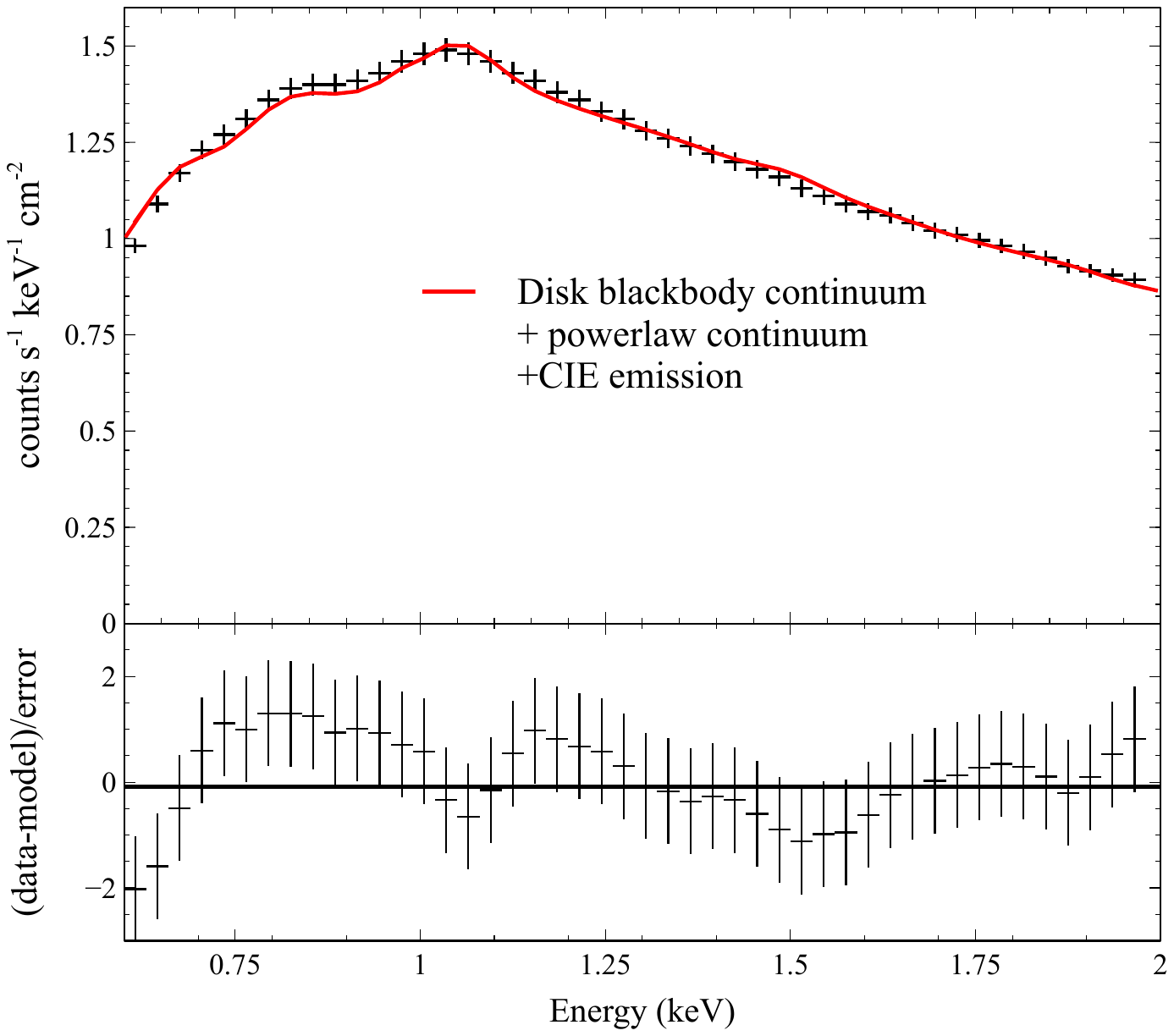} \\
        (a) & (b) \\
        \hspace{-0.3in}\includegraphics[width=0.55\textwidth]{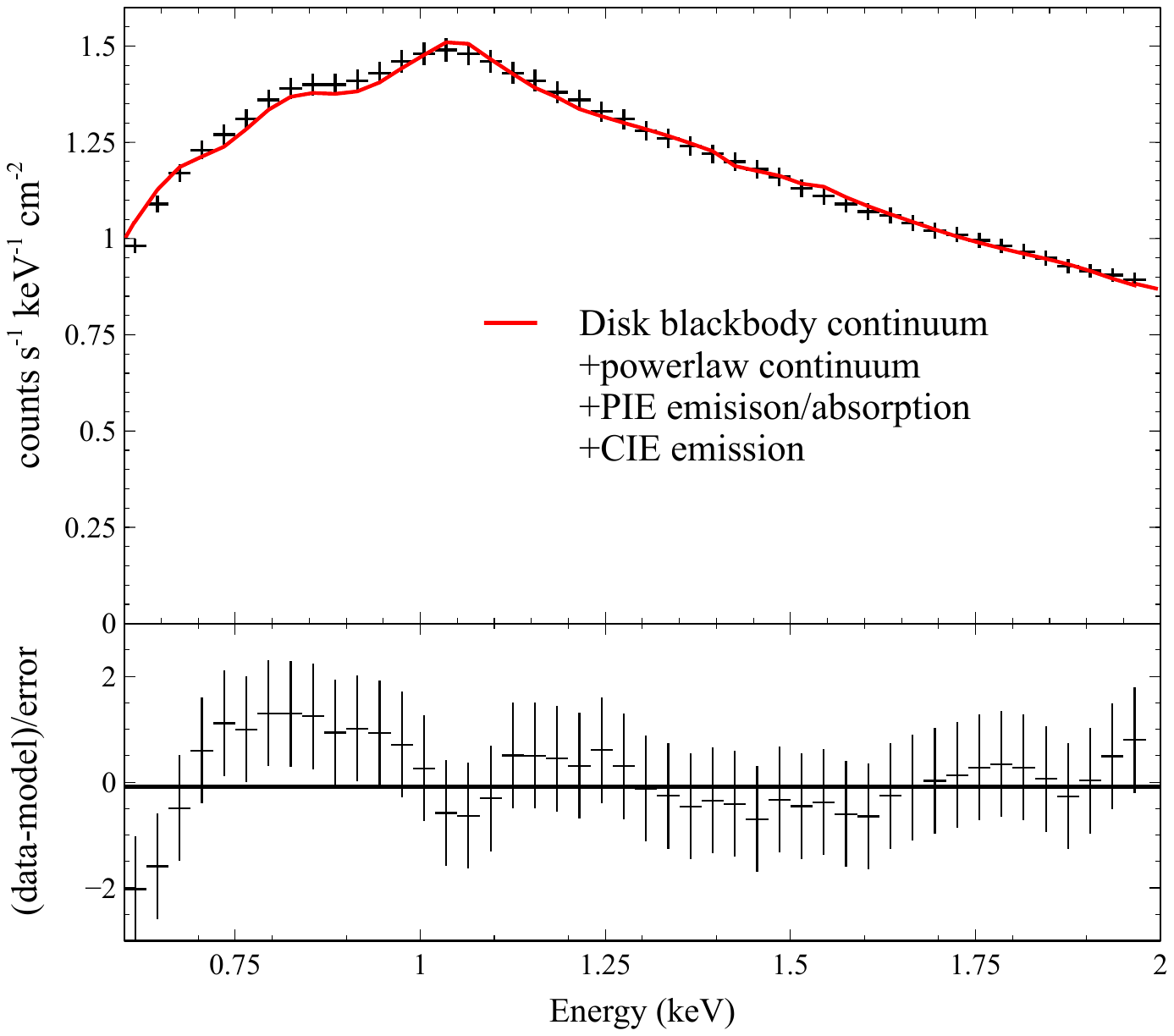} & 
        \hspace{-0.5in}\includegraphics[width=0.55\textwidth]{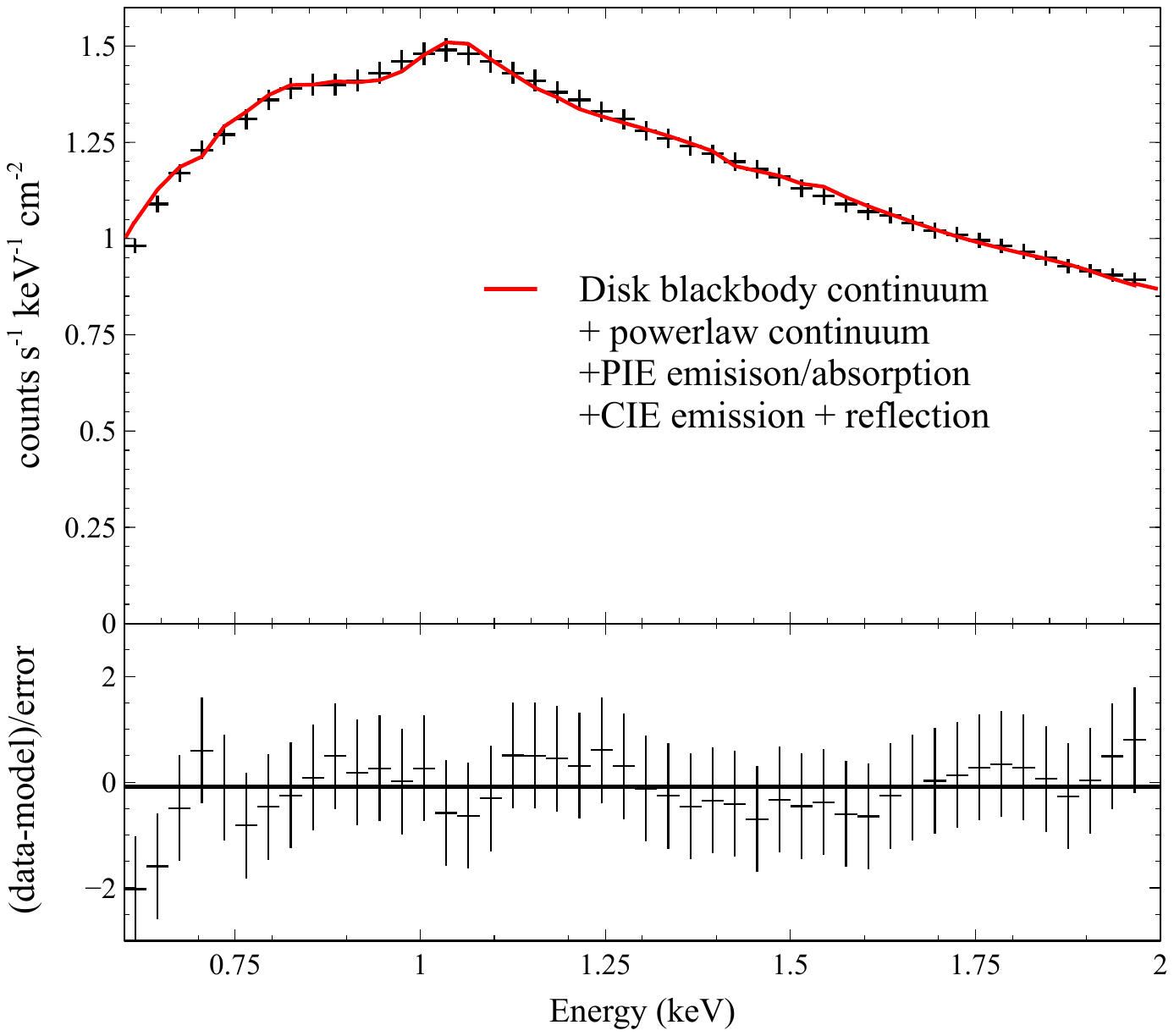} \\
        (c) & (d) \\
    \end{tabular}
    \caption{a) {\textit{NICER}  spectrum of Cyg X-2, overlaid with best-fitting continuum model, the 1 keV residual is more prominent in emission, less noticeable in absorption. b) Adding CIE emission to the continuum model significantly improves the residuals, although some emission residuals persist in specific energy ranges (0.65-1 keV and 1.1-1.2 keV), with minor absorption residuals in the 1.4-1.6 keV range. c) The continuum + CIE emission + PIE emission/absorption effectively eliminates all the absorption residuals, small emission residual remains between 0.65-1.0 keV.  (d) The continuum + CIE emission + PIE emission/absorption + reflection model removes all emission and absorption residuals. Panel (d) yield the most complete and statistically favored fit, resolving all spectral residuals.}}
    \label{fig:cygx2}
\end{figure*}


\subsection{Serpens X-1}\label{serx1}
The continuum in Ser X-1 has been previously characterized using a blend of a disk blackbody component, a single-temperature blackbody component, and a power-law component for a broadband fit using \textit{NuSTAR} and \textit{NICER} observations \citep{2013ApJ...779L...2M, 2018ApJ...858L...5L}. 
In the narrower 0.5-2.0 keV energy range, our analysis revealed that a basic combination of a blackbody and disk black body was sufficient for fitting the continuum,  with the introduction of a power law showing no improvement in the fit. We therefore proceeded with the model \texttt{tbabs*(diskbb+bbody)}, with the Galactic absorption hydrogen column density set at $\sim$ 4.4 $\times$ 10$^{21}$ cm$^{-2}$ \citep{2016A&A...594A.116H} for fitting the continuum in Ser X-1.  The best-fit temperature for the \texttt{diskbb} component was 1.05 $\pm$ 0.05 keV, while the \texttt{bbody} component had a best-fit temperature of 1.90 $\pm$ 0.06 keV.
Figure \ref{fig:serx1}, Panel a, shows the continuum model overlaid on the spectrum, while the lower sub-figure shows a significant emission residual in the 0.65-0.9 keV range,  and a less pronounced emission residual in the 0.9-1.3 keV range.\\
Following that, we integrated a PIE plasma emission model into our spectral analysis, utilizing the SED described in Section \ref{cygser}, and combining it with the continuum model to form a continuum+PIE model, as illustrated in panel b of Figure \ref{fig:serx1}.  The best fit parameters were determined to be log~$\xi$ = 1.5 $\pm$ 0.1 and log$(N$$_{\rm H}$/cm$^{-2}$) = 20.2 $\pm$ 0.2. The best-fit temperatures for the continuum models remained unaltered. The measured widths of the PIE emission lines were found to be $\sim$ 4200 km/s.
 The addition of the PIE  model successfully eliminated the residuals within the 0.65-0.9 keV range, although the residual in the 0.9-1.3 keV range persisted (see the lower sub-figure of panel b, Figure \ref{fig:serx1}). The fit quality significantly improved compared to the continuum-only fit ($\Delta$C-stat $>$ 2000). \\
When hard X-rays from an external source interact with an accretion disk, one can anticipate the occurrence of disk reflection. In earlier investigations, reflection features were identified in Ser X-1, including both Fe-K and Fe-L features \citep{2013ApJ...779L...2M, 2018ApJ...858L...5L, 2020MNRAS.494.3177M}. To address both emission residuals, we incorporated a continuum+PIE+reflection model for modeling the spectrum, as displayed in panel c of Figure \ref{fig:serx1}. The reflection feature was self-consistently modeled with \textsc{Cloudy} (see appendix for model details) with the SED described in Section 
\ref{cygser}.
The electron density was set at log $(n_{\rm e}$/cm$^{-3}$) = 20, and log$(N$$_{\rm H}$/cm$^{-2}$) was set at 23.0 since previous reflection models have assumed high electron density and column density \citep{2019MNRAS.489.3436J}.
The best-fit parameters were determined to be log~$\xi$ = 3.2 $\pm$ 0.1, an inner radius of log($r_{in}$/cm) = 6.20 $\pm$ 0.15, and an illumination angle of  29$^{\circ}$ $\pm$ 5$^{\circ}$. The blackbody temperatures in the continuum model remained unchanged. The fit quality further improved compared to the continuum-only model, with a $\Delta$C-stat $>$ 2900. 
Both emission residuals were successfully eliminated, as demonstrated in the lower subfigure of panel c, Figure \ref{fig:serx1}. 


\subsection{Cyg X-2:}\label{cyg x-1}

To model the continuum emission in Cyg X-2 within the energy range of 0.5-2.0 keV, we utilized the XSPEC model \texttt{tbabs*(diskbb+powerlaw)}, with the Galactic absorption hydrogen column density fixed at 2.2 $\times$ 10$^{21}$ cm$^{-2}$. Although broader energy spectra would require an additional single-temperature blackbody component \citep{2022ApJ...927..112L}, our continuum model adequately fits the 0.5-2.0 keV range.
The best-fit temperature for the \texttt{diskbb} component was found to be 1.64 $\pm$ 0.03 keV, with a photon index for the power-law measured at 2.78 $^{+0.06}_{-0.07}$. In Figure \ref{fig:cygx2}, Panel a, we presented the overlaid continuum model with the spectrum. The lower sub-figure displays the residual, revealing a significant positive residual between 0.5 and 1.3 keV, with a less pronounced negative residual found between 1.4 and 2.0 keV.\\
Subsequently, we incorporated a CIE plasma emission model into our spectral analysis, integrating it with the continuum model, as shown in panel b of Figure \ref{fig:cygx2}. 
The best-fit temperature for the \texttt{diskbb} component was determined to be 1.61 $\pm$ 0.04 keV, accompanied by a photon index for the power-law measured at 2.80 $\pm$ 0.07.
The CIE component yielded a best-fit temperature of log$(T$/K) = 7.01 $\pm$ 0.03, and a best-fit column density of log$(N$$_{\rm H}$/cm$^{-2}$) = 20.9 $\pm$ 0.2. The addition of the CIE component notably improved the fit, although some emission residuals persisted between 0.65 and 1 keV and 1.1 and 1.2 keV (see the lower panel of \ref{fig:cygx2}b ). Minor absorption residuals persisted within the 1.4 and 1.6 keV range. 
The velocity broadening was determined to be v$_{\rm broad}$ $\sim$ 3100  km/s.
 The introduction of a CIE model greatly enhanced the goodness of fit, with $\Delta$C-stat $\sim$ 250 compared to the continuum-only model.\\
Next, we integrated a PIE plasma emission/absorption model with the continuum+CIE model to fit the observed spectrum, as shown in the bottom left panel of Figure \ref{fig:cygx2}. The best-fit parameters for the \texttt{diskbb} temperature and power-law index were 1.63 $\pm$ 0.05 keV and 2.80 $^{+0.08}_{-0.09}$, respectively. The CIE emission model parameters remained unchanged. For the PIE emission model, the best-fit parameters were log $\xi$ = 2.51 $\pm$ 0.25 and log$(N$$_{\rm H}$/cm$^{-2}$) = 21.5 $\pm$ 0.3, while for the PIE absorption model, the best-fit parameters were log $\xi$ = 3.50 $\pm$ 0.25 and log$(N$$_{\rm H}$/cm$^{-2}$) = 21.7 $\pm$ 0.3. Absorption features were identified at energies above 1.4 keV, with a marginal blueshift of $\sim$ 0.007c suggesting an associated wind outflow, consistent with the findings reported by \citet{2009ApJ...699.1223S} in their Suzaku observation of Cyg X-2. The line broadening of the emission/absorption lines was determined to be $\sim$ 3400  km/s.  The addition of the PIE model further improved the fit quality  with $\Delta$C-stat $>$ 350 compared to the continuum-only model. The fit successfully eliminated the absorption residuals, as demonstrated in the lower subfigure of  Figure \ref{fig:cygx2}c, but slight emission residual persisted between 0.65 and 1.0 keV.

In earlier studies, reflection features were detected in Cyg X-2 within the energy range of 0.5-0.8 keV.
For comprehensive modeling of the 1 keV feature in Cyg X-2, a continuum + CIE emission + PIE emission/absorption model + reflection model was utilized to fit the observed spectrum, as shown in the bottom right panel of Figure \ref{fig:cygx2}. The best-fit parameters for the \texttt{diskbb} and power-law index remained the same.
For the CIE model, we derived the best-fit temperature to be log$(T$/K) = 7.11 $\pm$ 0.03,  best-fit column density to be and log$(N$$_{\rm H}$/cm$^{-2}$) = 20.8 $\pm$ 0.2. The parameters of the PIE emission/absorption model and velocity broadening values remained unchanged from the continuum + CIE + PIE emission/absorption model. 
For the reflection model, we assumed high electron density (log $(n_{\rm e}$/cm$^{-3}$) = 20) and column density (log$(N$$_{\rm H}$/cm$^{-2}$) = 23)  and maintained them as fixed parameters. The best-fit parameters for the reflection model were determined to be log~$\xi$ = 2.1 $\pm$ 0.3, an inner radius of log($r_{in}$/cm) = 6.50 $\pm$ 0.25, and an illumination angle of  54$^{\circ}$ $\pm$ 8$^{\circ}$. The  fit further improved with $\Delta$C-stat $>$ 380 compared to the continuum-only model. The application of this combined model successfully eliminated both emission and absorption residuals, as demonstrated in the lower subfigure of  Figure \ref{fig:cygx2}d. 
The contour plots displayed in Fig \ref{fig:em1}, \ref{fig:em2}, \ref{fig:abs1},  \ref{fig:abs2}, \ref{fig:col}, and the linear plot displayed in  Fig \ref{fig:ref} illustrate the best-fit parameters of this final model, symbolized by red inverted squares.

\section{Conclusions}

In this paper, we focused on the 1 keV feature observed in XRBs, which offers vital insights into the mechanisms governing accretion onto neutron stars or black holes, and the characteristics of super-Eddington accretion. The primary objective of our study was to present a comprehensive theoretical framework that explains the origin of the 1 keV feature along with its variability, both in terms of its centroid and intensity.
We conducted a thorough analysis of emission and absorption lines under three specific conditions: photoionization equilibrium (PIE), collisional ionization equilibrium (CIE), and reflection of X-rays off the inner regions of an accretion disk. Using the spectral synthesis code \textsc{Cloudy}, we ran grid simulations by varying the ionization parameter and column density in PIE plasmas, temperature and column density in CIE plasmas, and the ionization parameter for modeling the reflection lines. Finally, to illustrate the application of this model, we utilized it to successfully fit and explain the physics behind the 1 keV features observed in 5 XRBs, two ULXs - NGC 247 ULX-1 and NGC 1313 X-1, one X-ray pulsar - Hercules X-1, and two typical LMXBs - Cyg X-2 and Serpens X-1.

\begin{itemize}
    \item 

In our \textsc{Cloudy} simulations, we used observed SEDs from NGC 1313 X-1, NGC 247 ULX-1, Hercules X-1, Cyg X-2, and Serpens X-1, covering X-ray, optical, and UV energy ranges, as radiation sources. This study aimed to understand how variations in plasma properties influence the intensity and centroid positions of the 1 keV feature across different types of plasmas. For PIE plasma, we constructed three sets of emission line blends---$\mathrm{Em}_{\text{blend}}$ (0.6--1.4 keV), $\mathrm{Em}_{\text{left}}$ (0.6--1.0 keV), and $\mathrm{Em}_{\text{right}}$ (1.0--1.4 keV)---and similarly, three absorption line blends---$\mathrm{Abs}_{\text{blend}}$ (0.5--2.0 keV), $\mathrm{Abs}_{\text{left}}$ (0.5--1.0 keV), and $\mathrm{Abs}_{\text{right}}$ (1.0--2.0 keV). These were studied using logarithmic grids for $\xi$ and $N_{\text{H}}$. For CIE plasma, we utilized identical energy ranges as the PIE emission model to construct emission line blends and analyzed them over grids for temperature and $N_{\text{H}}$. To model reflection lines, we generated three reflection line blends---$\mathrm{Reflect}_{\text{blend}}$ (0.5--2.0 keV), $\mathrm{Reflect}_{\text{left}}$ (0.5--1.0 keV), and $\mathrm{Reflect}_{\text{right}}$ (1.0--2.0 keV). These blends were analyzed across a range of $\xi$ values, with a particularly high column density ($N_{\text{H}}$ = 10$^{23}$ cm$^{-2}$) assigned to simulate conditions typical of dense astrophysical environments.

\end{itemize}

\begin{itemize}
    \item 
    We effectively describe the origin and variation of the 1 keV feature across a diverse range of X-ray binaries using \textit{XMM-Newton}/RGS and \textit{NICER} observations.
    The 1 keV feature in NGC 1313 X-1 was best modeled by a combination of a two-blackbody continuum and a PIE emission/absorption model, which yielded the most precise fit to the spectra. NGC 247 ULX-1
    necessitated a two-blackbody continuum + PIE emission/absorption + CIE emission model for effectively modeling the 1 keV feature. In the case of NGC 247 ULX-1, we found that an effective model for accurately representing the 1 keV feature required incorporating a combination of a two-blackbody continuum, PIE emission/absorption, and CIE emission. Similarly, in Hercules X-1, the integration of a blackbody+powerlaw continuum, CIE emission, and PIE emission/absorption models effectively eliminated the residuals at 1 keV. For Cyg X-2, achieving a thorough modeling of the 1 keV feature required integrating reflection lines from the inner accretion disk, PIE emission/absorption, and CIE emission components, in addition to the disk blackbody component and a power-law component utilized for modeling the continuum. 
    For Serpens X-1, effectively eliminating the 1 keV residual required a combination of blackbody and disk blackbody continuum components, along with PIE emission and reflection lines.
    
\end{itemize}

\begin{itemize}
    \item Table \ref{tab:1keV_fullsummary} summarizes the best-fit parameters characterizing the 1 keV feature, including component classification (CIE, PIE, and reflection). A complete list of all the lines within the 1 keV emission, absorption, and reflection feature across the entire parameter space explored in this paper, both bright and faint, has been listed in Table \ref{t:p2}. The instruction to generate PIE emission/absorption, CIE emission and reflection line blends have been discussed in the appendix.
    
\end{itemize}

\begin{itemize}
    \item 

This paper revisits systems previously analyzed, with the current study  focused on understanding the physical processes responsible for the origin and variability of the 1 keV feature, a novel aspect of this work. Unlike prior studies that detailed the properties and geometries of these systems, our research is specifically aimed at investigating the atomic origin of the 1 keV feature. 
All our fittings were conducted using \textsc{cloudy}, and we observed that our results were largely consistent with those of previous studies, which used codes such as \textsc{spex} \citep{1996uxsa.conf..411K} and \textsc{Relxill} \citep{2014ApJ...782...76G}, although there were some minor discrepancies.  
\citet{2020MNRAS.492.4646P} modeled the intermediate-bright state of NGC 1313 X-1 using a PIE emission/absorption model implemented through \textsc{spex}.  Similarly, our analysis 
of the same observation using the \textsc{cloudy} model, which also required a  PIE emission/absorption approach, yielded ionization parameters and column densities that closely align with those reported in the prior study. The data for NGC 247 ULX-1, previously fitted with separate CIE and PIE models by \citet{2021MNRAS.505.5058P}, highlighted the need for more complex models to interpret the 1 keV features.
Our results indicate that a combined CIE and PIE model approach fits the residuals more effectively, though this difference may be attributed to our deliberate omission of a CIE absorption model, as the absorption is predominantly linked to photoionized winds in XRBs \citep{2015ApJ...807..107H}. 
In their study of Her X-1, \citet{2022ApJ...936..185K} modeled the 1 keV spectral feature using a Gaussian profile with a width between 0.35 and 0.4 keV. They emphasized the necessity of applying physical modeling to fully comprehend the nature of this feature. In this work, we utilize a combination of  CIE emission and PIE emission/absorption models to adequately fit this spectral feature.
In modeling the Fe L complex reflection feature near 1 keV for Ser X-1, \citet{2018ApJ...858L...5L} used the \textsc{Relxillns} reflection model but noted persistent residuals between 0.5-0.9 keV. We addressed the residuals around 1 keV using a combination of reflection emission lines and a PIE emission model, finding our parameters for ionization and inner disk radius consistent with those reported by \citet{2018ApJ...858L...5L} for the reflection feature.
The higher illumination angle obtained in our analysis may 
 reflect differences in the underlying code assumptions adopted in each study, which are beyond the scope of this work.
For Cyg X-1, \citet{2022ApJ...927..112L} applied \textsc{Relxillns} and the CIE emission model \textsc{mekal} \citep{1995ApJ...438L.115L} to model the 1 keV feature, but  slight absorption residuals remained. We incorporated a PIE model to address the absorption, complementing it with CIE and reflection models to achieve a spectrum fit with minimal residuals.
The temperature of our CIE model was in agreement with that of the \textsc{mekal} model, and the parameters for ionization, illumination angle, and inner disk radius were in agreement with the findings reported by \citet{2022ApJ...927..112L} for the reflection model.

\end{itemize}

To confirm that the 1 keV emission feature can be disentangled from the continuum and assess its robustness while excluding model degeneracies or artificial correlations, we compare our results with previously reported broadband spectral fits. For NGC 1313 X-1, broadband continuum modeling with a two-blackbody approach was performed over the 0.3–20 keV range using \textit{XMM-Newton} RGS, EPIC-PN, and \textit{NuSTAR} spectra \citep{2020MNRAS.492.4646P}, supporting the detection of the 1 keV feature.  For NGC 247, \citet{2021MNRAS.505.5058P} performed spectral fits with EPIC and RGS data in the 0.3–10.0 keV range, detecting the 1 keV feature with a best-fit continuum model of two blackbody components. \citet{2022ApJ...927..112L} conducted continuum spectral modeling of Cygnus X-2 in the 0.5–30 keV range using \textit{NuSTAR} and \textit{NICER}, employing a disk blackbody and power-law model, and reported an emission feature near 1 keV. Using \textit{NICER}, \textit{NuSTAR}, and \textit{XMM-Newton} data, the continuum modeling of Ser X-1 with a disk blackbody, blackbody, and power-law component in the 0.4–10.0 keV range revealed residuals near 1 keV \citep{2018ApJ...858L...5L}. Using \textit{Suzaku} and \textit{NuSTAR} data, \citet{2013ApJ...779...69F} explored various continuum models in the 0.7–60 keV range and reported a strong residual around 1 keV in Her X-1.


\section*{Acknowledgements}
PC acknowledges support from NASA XRISM grant 80NSSC26K0317. GJF acknowledges support by JWST through AR6428, AR6419 GO5354, and GO5018.

\appendix

\section{How to create the 1 keV blend?}
Here, we describe the \textsc{Cloudy} commands used in our study of the 1 keV feature.
The SED for each source was determined using the command:\\
\noindent
\texttt{Table sed "input.txt"\\}
with the input SEDs corresponding to the observed SEDs discussed in Section 3.
We used the \texttt{blend} command, described in \citet{2017RMxAA..53..385F} in Section 2.3.2, to define line blends of the lines listed in Table \ref{t:p2}. 
We created two line blends \texttt{blend 11} for lines with energies $>$ 1 keV, and   \texttt{blend 13} for lines with energies $<$ 1 keV listed in table \ref{t:p2}.
For creating line blends the command we used in \textsc{Cloudy} was:\\
\noindent
\texttt{
set blend 11 \\
Mg12  6.58008 \\
Al12    6.63476 \\
Si13    6.64803 \\
Si13    6.68827 \\
Si13    6.74039 \\
Mg11    7.03734 \\
Mg12    7.10615 \\
Al13    7.17271 \\
Mg11    7.31028 \\
Mg11    7.47313 \\
Al12    7.75730 \\
Mg11    7.85052 \\
Fe20    7.86906 \\
Fe20    8.20998 \\
Fe20    8.27000 \\
Fe20    8.31000 \\
Mg12    8.42100 \\
Fe20    8.44003 \\
Fe20    8.46002 \\
Fe19    8.49473 \\
Fe20    8.50000 \\
Fe20    8.69996 \\
Fe22    8.71500 \\
Fe22    8.72200 \\
Fe20    8.74004 \\
Fe20    8.77001 \\
Fe19    8.81003 \\
Fe21    8.84000 \\
Fe21    8.85500 \\
Fe21    8.89800 \\
Fe20    8.89996 \\
Fe19    8.91997 \\
Fe19    8.91997 \\
Fe20    8.93001 \\
Fe19    8.93001 \\
Fe19    8.96001 \\
Fe22    8.97700 \\
Fe19    9.03000 \\
Fe20    9.06500 \\
Fe20    9.06900 \\
Fe20    9.06900 \\
Fe19    9.06997 \\
Fe19    9.06997 \\
Fe19    9.06997 \\
Fe22    9.07300 \\
Fe20    9.11000 \\
Fe20    9.11000 \\
Fe22    9.12200 \\
Fe21    9.14000 \\
Fe22    9.14800 \\
Fe20    9.16300 \\
Mg11    9.16875 \\
Fe20    9.18400 \\
Fe19    9.18999 \\
Fe19    9.19997 \\
Fe19    9.19997 \\
Fe20    9.21600 \\
Fe20    9.21600 \\
Fe19    9.21999 \\
Fe20    9.22000 \\
Fe20    9.22000 \\
Fe19    9.23003 \\
Fe19    9.23003 \\
Mg11    9.23121 \\
Fe20    9.25800 \\
Fe20    9.28100 \\
Mg11    9.31434 \\
Fe19    9.32001 \\
Fe19    9.32001 \\
Fe20    9.32500 \\
Ni25    9.33000 \\
Fe19    9.33001 \\
Ne10    9.36162 \\
Ni20    9.37700 \\
Ni20    9.38500 \\
Fe19    9.45001 \\
Ni20    9.45500 \\
Fe21    9.47500 \\
Ne10    9.48075 \\
Fe21    9.48200 \\
Fe21    9.54200 \\
Fe21    9.54800 \\
Ni20    9.55800 \\
Ni20    9.55900 \\
Fe21    9.58200 \\
Fe21    9.58700 \\
Fe19    9.63900 \\
Fe21    9.69000 \\
Fe19    9.69100 \\
Fe19    9.69100 \\
Fe21    9.70000 \\
Fe21    9.70500 \\
Ne10    9.70818 \\
Fe19    9.72600 \\
Fe21    9.82100 \\
Fe19    9.84800 \\
Fe19    9.85200 \\
Fe19    9.88800 \\
Na11    10.02500 \\
Fe18    10.08000 \\
Ni19    10.11000 \\
Ne10    10.23890 \\
Fe18    10.41000 \\
Fe18    10.41000 \\
Fe23    10.50600 \\
Fe18    10.52600 \\
Fe24    10.61900 \\
Fe19    10.63200 \\
Fe19    10.63200 \\
Fe19    10.63300 \\
Fe19    10.65500 \\
Fe19    10.65700 \\
Fe24    10.66300 \\
Fe19    10.68400 \\
Fe19    10.70200 \\
Cr20    10.71200 \\
Fe19    10.74400 \\
Fe19    10.75800 \\
Fe19    10.76000 \\
Fe19    10.76000 \\
Ni20    10.77200 \\
Fe19    10.80500 \\
Fe19    10.80500 \\
Fe19    10.81600 \\
Fe19    10.82700 \\
Fe19    10.82700 \\
Fe19    10.88000 \\
Fe19    10.91600 \\
Fe19    10.93300 \\
Fe19    10.93300 \\
Fe23    10.98000 \\
Na10    11.00260 \\
Fe23    11.01800 \\
Ni20    11.13800 \\
Ni20    11.13800 \\
Ni20    11.15800 \\
Ni20    11.22600 \\
Ni20    11.22600 \\
Ni21    11.22700 \\
Ni21    11.24100 \\
Ni21    11.24200 \\
Ni21    11.27200 \\
Ni20    11.28200 \\
Ni21    11.30200 \\
Ni21    11.31900 \\
Fe18    11.32600 \\
Fe18    11.32600 \\
Ni21    11.38000 \\
Fe18    11.42000 \\
Fe22    11.44200 \\
Fe22    11.45900 \\
Ni21    11.46800 \\
Fe23    11.48500 \\
Fe22    11.51000 \\
Ni21    11.51600 \\
Ni21    11.51700 \\
Fe18    11.52500 \\
Fe18    11.52500 \\
Ni21    11.53900 \\
Ni19    11.53900 \\
Ne 9    11.54660 \\
Mn23	11.57660 \\
Ni21	11.59700 \\
Fe22	11.59900 \\
Fe22	11.66900 \\
Fe23	11.71800 \\
Fe23	11.73700 \\
Fe20	11.73900 \\
Fe22	11.76800 \\
Ni20	11.78700 \\
Ni20	11.83200 \\
Ni20	11.84100 \\
Fe23	11.84600 \\
Ni20	11.86500 \\
Ni20	11.87400 \\
Fe22	11.92100 \\
Fe20	11.93300 \\
Fe22	11.93400 \\
Fe21	11.93800 \\
Ni20	11.96100 \\
Mn22	11.97000 \\
Fe21	11.97500 \\
Ni20	11.97800 \\
Fe20	11.98700 \\
Ca18	11.98900 \\
Ni20	11.99100 \\
Ni20	11.99100 \\
Mn22	11.99800 \\
Ni20	12.00600 \\
Ti19	12.01000 \\
Fe23	12.02700 \\
Fe21	12.04400 \\
Ni20	12.04700 \\
Ni20	12.08100 \\
Fe21	12.08200 \\
Fe21	12.10700 \\
Ni20	12.11200 \\
Ni20	12.13000 \\
Ne10	12.13390 \\
end of blend 11\\
}

\noindent
\texttt{\\
\\
set blend 13\\
Fe21	12.14600 \\
Ni20	12.15700 \\
Fe23	12.16100 \\
Ni21	12.16500 \\
Fe22	12.19300 \\
Ni21	12.20900 \\
Fe21	12.26100 \\
Ni21	12.27600 \\
Fe21	12.28200 \\
Fe21	12.29700 \\
Fe21	12.32700 \\
Fe21	12.32700 \\
Fe23	12.35100 \\
Fe21	12.39500 \\
Fe21	12.42200 \\
Fe20	12.42600 \\
Fe20	12.42600 \\
Ni19	12.43500 \\
Fe23	12.44400 \\
Ni21	12.44600 \\
Fe21	12.46200 \\
Ni21	12.47200 \\
Fe21	12.49000 \\
Fe21	12.49200 \\
Fe23	12.49300 \\
Fe21	12.49900 \\
Fe21	12.50000 \\
Fe21	12.52300 \\
Fe21	12.52300 \\
Fe21	12.53300 \\
Ni21	12.53300 \\
Fe20	12.56600 \\
Fe21	12.56800 \\
Fe20	12.58100 \\
Fe20	12.58100 \\
Ni21	12.59100 \\
Cr22	12.61300 \\
Fe21	12.62300 \\
Ca18	12.63600 \\
Ca18	12.63600 \\
Ni21	12.64800 \\
Fe23	12.65300 \\
Cr22	12.65500 \\
Ni19	12.65600 \\
Fe21	12.66300 \\
Fe21	12.69100 \\
Fe21	12.70100 \\
Fe23	12.70300 \\
Fe22	12.74300 \\
Fe20	12.75300 \\
Fe21	12.77200 \\
Fe20	12.80400 \\
Fe20	12.80400 \\
Fe20	12.81200 \\
Fe18	12.81800 \\
Fe21	12.82200 \\
Fe20	12.82400 \\
Fe20	12.82700 \\
Fe20	12.84500 \\
Fe21	12.87000 \\
Fe20	12.90500 \\
Fe19	12.92400 \\
Fe19	12.92400 \\
Fe19	12.92400 \\
Ni20	12.92700 \\
Fe22	12.93600 \\
Fe20	12.95100 \\
Fe20	12.96600 \\
Fe20	12.96600 \\
Fe20	12.98200 \\
Fe20	12.99100 \\
Fe19	13.01800 \\
Ni20	13.03200 \\
Fe19	13.03900 \\
Fe20	13.04400 \\
Fe20	13.04600 \\
Fe19	13.05100 \\
Fe20	13.05200 \\
Fe21	13.05200 \\
Fe20	13.05900 \\
Fe19	13.07500 \\
Fe19	13.07500 \\
Ni20	13.07500 \\
Fe20	13.07800 \\
Fe20	13.09100 \\
Fe19	13.09100 \\
Fe20	13.09100 \\
Fe19	13.09100 \\
Fe20	13.11400 \\
Fe20	13.11400 \\
Fe20	13.12300 \\
Cr21	13.12300 \\
Fe20	13.14000 \\
Fe20	13.14300 \\
Fe21	13.17900 \\
Fe20	13.18800 \\
Fe20	13.20300 \\
Fe20	13.20600 \\
Fe19	13.21200 \\
Fe19	13.21200 \\
Fe19	13.21200 \\
Fe20	13.25300 \\
Fe20	13.25400 \\
Fe19	13.25400 \\
Fe21	13.25500 \\
Ni20	13.25600 \\
Fe20	13.26700 \\
Fe20	13.26700 \\
Fe20	13.26900 \\
Fe20	13.27000 \\
Ni20	13.28200 \\
Fe20	13.29200 \\
Fe20	13.30100 \\
Fe19	13.31100 \\
Fe18	13.31900 \\
Fe18	13.31900 \\
Fe19	13.32700 \\
Fe20	13.33300 \\
Fe19	13.33600 \\
Fe18	13.35500 \\
Fe20	13.35900 \\
Fe22	13.36100 \\
Fe20	13.36600 \\
Fe18	13.37400 \\
Fe20	13.37900 \\
Fe20	13.38200 \\
Fe18    13.39700 \\
Fe20    13.40200 \\
Fe20    13.40500 \\
Fe20    13.41900 \\
Fe18    13.42400 \\
Fe19    13.43000 \\
Ne 9    13.44710 \\
Fe19    13.45600 \\
Ca17    13.46000 \\
Fe19    13.46200 \\
Fe18    13.46400 \\
Fe20    13.46700 \\
Fe19    13.47100 \\
Fe21    13.48200 \\
Fe19    13.50600 \\
Fe19    13.50700 \\
Fe20    13.51700 \\
Fe19    13.52500 \\
Fe20    13.53300 \\
Fe20    13.55300 \\
Fe19    13.55400 \\
Fe19    13.55500 \\
Fe19    13.55700 \\
Fe21    13.57400 \\
Fe22    13.61700 \\
Fe19    13.62000 \\
Fe19    13.62100 \\
Fe19    13.63400 \\
Fe19    13.63600 \\
Fe19    13.64300 \\
Fe19    13.64600 \\
Fe19    13.64800 \\
Fe20    13.67000 \\
Fe19    13.67200 \\
Fe19    13.67200 \\
Fe19    13.67300 \\
Fe22    13.67400 \\
Fe19    13.69100 \\
Fe19    13.69400 \\
Fe21    13.71700 \\
Fe19    13.72000 \\
Fe19    13.72100 \\
Fe19    13.72100 \\
Fe19    13.74100 \\
Fe19    13.75400 \\
Fe19    13.76200 \\
Fe19    13.76900 \\
Fe22    13.77100 \\
Ni19    13.77800 \\
Fe20    13.78100 \\
Fe19    13.79200 \\
Fe19    13.79900 \\
Fe19    13.82200 \\
Fe19    13.82200 \\
Fe19    13.84100 \\
Fe19    13.84300 \\
Fe19    13.84400 \\
Fe20    13.84400 \\
Fe19    13.87100 \\
Fe19    13.87200 \\
Fe19    13.93600 \\
Fe19    13.93800 \\
Fe19    13.93800 \\
Fe19    13.93800 \\
Fe19    13.94200 \\
Fe19    13.95700 \\
Fe18    13.96200 \\
Fe19    13.96400 \\
Fe19    13.97000 \\
Fe20    13.97200 \\
Fe21    14.00800 \\
Fe19    14.01700 \\
Ni19    14.04000 \\
Ca18    14.04900 \\
Ca18    14.05900 \\
Fe19    14.07100 \\
Fe19    14.08600 \\
Fe19    14.08600 \\
Fe19    14.11400 \\
Fe19    14.11400 \\
Fe20    14.11500 \\
Fe20    14.12300 \\
Fe18    14.12400 \\
Fe19    14.12700 \\
Fe19    14.12800 \\
Fe18    14.13600 \\
Fe18    14.14400 \\
Fe19    14.17500 \\
Fe20    14.19000 \\
Fe18    14.20400 \\
Fe18    14.20900 \\
Cr21    14.24500 \\
Fe20    14.24800 \\
Fe19    14.25200 \\
Fe18    14.25800 \\
Fe18    14.25800 \\
Fe18    14.34400 \\
Fe18    14.35200 \\
V 20    14.35960 \\
Fe18    14.37300 \\
Fe18    14.41900 \\
Fe18    14.41900 \\
Fe18    14.45300 \\
Fe18    14.47000 \\
Fe18    14.48700 \\
Fe20    14.49300 \\
Fe20    14.50000 \\
Fe18    14.53700 \\
Fe18    14.55100 \\
Fe19    14.57000 \\
Fe18    14.58000 \\
Fe18    14.61000 \\
O  8    14.63430 \\
Fe19    14.66400 \\
Fe19    14.66900 \\
Fe18    14.67100 \\
Fe19    14.69400 \\
Fe19    14.73800 \\
Fe19    14.74100 \\
Fe20    14.76400 \\
Fe18    14.77100 \\
O  8    14.82060 \\
Fe19    14.86800 \\
Fe20    14.91300 \\
Fe21    14.91600 \\
Fe19    14.93200 \\
Fe20    14.93200 \\
Fe19    14.93500 \\
Ca17    14.94000 \\
Fe19    14.99200 \\
Fe19    14.99200 \\
Fe17    15.01300 \\
Fe19    15.04200 \\
Fe20    15.06000 \\
Fe20    15.06300 \\
Fe19    15.08100 \\
Fe19    15.11400 \\
Fe19    15.16300 \\
O  8    15.17620 \\
Ar16    15.19000 \\
Fe19    15.19600 \\
Fe19    15.20800 \\
Ti20    15.21100 \\
Ti20    15.25300 \\
Fe17    15.26200 \\
Fe19    15.33000 \\
Fe19    15.35200 \\
Fe20    15.51500 \\
Fe18    15.62200 \\
Fe18    15.76600 \\
Fe18    15.82800 \\
Ti19    15.86500 \\
Ar16    15.93300 \\
Fe18    16.00500 \\
O  8    16.00590 \\
Fe18    16.02600 \\
Fe18    16.07200 \\
Fe19    16.11000 \\
Fe19    16.27200 \\
Fe19    16.34000 \\
Fe17    17.05100\\
Fe18    17.62180 \\
Ar16    17.73700 \\
Ar16    17.74700 \\
Ca18    18.69100 \\
S 14    18.72000 \\
Ca18    18.73200 \\
O  8    18.96890 \\
Ca17    19.55800 \\
S 14    19.69000 \\
N  7    19.82580 \\
Ca17    20.4340 \\
Ca16    20.8590 \\
K 17    20.8990 \\
N  7    20.9098 \\
K 17    20.9310 \\
Ca16    20.9510 \\
Ca16    21.0200 \\
Ca16    21.1130 \\
Ca16    21.4500 \\
O  7    21.6020 \\
Ca16    21.6100 \\
Ca16    21.6190 \\
S 14    21.8192 \\
K 16    21.9110 \\
Ca15    22.7300 \\
Ca15    22.7300 \\
Ca15    22.7590 \\
Ca15    22.7770 \\
Ca15    22.8210 \\
S 14    23.0050 \\
S 14    23.0150 \\
Ca17    23.1750 \\
Ca17    23.1750 \\
Ar16    23.5060 \\
Ar16    23.5460 \\
Ne10    24.2678 \\
S 13    24.5900 \\
Ar15    24.7370 \\
N  7    24.7810 \\
end of blend 13\\
}
The blend appears as a new emission
lines with the label \texttt{blnd} and  wavelengths of 11\AA \hspace{0.3 mm} and 13\AA \hspace{0.3 mm}.
We chose the blends' wavelengths of 
11\AA \hspace{0.3 mm} and 13\AA \hspace{0.3 mm}
rather than the energy 1 keV to make it easier to identify in the
emission-line output.

In previous versions of \textsc{Cloudy} 
(up to C23 or earlier), we only used experimental energy values sourced from the Chianti database due to their superior accuracy \citep{2013MNRAS.429.3133L}.
In the upcoming release, C24, there will be an option to incorporate theoretical energy values for instances where experimental data are absent. The The \textsc{Cloudy} command to use such a 
 `mixed' case is:\\
\texttt{Database Chianti mixed} \\
Nevertheless, for the 1 keV feature, the impact of this change appears to be minimal, as the newly introduced lines in the 1 keV range are faint, making little difference in their overall intensities.\\
The emitted line blends were saved with the command:\\
\texttt{save emitted continuum} \\
The reflected line blends were saved with the command:\\
\texttt{save reflected continuum} \\
The inner accretion disk radius can be specified using the following command (for example):\\
\texttt{radius 6.5     \# log of inner radius in cm}\\
The illumination angle can be specified using the  following command (for example):\\
\texttt{illumination angle 45 deg}

\begin{figure*}
\centering

\begin{minipage}{0.8\textwidth}
    \centering
    \includegraphics[width=\textwidth]{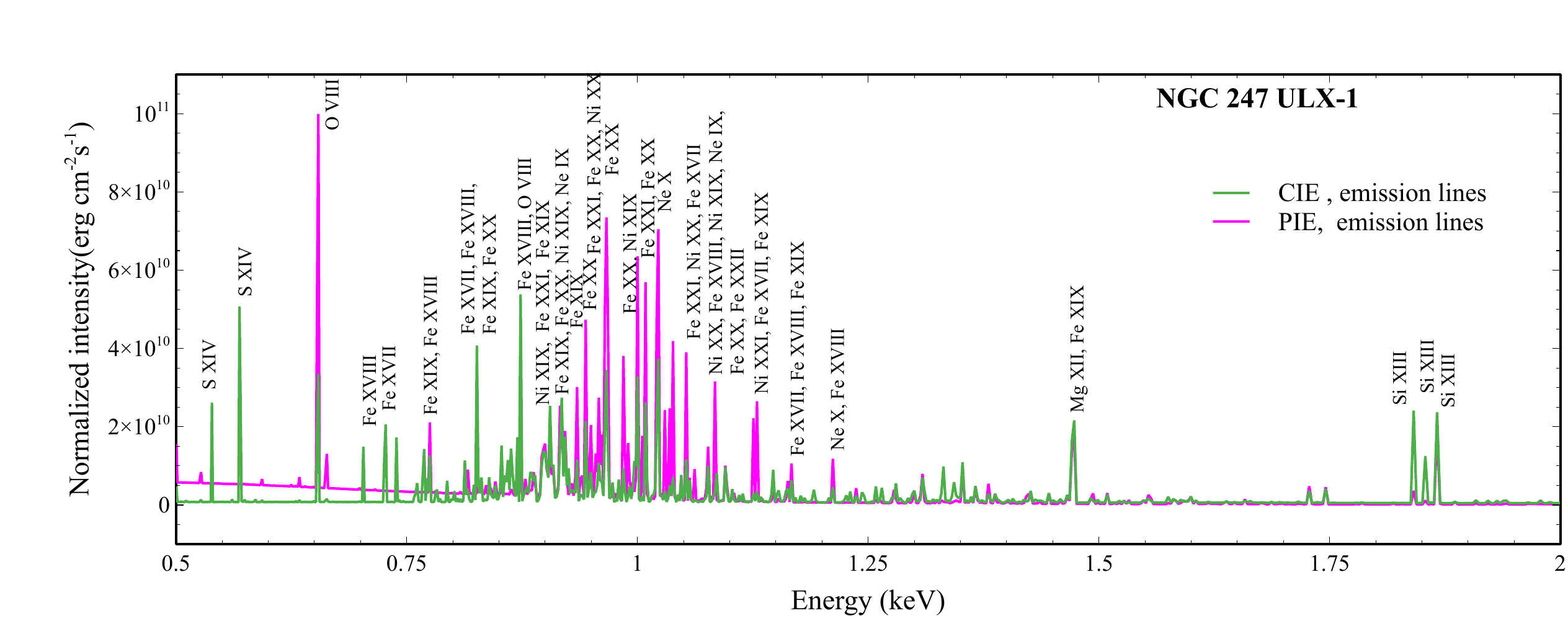}
\end{minipage}


\begin{minipage}{0.8\textwidth}
    \centering
    \includegraphics[width=\textwidth]{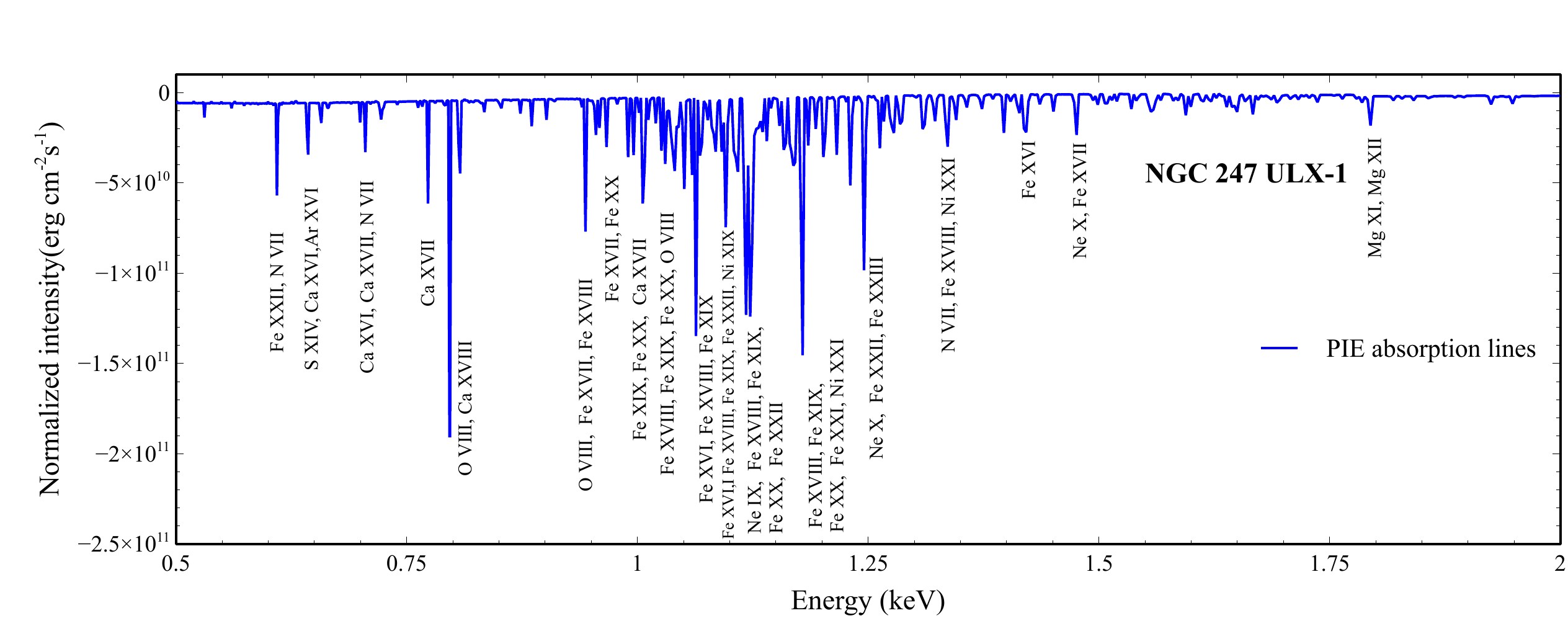}
\end{minipage}


\begin{minipage}{0.4\textwidth} 
    \centering
    \includegraphics[width=\textwidth]{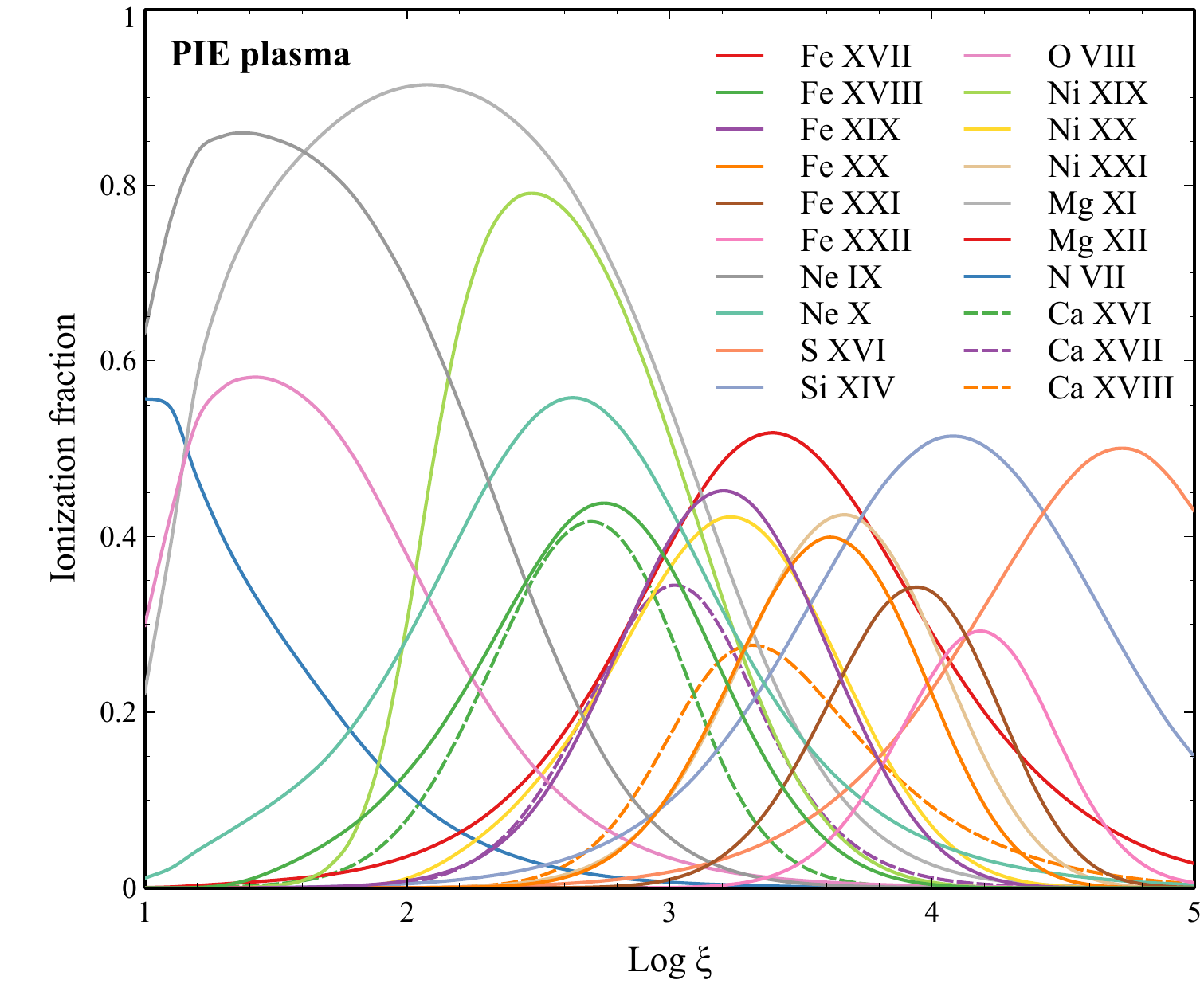}
\end{minipage}
\begin{minipage}{0.4\textwidth}
    \centering
    \includegraphics[width=\textwidth]{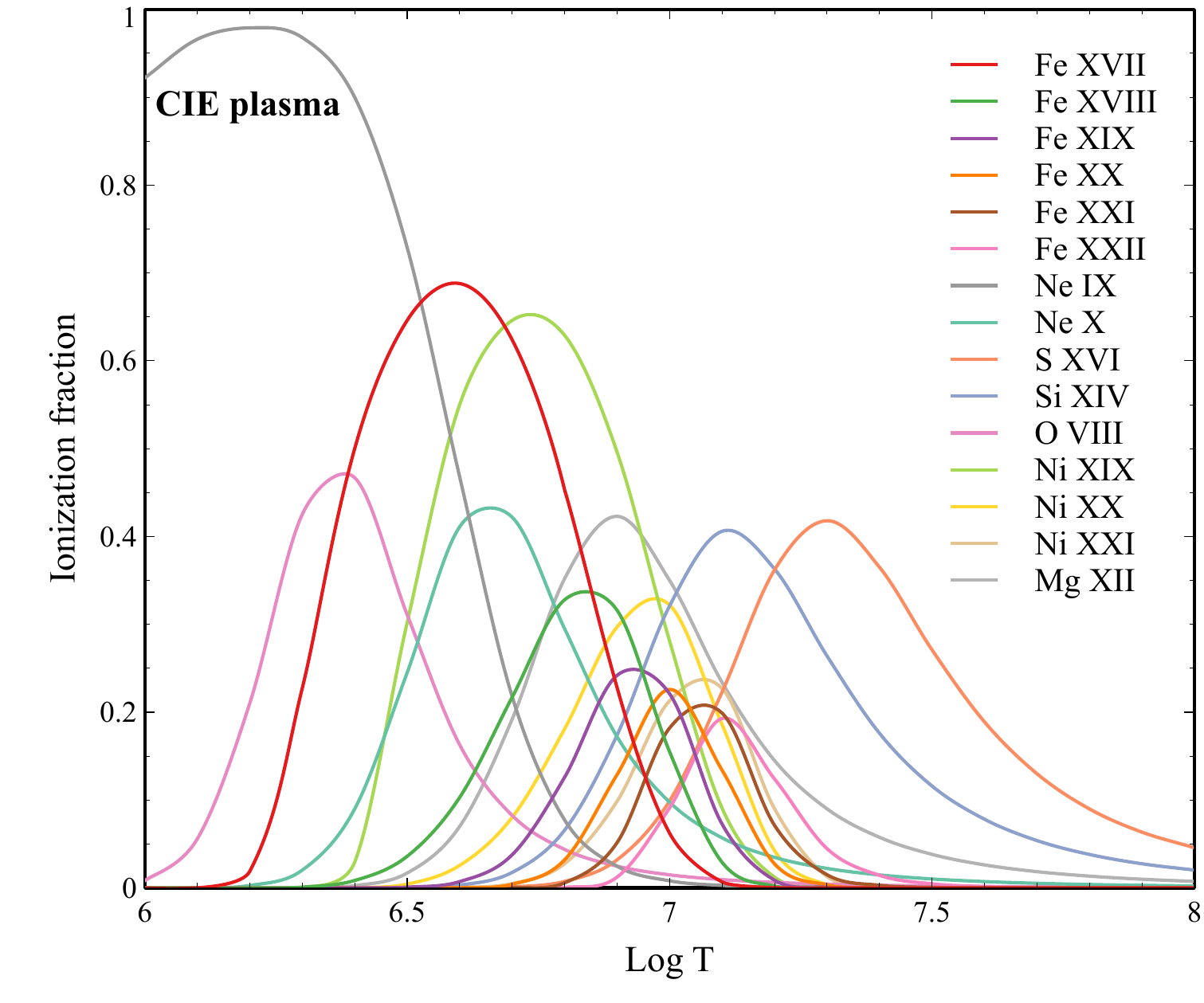}
\end{minipage}

 \caption{High-resolution \textsc{Cloudy} model of the 1 keV feature for NGC 247 ULX-1 at the spectral resolution of \textit{NewAthena}. 
Top panel: PIE and CIE emission lines for the best-fit parameter values reported in Section \ref{ngc247}. 
Middle panel: PIE absorption lines for the best-fit parameter values reported in Section \ref{ngc247}. 
Bottom left: Charge state distribution for the PIE plasma. 
Bottom right: Charge state distribution for the CIE plasma.}

\label{fig:hires}

\end{figure*}

To provide a detailed view of the high-resolution spectral lines contributing to the 1 keV complex, Figure \ref{fig:hires} presents a \textsc{Cloudy} model of NGC 247 ULX-1, constructed using the SED from Figure \ref{fig:sed} and adjusted to the spectral resolution of \textit{NewAthena} for the best-fit parameters reported in Section \ref{ngc247}. This model quantifies the atomic line contributions to the observed spectra. The spectrum is decomposed into its individual CIE and PIE components, with strong lines from Table \ref{t:p2} labeled for clarity.
The top panel displays emission lines from both CIE and PIE, while the middle panel highlights PIE absorption lines. The bottom panel examines the charge state distribution of the contributing ions in CIE and PIE, providing a quantitative assessment of their role in shaping the 1 keV feature.

\section{Impact of Column Density on Emission Line Strengths in CIE plasma}

 Although the optically thin approximation is commonly adopted in CIE modeling, \textsc{Cloudy} includes optical depth effects by default.  Line emission is computed by solving a fully coupled system of equations governing level populations and radiative transitions, as described in Section~III of \citet{1989ApJ...347..640R}. An in-depth discussion of optical depth effects on soft X-ray spectra  is presented in Sections~2 and 4 of \citet{2022ApJ...935...70C}.

The mathematical framework describing the dependence of line intensities on optical depth, and thus on hydrogen column density, is presented below.

\noindent Optical depth scales with  $N_{\mathrm{H}}$ and can be expressed as:
\begin{equation}
\tau = N_{\mathrm{H}}\, \alpha_\nu,
\end{equation}
where \( \alpha_\nu \) is the absorption cross-section at frequency \( \nu \) 
(see Equations~2 and 3 in \citealt{2020ApJ...901...69C}).

Thus, the impact of increasing column density can be understood in terms of how line photons are attenuated through photoelectric absorption and electron scattering. These processes can significantly suppress the observed line intensities at higher column densities. To estimate the fraction of photons that escape such an environment, \citet{2022ApJ...935...70C} introduced the \textit{line modification factor}, \( f_{\rm mod} \), which represents the survival probability of a line photon subject to N number of scatterings:
\begin{equation}
f_{\rm mod} = (1 - P_{\rm photoelectric} - P_{\rm scattering})^N,
\end{equation}
where \( N \) is the mean number of scatterings a photon experiences before escaping. \( N \) depends on  \( \tau \) (and therefore $N_{\mathrm{H}}$ ) and is approximated by \citep{1979ApJ...229..274F}:
\begin{equation}
N = \frac{1.11\, \tau^{1.071}}{1 + (\log \tau / 5.5)^5}.
\end{equation}

The probability of photoelectric absorption per scattering, \( P_{\rm photoelectric} \), is given by the ratio of the photoelectric opacity (\( \kappa_{\rm photoelectric} \)) to the total opacity (\( \kappa_{\rm total} \)):
\begin{equation}
\label{eqn:Pphoto}
P_{\rm photoelectric} = \frac{\kappa_{\rm photoelectric}}{\kappa_{\rm total}}.
\end{equation}

The probability that a line photon is scattered by free electrons, \( P_{\rm scattering} \), is given by the ratio of the electron scattering opacity (\( \kappa_{\rm scattering} \)) to the total opacity:
\begin{equation}
\label{eqn:Pscatt}
P_{\rm scattering} = \frac{\kappa_{\rm scattering}}{\kappa_{\rm total}}.
\end{equation}

The total opacity  is the sum of the line opacity (\( \kappa_{\rm line} \)), photoelectric opacity , and electron scattering opacity :
\begin{equation}
\kappa_{\rm total} = \kappa_{\rm line} + \kappa_{\rm photoelectric} + \kappa_{\rm scattering}.
\end{equation}

The line opacity (\( \kappa_{\rm line} \)) is expressed as the product of the ionic number density (\( n_{\rm ion} \)) and the absorption cross-section (\( \alpha_\nu \)) for the corresponding transition:
\begin{equation}
\kappa_{\rm line} = n_{\rm ion} \alpha_\nu
\end{equation}

where the absorption cross-section, \( \alpha_\nu \), is described by:
\begin{equation}
\alpha_\nu(x) = 2.24484 \times 10^{-14} A_{u,l} \lambda_{\mu m}^3 \frac{g_u}{g_l} \frac{\varphi_\nu(x)}{u_{\rm Dop}}.
\end{equation}

Here, \( A_{u,l} \) is the transition probability from level \( u \) to level \( l \), \( \lambda_{\mu m} \) is the wavelength in microns, \( g_u \) and \( g_l \) are the statistical weights of the upper and lower levels, \( u_{\rm Dop} \) is the Doppler velocity, and \( \varphi_\nu(x) = 1 \) at line center.

\begin{table*}

\centering{ 
\setlength{\tabcolsep}{-2 pt} 
\caption{List of spectral lines in the 1 keV blend. Line intensities vary depending on the shape of the SED, $\xi$, $N$$_{\rm H}$, and $T$.}\label{t:p2} }
\scriptsize
\begin{tabular}{c|c|c|c|c|c|c|c|c|c|c|c}
\hline
 Ion  &$\lambda$ (\AA) & Ion  &$\lambda$ (\AA) & Ion &$\lambda$ (\AA) &Ion  &$\lambda$ (\AA) &Ion  &$\lambda$ (\AA) &Ion  &$\lambda$ (\AA)\\
\hline
  Mg\,XII & 6.58008 & Al\,XII & 6.63476 & Si\,XIII & 6.64803 & Si\,XIII & 6.68827 & Si\,XIII & 6.74039 &Mg\,XII & 7.10615  \\
  Al\,XIII & 7.17271 & Mg\,XI & 7.31028 & Mg\,XI & 7.47313 & Al\,XII & 7.75730  & Mg\,XI & 7.85052   & Fe\,XX & 8.20998  \\
 Fe\,XX & 8.27000 & Fe\,XX & 8.31000 &Mg\,XII & 8.42100 & Fe\,XX & 8.44003 &  Fe\,XX & 8.46002 & Fe\,XX & 8.50000  \\
    Fe\,XX & 8.69996 & Fe XXII & 8.71500 & Fe XXII  & 8.72200 & Fe XX	& 8.74004 & Fe XX & 8.77001 & Fe XIX & 8.81003\\

Fe XXI & 8.84000 & Fe XXI	&8.85500 & Fe XXI & 8.89800 & Fe XX	& 8.89996 &Fe XX	& 8.91997  & Fe XX	& 8.93001\\

Fe XIX & 8.96001 & Fe XXII & 8.97700 & Fe XIX 	& 9.03000 & Fe XX	 &9.06500 &  Fe XIX & 	9.06997 & Fe XXII & 9.07300\\
Fe XX	& 9.11000 & Fe XXII & 9.12200& Fe XXI	 & 9.14000& Fe XXII	& 9.14800 & Fe XX	& 9.16300 & Mg XI	&9.16875\\
Fe XX & 9.18400 & Fe XIX & 9.18999 & Fe XIX &	9.19997 & Fe XX & 9.21600 & Fe XIX & 9.21999 & Fe XX & 9.22000\\
Fe XIX &9.23003 &MgXI &9.23121 &Fe XX	&9.25800 &Fe XX	&9.28100 & Mg XI	& 9.31434 & Fe XIX & 9.32001\\
Fe XX & 9.32500 & Ni XXV	& 9.33000  & Fe XIX &	9.33001 & Ne X & 9.36162 &Ni XX& 9.37700& Ni XX	 & 9.38500 \\
Fe XIX& 9.45001 & Ni XX	& 9.45500 & Fe XXI& 9.47500 & Ne X& 9.48075 & Fe XXI	& 9.48200 & Fe XXI	& 9.54200 \\
Fe XXI	& 9.54800 & Ni XX & 9.55800 & Ni XX & 9.55900 & Fe XXI	& 9.58200 & Fe XXI	& 9.58700 & Fe XIX	&9.63900\\
Fe XXI & 9.69000 & Fe XIX	& 9.69100 &Fe XXI	&9.70000& Fe XXI	&9.70500& Ne X	&9.70818& Fe XIX	&9.72600\\
Fe XXI& 9.82100 & Fe XIX	 & 9.84800 & Fe XIX & 9.85200  & Fe XIX & 9.88800 & Na XI& 10.02500 & Fe XVIII &10.08000\\
Ni XIX & 10.11000 & Ne X	& 10.23890 & Fe XVIII & 10.41000 & Fe XXIII	&10.50600& Fe XVIII	&10.52600& Fe XXIV	&10.61900\\
Fe XIX	&10.63200 &Fe XIX	&10.63300 &Fe XIX	&10.65500& Fe XIX	&10.65700& Fe XXIV	&10.66300& Fe XIX &10.68400\\
Fe XIX	&10.70200 & Cr XX	& 10.71200 & Fe XIX	& 10.74400  & Fe XIX	& 10.75800 & Fe XIX & 10.76000  &Ni XX	&10.77200 \\
Fe XIX &10.80500 &Fe XIX	&10.81600 &Fe XIX &10.82700& Fe XIX &10.88000 &Fe XIX &10.91600& Fe XIX & 10.93300\\
Fe XXIII&10.98000 &Na X	&11.00260& Fe XXIII &11.01800 &Ni XX	&11.13800 &Ni XX	&11.15800 &Ni XX	&11.22600\\
Ni XXI	&11.22700 &Ni XXI	&11.24100& Ni XXI&11.24200& Ni XXI &11.27200& Ni XX	&11.28200& Ni XXI	&11.30200\\
Ni XXI	&11.31900 &Fe XVIII	&11.32600& Ni XXI	&11.38000& Fe XVIII	&11.42000 &Fe XXII&	11.44200& Fe XXII	&11.45900\\
Ni XXI	&11.46800& Fe XXIII &11.48500& Fe XX	&11.51000& Ni XXI	&11.51600 &Ni XXI &11.51700& Fe XVIII	&11.52500\\
Ni XXI&	11.53900 &Ne IX&	11.54660& Mn XXIII	&11.57660 &Ni XXI	&11.59700& Fe XXII&	11.59900& Fe XXII	&11.66900\\
Fe XXIII&11.71800& Fe XXIII	&11.73700& Fe XX&	11.73900& Fe XXII &11.76800& Ni XX	&11.78700& Ni XX	&11.83200\\ 

Fe XXIII	&11.84600& Ni XX	&11.86500 & Ni XX	& 11.87400 & Fe XXII& 11.92100 & FeXX	& 11.93300 & Fe XXII&11.93400\\
Fe \, XXI & 11.93800 & Ni \, XX & 11.96100 & Mn \, XXII & 11.97000 & Fe \, XXI & 11.97500 & Ni \, XX & 11.97800 & Fe \, XX & 11.98700 \\
Ca\,XVIII & 11.98900 & Ni\,XX & 11.99100 & Mn\,XXII & 11.99800 & Ni\,XX & 12.00600 & Ti\,XIX & 12.01000 & Fe\,XXIII & 12.02700 \\
Fe\,XXI & 12.04400 & Ni\,XX & 12.04700 & Ni\,XX & 12.08100 & Fe\,XXI & 12.08200 & Fe\,XXI & 12.10700 & Ni\,XX & 12.11200 \\
Ni\,XX & 12.13000 & Ne\,X & 12.13390 & Fe\,XXI & 12.14600 & Ni\,XX & 12.15700 & Fe\,XXIII & 12.16100 & Ni\,XXI & 12.16500 \\
Fe\,XXII & 12.19300 & Ni\,XXI & 12.20900 & Fe\,XXI & 12.26100 & Ni\,XXI & 12.27600 & Fe\,XXI & 12.28200 & Fe\,XXI & 12.29700\\
Fe\,XXI & 12.32700 & Fe\,XXIII & 12.35100 & Fe\,XXI & 12.39500 & Fe\,XXI & 12.42200 & Fe\,XX & 12.42600 & Ni\,XIX & 12.43500\\
Fe\,XXIII & 12.44400 & Ni\,XXI & 12.44600 & Fe\,XXI & 12.46200 & Ni\,XXI & 12.47200 & Fe\,XXI & 12.49000 & Fe\,XXI & 12.49200\\
Fe\,XXIII & 12.49300 & Fe\,XXI & 12.49900 & Fe\,XXI & 12.50000 & Fe\,XXI & 12.52300 & Ni\,XXI & 12.53300 & Fe\,XX & 12.56600\\
Fe\,XXI & 12.56800 & Fe\,XX & 12.58100 & Ni\,XXI & 12.59100 & Cr\,XXII & 12.61300 & Fe\,XXI & 12.62300 & Ca\,XVIII & 12.63600\\
Ni\,XXI & 12.64800 & Fe\,XXIII & 12.65300 & Cr\,XXII & 12.65500 & Ni\,XIX & 12.65600 & Fe\,XXI & 12.66300 & Fe\,XXI & 12.69100\\
Fe\,XXI & 12.70100 & Fe\,XXIII & 12.70300 & Fe\,XXII & 12.74300 & Fe\,XX & 12.75300 & Fe\,XXI & 12.77200 & Fe\,XX & 12.80400\\
Fe\,XX & 12.81200 & Fe\,XVIII & 12.81800 & Fe\,XXI & 12.82200 & Fe\,XX & 12.82400 & Fe\,XX & 12.82700  & Fe\,XX & 12.84500\\
Fe\,XXI & 12.87000 & Fe\,XX & 12.90500 & Fe\,XIX & 12.92400 & Ni\,XX & 12.92700 & Fe\,XXII & 12.93600 & Fe\,XX & 12.95100\\
Fe\,XX & 12.96600 & Fe\,XX & 12.98200 & Fe\,XX & 12.99100 & Fe\,XIX & 13.01800 & Ni\,XX & 13.03200 & Fe\,XIX & 13.03900\\

Fe\,XX & 13.04400 & Fe\,XX & 13.04600 & Fe\,XIX & 13.05100 & Fe\,XX & 13.05200 & Fe\,XXI & 13.05200 & Fe\,XX & 13.05900\\
Fe\,XIX & 13.07500 & Ni\,XX & 13.07500 & Fe\,XX & 13.07800 & Fe\,XX & 13.09100 & Fe\,XIX & 13.09100 & Fe\,XX & 13.09100\\
Fe\,XX & 13.11400 & Fe\,XX & 13.12300 & Cr\,XXI & 13.12300 & Fe\,XX & 13.14000 & Fe\,XX & 13.14300 & Fe\,XXI & 13.17900\\
Fe\,XX & 13.18800 & Fe\,XX & 13.20300 & Fe\,XX & 13.20600 & Fe\,XIX & 13.21200 & Fe\,XIX & 13.21200 & Fe\,XX & 13.25300\\
Fe\,XX & 13.25400 & Fe\,XIX & 13.25400 & Fe\,XXI & 13.25500 & Ni\,XX & 13.25600 & Fe\,XX & 13.26700 & Fe\,XX & 13.26900\\
Fe\,XX & 13.27000 & Ni\,XX & 13.28200 & Fe\,XX & 13.29200 & Fe\,XX & 13.30100 & Fe\,XIX & 13.31100 & Fe\,XVIII & 13.31900\\
Fe\,XIX & 13.32700 & Fe\,XX & 13.33300 & Fe\,XIX & 13.33600 & Fe\,XVIII & 13.35500 & Fe\,XX & 13.35900 & Fe\,XXII & 13.36100\\
Fe\,XX & 13.36600 & Fe\,XVIII & 13.37400 & Fe\,XX & 13.37900 & Fe\,XX & 13.38200 & Fe\,XVIII & 13.39700 & Fe\,XX & 13.40200\\
Fe\,XX & 13.40500 & Fe\,XX & 13.41900 & Fe\,XVIII & 13.42400 & Fe\,XIX & 13.43000 & Ne\,IX & 13.44710 & Fe\,XIX & 13.45600\\
Ca\,XVII & 13.46000 & Fe\,XIX & 13.46200 & Fe\,XVIII & 13.46400 & Fe\,XX & 13.46700 & Fe\,XIX & 13.47100 & Fe\,XXI & 13.48200 \\
Fe\,XIX & 13.50600 & Fe\,XIX & 13.50700 & Fe\,XX & 13.51700 & Fe\,XIX & 13.52500 & Fe\,XX & 13.53300 & Fe\,XIX & 13.55400\\
Fe\,XIX & 13.55500 & Fe\,XIX & 13.55700 & Fe\,XXI & 13.57400 & Fe\,XXII & 13.61700 & Fe\,XIX & 13.62000 & Fe\,XIX & 13.62100\\
Fe\,XIX & 13.63400 & Fe\,XIX & 13.63600 & Fe\,XIX & 13.64300 & Fe\,XIX & 13.64600 & Fe\,XIX & 13.64800 & Fe\,XX & 13.67000\\
Fe\,XIX & 13.67200 & Fe\,XIX & 13.67300 & Fe\,XXII & 13.67400 & Fe\,XIX & 13.69100 & Fe\,XIX & 13.69400 & Fe\,XXI & 13.71700 \\
Fe\,XIX & 13.72000 & Fe\,XIX & 13.72100 & Fe\,XIX & 13.74100 & Fe\,XIX & 13.75400 & Fe\,XIX & 13.76200 & Fe\,XIX & 13.76900\\
Fe\,XXII & 13.77100 & Ni\,XIX & 13.77800 & Fe\,XX & 13.78100 & Fe\,XIX & 13.79200 & Fe\,XIX & 13.79900 & Fe\,XIX & 13.82200 \\
Fe\,XIX & 13.84100 & Fe\,XIX & 13.84300 & Fe\,XIX & 13.84400 & Fe\,XX & 13.84400 & Fe\,XIX & 13.87100 & Fe\,XIX & 13.87200 \\
Fe\,XIX & 13.93600 & Fe\,XIX & 13.93800 & Fe\,XIX & 13.94200 & Fe\,XIX & 13.95700 & Fe\,XVIII & 13.96200 & Fe\,XIX & 13.96400 \\
Fe\,XIX & 13.97000 & Fe\,XX & 13.97200 & Fe\,XXI & 14.00800 & Fe\,XIX & 14.01700 & Ni\,XIX & 14.04000 & Ca\,XVIII & 14.04900 \\
Ca\,XVIII & 14.05900 & Fe\,XIX & 14.07100 & Fe\,XIX & 14.08600 & Fe\,XIX & 14.11400 & Fe\,XX & 14.11500 & Fe\,XX & 14.12300 \\
Fe\,XVIII & 14.12400 & Fe\,XIX & 14.12700 & Fe\,XIX & 14.12800 & Fe\,XVIII & 14.13600 & Fe\,XVIII & 14.14400 & Fe\,XIX & 14.17500 \\
Fe\,XX & 14.19000 & Fe\,XVIII & 14.20400 & Fe\,XVIII & 14.20900 & Cr\,XXI & 14.24500 & Fe\,XX & 14.24800 & Fe\,XIX & 14.25200 \\
Fe\,XVIII & 14.25800 & Fe\,XVIII & 14.34400 & Fe\,XVIII & 14.35200 & V\,XX & 14.35960 & Fe\,XVIII & 14.37300 & Fe\,XVIII & 14.41900 \\
Fe\,XVIII & 14.45300 & Fe\,XVIII & 14.47000 & Fe\,XVIII & 14.48700 & Fe\,XX & 14.49300 & Fe\,XX & 14.50000 & Fe\,XVIII & 14.53700 \\
Fe\,XVIII & 14.55100 & Fe\,XIX & 14.57000 & Fe\,XVIII & 14.58000 & Fe\,XVIII & 14.61000 & O\,VIII & 14.63430 & Fe\,XIX & 14.66400 \\

\hline
\end{tabular}
\end{table*}
\begin{table*}
\centering{ 
\caption*{(Continued) List of spectral lines in the 1 keV blend. Line intensities vary depending on the shape of the SED, $\xi$, $N$$_{\rm H}$, and $T$.} }
\scriptsize
\begin{tabular}{c|c|c|c|c|c|c|c|c|c|c|c}
\hline
 Ion  &$\lambda$ (\AA) & Ion  &$\lambda$ (\AA) & Ion &$\lambda$ (\AA) &Ion  &$\lambda$ (\AA) &Ion  &$\lambda$ (\AA) &Ion  &$\lambda$ (\AA)\\
\hline
Fe\,XIX & 14.66900 & Fe\,XVIII & 14.67100 & Fe\,XIX & 14.69400 & Fe\,XIX & 14.73800 & Fe\,XIX & 14.74100 & Fe\,XX & 14.76400 \\
Fe\,XVIII & 14.77100 & O\,VIII & 14.82060 & Fe\,XIX & 14.86800 & Fe\,XX & 14.91300 & Fe\,XXI & 14.91600 & Fe\,XIX & 14.93200 \\
Fe\,XX & 14.93200 & Fe\,XIX & 14.93500 & Ca\,XVII & 14.94000 & Fe\,XIX & 14.99200 & Fe\,XVII & 15.01300 & Fe\,XIX & 15.04200 \\
Fe\,XX & 15.06000 & Fe\,XX & 15.06300 & Fe\,XIX & 15.08100 & Fe\,XIX & 15.11400 & Fe\,XIX & 15.16300 & O\,VIII & 15.17620 \\
Ar\,XVI & 15.19000 & Fe\,XIX & 15.19600 & Fe\,XIX & 15.20800 & Ti\,XX & 15.21100 & Ti\,XX & 15.25300 & Fe\,XVII & 15.26200 \\
Fe\,XIX & 15.33000 & Fe\,XIX & 15.35200 & Fe\,XX & 15.51500 & Fe\,XVIII & 15.62200 & Fe\,XVIII & 15.76600 & Fe\,XVIII & 15.82800 \\
Ti\,XIX & 15.86500 & Ar\,XVI & 15.93300 & Fe\,XVIII & 16.00500 & O\,VIII & 16.00590 & Fe\,XVIII & 16.02600 & Fe\,XVIII & 16.07200 \\
Fe\,XIX & 16.11000 & Fe\,XIX & 16.27200 & Fe\,XIX & 16.34000 & Fe\,XVII & 17.05100 & Fe\,XVIII & 17.62180 &Ar\,XVI & 17.73700 \\
Ar\,XVI & 17.74700 &Ca\,XVIII & 18.69100 & S\,XIV & 18.72000 & Ca\,XVIII & 18.73200 & O\,VIII & 18.96890 & Ca\,XVII & 19.55800 \\
S\,XIV & 19.69000 & N\,VII & 19.82580 & Ca\,XVII & 20.43400 & Ca\,XVI & 20.85900 & K\,XVII & 20.89900 & N\,VII & 20.90980 \\
K\,XVII & 20.93100 & Ca\,XVI & 20.95100 & Ca\,XVI & 21.02000 & Ca\,XVI & 21.11300 & Ca\,XVI & 21.45000 & O\,VII & 21.60200 \\
Ca\,XVI & 21.61000 & Ca\,XVI & 21.61900 & S\, XIV &  21.8192 & K\,XVI & 21.91100 & Ca\,XV & 22.73000 & Ca\,XV & 22.73000 \\
Ca\,XV & 22.75900 & Ca\,XV & 22.77700 & Ca\,XV & 22.82100 & S\,XIV & 23.00500 & S\,XIV & 23.01500 & Ca\,XVII & 23.17500 \\
Ca\,XVII & 23.17500 & Ar\,XVI & 23.50600 & Ar\,XVI & 23.54600 &  S\,XIII & 24.59000 & Ar\,XV & 24.73700 & N\,VII & 24.78100 \\

\hline
\end{tabular}
\end{table*}

\bibliography{sample701}{}
\bibliographystyle{aasjournalv7}



\end{document}